\documentclass[%
 reprint,
superscriptaddress,
 amsmath,amssymb,
 aps,
showkeys,
]{revtex4-2}

\usepackage{graphicx}
\usepackage{dcolumn}
\usepackage{bm}
\usepackage{hyperref}
\usepackage[T1]{fontenc}

\usepackage{scalerel,tikz}
\usetikzlibrary{svg.path}
\definecolor{orcidlogocol}{HTML}{A6CE39}
\tikzset{orcidlogo/.pic={\fill[orcidlogocol] svg{M256,128c0,70.7-57.3,128-128,128C57.3,256,0,198.7,0,128C0,57.3,57.3,0,128,0C198.7,0,256,57.3,256,128z}; \fill[white] svg{M86.3,186.2H70.9V79.1h15.4v48.4V186.2z} svg{M108.9,79.1h41.6c39.6,0,57,28.3,57,53.6c0,27.5-21.5,53.6-56.8,53.6h-41.8V79.1z M124.3,172.4h24.5c34.9,0,42.9-26.5,42.9-39.7c0-21.5-13.7-39.7-43.7-39.7h-23.7V172.4z} svg{M88.7,56.8c0,5.5-4.5,10.1-10.1,10.1c-5.6,0-10.1-4.6-10.1-10.1c0-5.6,4.5-10.1,10.1-10.1C84.2,46.7,88.7,51.3,88.7,56.8z};}}
\newcommand\orcidicon[1]{\href{https://orcid.org/#1}{\mbox{\scalerel*{
\begin{tikzpicture}[yscale=-1,transform shape]\pic{orcidlogo};
\end{tikzpicture}}{|}}}}

\usepackage{fancyvrb}
\newcommand{\mbf}[1]{\bm{#1}}
\newcommand{\del}{\nabla}
\usepackage[flushleft]{threeparttablex}
\usepackage{subcaption}
\usepackage{cprotect}
\usepackage{hyperref}
\usepackage{caption}
\newcommand{\sech}{\text{sech}}

\newcommand\aap{A\&A}                
\renewcommand\apj{ApJ}                 
\newcommand\apjl{ApJ}                
\newcommand\apjs{ApJS}               
\newcommand\araa{ARA\&A}             
\newcommand\grl{Geophys. Res. Lett.} 
\newcommand\jgr{J.~Geophys.~Res.}    
\newcommand\mnras{MNRAS}             
\renewcommand\nat{Nature}              
\renewcommand\prd{Phys. Rev.~D}        
\renewcommand\pre{Phys. Rev.~E}        
\renewcommand\prl{Phys. Rev.~Lett.}    
\newcommand\solphys{Sol.~Phys.}      
\newcommand\ssr{Space Sci. Rev.}     

\newcommand{\vecE}{\bm{E}}
\newcommand{\vecB}{\bm{B}}
\newcommand{\vecQ}{\bm{Q}}
\newcommand{\vecR}{\bm{R}}

\begin{document}

\preprint{APS/123-QED}

\title{Comparison of magnetic diffusion and reconnection in ideal and resistive relativistic magnetohydrodynamics, ideal magnetodynamics, and resistive force-free electrodynamics}

\author{Michael P. Grehan\orcidicon{0009-0003-1842-192X}} \email{michael.grehan@mail.utoronto.ca}
\affiliation{Department of Physics, University of Toronto, 60 St. George Street, Toronto, ON M5S 1A7, Canada}
\affiliation{Canadian Institute for Theoretical Astrophysics, 60 St. George Street, Toronto, ON M5S 3H8, Canada}

\author{ Tanisha Ghosal\orcidicon{0009-0009-4853-4670}}
\affiliation{Department of Physics, University of Toronto, 60 St. George Street, Toronto, ON M5S 1A7, Canada}
\affiliation{Canadian Institute for Theoretical Astrophysics, 60 St. George Street, Toronto, ON M5S 3H8, Canada}

\author{James R. Beattie\orcidicon{0000-0001-9199-7771}}
\affiliation{Department of Astrophysical Sciences, Princeton University, Princeton, NJ 08544, USA}
\affiliation{Canadian Institute for Theoretical Astrophysics, 60 St. George Street, Toronto, ON M5S 3H8, Canada}

\author{Bart Ripperda\orcidicon{0000-0002-7301-3908}}
\affiliation{Department of Physics, University of Toronto, 60 St. George Street, Toronto, ON M5S 1A7, Canada}
\affiliation{Canadian Institute for Theoretical Astrophysics, 60 St. George Street, Toronto, ON M5S 3H8, Canada}
\affiliation{D. A. Dunlap Department of Astronomy, University of Toronto, Toronto, ON M5S 3H4, Canada}
\affiliation{Perimeter Institute for Theoretical Physics, Waterloo, ON N2L 2Y5, Canada}

\author{Oliver Porth\orcidicon{0000-0002-4584-2557}}
\affiliation{Anton Pannekoek Institute for Astronomy, University of Amsterdam, Science Park 904, 1098 XH Amsterdam, The Netherlands}

\author{Fabio Bacchini\orcidicon{0000-0002-7526-8154}}
\affiliation{Centre for mathematical Plasma Astrophysics, Department of Mathematics, KU Leuven, Leuven, Belgium}
\affiliation{Royal Belgian Institute for Space Aeronomy, Solar-Terrestial Centre of Excellence, Brussels, Belgium}

\date{\today}

\begin{abstract}
    High-energy astrophysical systems and compact objects are frequently modeled using ideal relativistic magnetohydrodynamic (MHD) or force-free electrodynamic (FFE) simulations, with the underlying assumption that the discretisation from the numerical scheme introduces an effective (numerical) magnetic resistivity that adequately resembles an explicit resistivity. However, it is crucial to note that numerical resistivity can fail to replicate essential features of explicit resistivity. In this study, we compare the 1D resistive decay and 2D reconnection properties of four commonly used physical models. We demonstrate that the 1D Ohmic decay of current sheets via numerical dissipation in both ideal MHD and magnetodynamics (MD) is subdiffusive (i.e., sub-linear in time), whereas explicit resistive FFE and resistive MHD simulations match the predictions of resistive theory adequately. For low-resolution, reconnecting current sheets in 2D, we show that ideal MHD and MD have an analogue to the Sweet--Parker regime where the scaling of the reconnection rate depends directly on the resolution. At high resolutions, ideal MHD and MD have an asymptotic reconnection rate similar to resistive MHD. Furthermore, we find that guide field-balanced current sheets in ideal MHD and MD have a qualitative structure similar to that of one in resistive MHD. Similarly, a pressure-balanced current sheet in ideal MHD is found to have a qualitative structure similar to that of one in resistive MHD. For a guide field-balanced sheet, resistive FFE is found to have a nearly identical Sweet--Parker regime compared to resistive MHD and a similar asymptotic reconnection rate for large enough Lundquist numbers, but differs in the timescale for reconnection onset in the asymptotic regime. We discuss the implications of our findings for global simulations. 
\end{abstract}


\maketitle


\section{Introduction}\label{sec:intro}
Relativistic magnetohydrodynamic (MHD) simulations have become an essential tool in high-energy astrophysics for modeling the behavior of plasmas in a variety of astrophysical environments, ranging from the dynamics of accretion disks around black holes (e.g. \cite{Porth-short:2019ApJS..243...26P, white:2019ApJ...874..168W, Ripperda:2020ApJ...900..100R,Ripperda_2022}) to the evolution of magnetized outflows in stellar and galactic contexts (e.g. \cite{ 2010ApJ...709.1100P, 2016MNRAS.456.1739B, 2023A&A...679A..49M, 2022A&A...659A.139M, 2022ApJ...933L...9G, 2024ApJ...964...79L}). Similarly, force-free electrodynamics (FFE) has been used to perform global simulations of magnetospheres (e.g. \cite{2006ApJ...648L..51S, Timokhin_2006, 2019MNRAS.487.4114Y, Mahlmann2022, 2023MNRAS.524.6024S}) and black hole mergers (e.g. \cite{Alic2012_FFR}). Many of these global simulations often assume ideal frameworks, i.e. lack any explicit non-ideal, dissipative operators in the models, where the plasma is treated as a perfectly conducting fluid, or in the case of FFE, the plasma inertia is entirely neglected. Hence, any dissipation occurs only at a small kernel of the order of the grid scale due to numerical discretisation (typically of the order of 10 grid cells
for second-order spatial reconstruction methods; \cite{Kriel2022_turbulence_relation,Grete2023_transfer_function_dissipation,Shivakumar2025_numerical_dissipation}) of the model. This approach implicitly introduces an effective numerical resistivity, with an unknown diffusion operator, which is widely used to approximate the effects of explicit resistivity without explicitly solving the resistive equations.

Resistive processes play a fundamental role in astrophysical plasmas, influencing phenomena across a broad range of energies. In Newtonian settings characterized by low magnetization, where the rest mass energy of the plasma greatly exceeds the energy in the magnetic field, magnetic reconnection is a pivotal mechanism underpinning the dynamics of solar flares \cite{Sweet:1958IAUS....6..123S, Parker:1963ApJS....8..177P, 1976SoPh...50...85K}, geomagnetic storms \cite{2024GeoRL..5112730L}, the structure of the geomagnetic tail \cite{1981JGR....86.6802B, 1985GeoRL..12..105L}, and MHD turbulence \cite{2024arXiv240516626B}. In the relativistically magnetized regime, where the magnetic energy greatly exceeds the rest mass energy of the plasma, reconnection is integral to the behavior of plasmas in the equatorial current sheets of rotating neutron stars \cite{2022ARA&A..60..495P,2013MNRAS.435L...1T}, in accretion flows around black holes \cite{ Ripperda:2020ApJ...900..100R, 2021MNRAS.508.1241C, Ripperda_2022, 2022MNRAS.513.4267N, 2025ApJ...979..199S}, and has even been proposed as a potential mechanism behind fast radio bursts \cite{2020ApJ...897....1L,2022ApJ...932L..20M} and fast radio transients \cite{2023ApJ...956L..33M}. There are two distinct regimes of magnetic reconnection which we explore, parametrized by the Lundquist number
\begin{equation}
    S = \frac{4\pi v_{\rm{A}} \Delta}{\eta c^2},\label{eq:lundquist}
\end{equation}
where $v_{\rm{A}}$ is the Alfv\'{e}n speed \cite{Alfven:1942Natur.150..405A}, $\Delta$ is the current sheet half length, $\eta$ is the resistivity of the plasma, and $c$ is the speed of light.
In the Sweet--Parker regime \cite{Sweet:1958IAUS....6..123S, Parker:1963ApJS....8..177P}, the reconnection rate, i.e. the speed ratio between the inflow to the outflow of the current sheet, is directly dependent on the Lundquist number,
\begin{equation}
    v_{{\rm{rec}}} = \frac{v_{\rm{in}}}{v_{\rm{out}}}  = \sqrt{S^{-1}}, \label{eq:vrec_SP}
\end{equation}
where $v_{\rm{in}}$ and $v_{\rm{out}}$ are the inflow and outflow speeds respectively. This result holds in the relativistic regime \cite{Lyubarsky_2005MNRAS}. In the asymptotic, plasmoid-dominated regime where the Lundquist number exceeds the critical threshold of $S \gtrsim 10^4$, the reconnection rate becomes independent of the Lundquist number and plateaus at a value of $v_{{\rm{rec}}} = \mathcal{O}(0.01)$ in MHD \cite{Biskamp_1993, Lapenta.PhysRevLett.100.235001, 2007PhPl...14j0703L, Bhattacharjee.2009PhPl...16k2102B, Cassak.PhysRevLett.95.235002, Huang_2010, Uzdensky_2010, Loureiro_2012, Ni2012, Huang.10.1063/1.4802941, Takamoto_2013, Comisso.10.1063/1.4918331, Loureiro_2016, Huang_2017, Comisso_2018}. We expect realistic Lundquist numbers in astrophysical plasmas to be in the asymptotic regime, without however restricting the value of $v_{{\rm{rec}}}$ to MHD expectations, due to beyond-MHD effects occurring in real astrophysical systems. 

Resistive processes also play a key role in particle energization by facilitating the dissipation of magnetic field energy into thermal and kinetic energy \cite{2025arXiv250602101S, 2025arXiv250808533R}, and the non-thermal acceleration of particles \cite{2001ApJ...562L..63Z, 2012SSRv..173..521H,2014PhRvL.113o5005G, 2014ApJ...783L..21S, Werner2017ApJ, 2024SSRv..220...43G}. This is often associated with the formation of power-law energy distributions, characteristic of accelerated particle populations, as observed in relativistic astrophysical systems such as pulsar-wind nebulae, blazer jets, black hole flares, and gamma ray bursts \cite{2003astro.ph.12347L, 2011ApJ...741...39S, 2011MNRAS.410..381B, 2013ApJ...770..147C, 2014PhPl...21e6501C, 2019ApJ...886..122C}. A key open question in reconnection-driven particle acceleration is the relative importance of non-ideal electric fields, necessitating a deeper understanding of how resistivity governs energy dissipation \cite{2025arXiv250100979G,2003ApJ...586...72L}. This non-ideal electric field is absent in ideal MHD and magnetodynamics (MD), hence it is essential to understand how these models represent the heating of plasma by reconnection. 

FFE \cite{1999astro.ph..2288G, 2006astro.ph..4364G, McKinney_2006MNRAS.367.1797M, 2012MNRAS.423.1416P, 2013PhRvD..88j4031P, 2017CQGra..34u5001E, Mahlmann2021, Mahlmann2022} and MD \cite{Komissarov2002, Komissarov2004c, 2007A&A...473...11D} are equivalent descriptions of the infinite magnetization limit of ideal MHD, where the plasma inertia is entirely neglected and the dynamics are described solely by the evolution of the electric and magnetic field. FFE and MD differ in the choice of how the system is numerically evolved: FFE solves
Faraday’s law and Amp\'{e}re’s law to evolve the magnetic
field, $\bm{B}$, and the electric field, $\bm{E}$; while MD solves Faraday’s law and the momentum equation for the Poynting
flux to evolve the magnetic field, $\bm{B}$, and the Poynting
flux, $\bm{S} = \bm{E}\times\bm{B}$.
These models are necessary because MHD becomes numerically unstable at very high magnetization \cite{2006ApJ...641..626N, 2018ApJ...859...71S, Ripperda:2019lsi, 2021PhRvD.103b3018K}.
Modeling the infinite magnetization limit is appropriate in astrophysical systems where the magnetization is very high (e.g. neutron star magnetospheres and jets), but fails to capture dissipation where the magnetization drops and the plasma heats up (e.g. magnetic reconnection).
In this work, we investigate the nature of numerical resistivity in ideal MHD and MD simulations and compare with explicit resistive prescriptions in MHD and FFE. We focus on the impact of numerical resistivity on the evolution of thin current sheets -- a controlled numerical experiment representative for most of the astrophysical reconnection-driven processes mentioned above. 
Our analysis demonstrates that numerical resistivity leads to subdiffusive behavior in the decay of current sheets, with implications for the stability and evolution of astrophysical plasmas in simulations. As well, we show that ideal MHD and MD demonstrate an analogue to the Sweet--Parker regime at low resolutions, and an analogue to the asymptotic regime corresponding to high-resolution simulations. Through direct comparison with resistive simulations, where the 
reconnection rate is well-resolved, 
we highlight the importance of carefully considering the role of numerical dissipation in ideal models.

Comparisons between the resistive FFE scheme considered in this paper \cite{Alic2012_FFR,Ripperda2021_FFR} and ideal MHD and other resistive FFE schemes, which use alternative approaches to damp FFE violations, have been performed in the past \cite{Mahlmann2021}. In our analysis, we additionally compare to resistive MHD, clearly identifying how and where ideal MHD, MD, and resistive FFE fail to reproduce physical resistive processes. Our analysis is agnostic to the effective diffusive operator present due to numerical dissipation. We consider tests of resistive processes for which the evolution of quantities can be measured and compared to the resistive predictions. This differs from previous work which assumed a particular form of the numerical dissipation operator, typically a Laplacian operator in the induction equation $\sim \eta_{\rm{num}} \del^2 B$ (e.g. \cite{2025MNRAS.536.1268K}).

The asymptotic plasmoid-unstable regime of reconnection has been previously identified in ideal MHD in the study of black hole accretion disks (e.g. \cite{Nathanail_2020,Chashkina_2021,Ripperda_2022,Nathanail_2022,salas2024resolutionanalysismagneticallyarrested}) and in ideal MHD and FFE in the study of neutron star magnetosphere (e.g. \cite{2012ApJ...754L..12P, 2016MNRAS.457.3384T, 2023ApJ...947L..34M, 2023MNRAS.524.6024S}).
We carefully study this regime with ideal MHD and MD methods, and compare to resistive MHD and FFE schemes. In the ideal cases (i.e. without explicit resistivity), we progressively increase numerical resolution to achieve smaller numerical resistivity and reach the asymptotic reconnection regime.
We specifically diagnose the reconnection rate, diffusion rate, time scales, and visual appearance and resolution of the current sheets.

This paper is organized as follows: In \autoref{sec:Numerical Simulations and Models} we briefly describe the four physical models which will be compared, and in \autoref{sec:initial conditions}, we present the initial conditions of the setups used to simulate magnetic diffusion and reconnection. In \autoref{sec:Comparison of Magnetic Diffusion}, we outline the two tests of magnetic diffusion: (1) the decay of the current density amplitude; and (2), the growth of the second moment of the current density. These two tests are then compared with resistive MHD theory to study the accuracy of the four physical models. In \autoref{sec:Comparison of Magnetic Reconnection}, we outline the expected scaling with resistivity and the asymptotic regime of the reconnection rate in a guide field-supported current sheet and in a pressure-balanced current sheet. Lastly, in \autoref{sec:Discussion} we summarize the findings of this paper and consider the implications for global simulations.

\section{Numerical Simulations and Models}\label{sec:Numerical Simulations and Models}

All simulations in this study are performed using the Black Hole Accretion Code (\Verb+BHAC+) \cite{Porth:2016rfi, Olivares2019} which is capable of solving the multidimensional general-relativistic magnetohydrodynamic equations in both ideal and resistive \citep{Ripperda:2019lsi} forms. As well, \Verb+BHAC+ is able to perform ideal magnetodynamic and resistive force-free electrodynamic simulations \cite{Ripperda2021_FFR} using the scheme described in \cite{Alic2012_FFR} and using the constrained transport method to evolve the magnetic flux and preserve $\bm{\del} \cdot \mbf{B}=0$ to machine precision \cite{Olivares2019}, where $B$ is the magnetic field. We present, for the first time, the magnetodynamic scheme in \verb+BHAC+. In our scheme, we follow the reasoning of \cite{McKinney_2006MNRAS.367.1797M} and limit the drift velocity before it becomes superluminal. The physical motivation for this choice is that plasma inertia (and thus effects not modeled within FFE) must eventually become important and limit the achievable Lorentz factor of the plasma. Detailed descriptions of the magnetodynamic and resistive force-free electrodynamic schemes are presented in \autoref{sec:md} and \autoref{sec:ffr}. In total, we use four different models, which we outline in the following subsections. Note that all simulations use a third-order spatial reconstruction scheme but the method remains overall second-order accurate in space, and a second-order time integration method. We discuss the effects of the order of the spatial reconstruction scheme and the time integration method in \autoref{app:order}, where we show that results in the main text are robust to a change in order of the spatial reconstruction scheme or the time integration method.

In \autoref{sec:Numerical Simulations and Models}, we write our models and initial conditions in units with $c=k_B=1$, $\bm{B}/\sqrt{4\pi} \rightarrow \bm{B}$, $\bm{E}/\sqrt{4\pi} \rightarrow \bm{E}$, $\sqrt{4\pi} \bm{J} \rightarrow \bm{J}$, $\sqrt{4\pi} q \rightarrow q$, and $\eta/4\pi \rightarrow \eta$, where $\bm{E}$ is the electric field, $\bm{J}$ is the current density, $q$ is the charge density, and $k_B$ is the Boltzmann constant. All sections following this one use cgs units. Note that we use Latin indices to indicate spatial indices (e.g. $i=1,2,3$) and Greek indices to denote spacetime indices (e.g. $\mu=0,1,2,3$). For completeness, we state the general models in the 3+1 Arnowitt--Deser--Misner (ADM) formalism \cite{1962gicr.book..227A}, with metric tensor
\begin{equation}\label{eq:metric_tensor}
    g^{\mu \nu} = \left(  
    \begin{array}{cc}
        -1/\alpha^2 & \beta^j/\alpha^2 \\
        \beta^j/\alpha^2 & \gamma^{ij} - \beta^i \beta^j/\alpha^2 
    \end{array}
    \right),
\end{equation}
where $\alpha$ is the lapse function, $\beta^i$ is the shift vector, and $\gamma^{ij}$ is the emergent spatial metric on the spacelike hypersurfaces. The simulations presented in this paper are in flat Cartesian Minkowski spacetime with $\alpha =1$, $\beta^i = 0$, and $\gamma^{ij} = \delta^{ij}$.  

Each of the four models can be written in conservative form,
\begin{equation}
    \partial_t \left( \sqrt{\gamma} \bm{U} \right) + \partial_i \left( \sqrt{\gamma} \bm{F}^i \right) = \sqrt{\gamma} \bm{S},
\end{equation}
with conserved variables $\bm{U}$, conserved fluxes $\bm{F}$, and non-conserved fluxes including source terms, $\bm{S}$. We will list $\bm{U}$,  $\bm{F}$, and  $\bm{S}$ for each of the models in the following subsections.

\subsection{Ideal and Resistive Relativistic Magnetohydrodynamics (RMHD)} 

The magnetohydrodynamic formulations in \verb+BHAC+ solve the multidimensional general-relativistic MHD equations in both ideal and resistive forms. The resistivity is introduced via a relativistic resistive Ohm's law and requires the evolution of the full resistive electric field. 

\subsubsection{Ideal MHD}

For the ideal formulation, the conserved variables are defined as
\begin{align}
    \bm{U} & = \left[ 
    \begin{array}{c}
        D   \\
        S_j  \\
        \tau \\
        B^j  \\
    \end{array}
    \right],  
\end{align}
where $D = \Gamma \rho$ is the proper mass density, $\Gamma$ is the bulk Lorentz factor, $\rho$ is the mass density, and $\tau$ is the (rescaled) conserved energy density $\tau = U - D$.
The energy density is
\begin{equation}
    U = \rho h \Gamma^2 - p + \frac{1}{2} \left( B^2(1 + v^2) - (B^j v_j)^2 \right),
\end{equation}
where $v^i$ is the spatial three-velocity, $p$ is the gas pressure, $B^2 = B_i B^i$, and
the covariant momentum density is
\begin{equation}
    S_i = \rho h \Gamma^2 v_i + B^2 v_i - (B^j v_j) B_i.
\end{equation}
The fluxes are defined as
\begin{align}
    \bm{F}^i & = \left[ 
    \begin{array}{c}
        \mathcal{V}^i D\\
        \alpha W^i_j - \beta^i S_j\\
        \alpha (S^i - v^i D) - \beta^i \tau\\
        \mathcal{V}^i B^j - B^i \mathcal{V}^j
    \end{array}
    \right], 
\end{align}
where we define the transport velocity $\mathcal{V}^i = \alpha v^i - \beta^i$.
The sources are defined as
\begin{equation}
    \bm{S} = \left[ 
    \begin{array}{c}
        0\\
        \frac{1}{2} \alpha W^{ik} \partial_j \gamma_{ik} + S_i \partial_j \beta^i - U \partial_j \alpha\\
        \frac{1}{2} W^{ik} \beta^j \partial_j \gamma_{ik} + W^j_i \partial_j \beta^i - S^j \partial_j \alpha\\
        0^j\\
    \end{array}
    \right].
\end{equation}
The spatial variant part of the stress--energy tensor is
\begin{equation}
\begin{split}
    W^{ij} = S^i v^j + p_{\rm{tot}} \gamma^{ij} - \frac{B^i B^j}{\Gamma^2} - (B^k v_k) v^i B^j,
\end{split}
\end{equation}
where
\begin{equation}
    p_{\rm{tot}} = p +b^2/2,
\end{equation}
with $b^2 = B^2 - E^2$ the
square of the fluid frame magnetic field strength, and $E^i = - \gamma^{1/2} \epsilon^{ijk} v_j B_k$ is the ideal electric field and where $\epsilon^{ijk}$ is the Levi--Civita tensor. The specific enthalpy is
\begin{equation}
    h = 1 + \frac{\hat{\gamma}}{\hat{\gamma}-1} \frac{p}{\rho},
\end{equation}
where $\hat{\gamma}$ is the adiabatic index.
See \cite{Porth:2016rfi, Olivares2019} for further details on the ideal model and numerical scheme.

\subsubsection{Resistive MHD}
For the resistive formulation, the conserved variables are defined as
\begin{align}
    \bm{U} & = \left[ 
    \begin{array}{c}
        D   \\
        S_j  \\
        \tau \\
        B^j  \\
        E^j
    \end{array}
    \right],  
\end{align}
and fluxes are defined as
\begin{align}
    \bm{F}^i & = \left[ 
    \begin{array}{c}
        \mathcal{V}^i D\\
        \alpha W^i_j - \beta^i S_j\\
        \alpha (S^i - v^i D) - \beta^i \tau\\
        \beta^j B^i - \beta^i B^j + \gamma^{-1/2} \epsilon^{jik} \alpha E_k\\
        \beta^j E^i - \beta^i E^j - \gamma^{-1/2} \epsilon^{jik} \alpha B_k
    \end{array}
    \right].
\end{align}
The sources are defined as
\begin{equation}
    \bm{S} = \left[ 
    \begin{array}{c}
        0\\
        \frac{1}{2} \alpha W^{ik} \partial_j \gamma_{ik} + S_i \partial_j \beta^i - U \partial_j \alpha\\
        \frac{1}{2} W^{ik} \beta^j \partial_j \gamma_{ik} + W^j_i \partial_j \beta^i - S^j \partial_j \alpha\\
        0^j\\
        -\alpha J^i + \beta^j q
    \end{array}
    \right].
\end{equation}
The current density is found from the relativistic resistive Ohm's law \cite{1993PhRvL..71.3481B, 2003ApJ...589..893L, 2007MNRAS.382..995K}
\begin{equation}
    J^i = q v^i + \frac{\Gamma}{\eta} \left[ E^i + \gamma^{-1/2} \epsilon^{ijk} v_j B_k - (v_k E^k) v^i \right],
\end{equation}
the covariant momentum density is
\begin{equation}
    S_i = \rho h \Gamma^2 v_i + \gamma^{1/2} \epsilon_{ijk} E^j B^k,
\end{equation}
the energy density is
\begin{equation}
    U = \rho h \Gamma^2 - p + \frac{1}{2} \left( E^2 + B^2 \right),
\end{equation}
the spatial variant of the stress--energy tensor is
\begin{equation}
\begin{split}
    W^{ij} = & \rho h \Gamma^2 v^i v^j - E^i E^j - B^i B^j \\ & + \left[ p + \frac{1}{2} \left( E^2 + B^2 \right) \right] \gamma^{ij},
\end{split}
\end{equation}
and the charge density is
\begin{equation}
    \gamma^{-1/2} \partial_i \left( \gamma^{1/2} E^i \right) =q,
\end{equation}
which is not evolved nor conserved here. In \cite{2019MNRAS.486.4252M} the charge density is conserved from evolving the electric field with the constrained transport formalism which conserves $\bm{\del}\cdot \bm{E}=q$.
See \citep{Ripperda:2019lsi} for further details regarding the resistive model and numerical scheme. Note that the ideal limit is achieved by replacing $E^i$ in the above with $E^i = - \gamma^{1/2} \epsilon^{ijk}v_j B_k$.

\subsection{Magnetodynamics (MD)}

The magnetodynamic formulation in \verb+BHAC+ solves Faraday’s law and the momentum equation for the Poynting flux to evolve the magnetic field, $\mbf{B}$, and the Poynting flux, $\mbf{S} = \mbf{E}\times \mbf{B}$ \cite{Komissarov2002, Komissarov2004c, 2007A&A...473...11D}. Note, this is equivalent to evolving $\bm{E}$ and $\bm{B}$. An advantage of the $(\mbf{S}, \mbf{B})$ formulation presented here, over the one in terms of $(\mbf{E}, \mbf{B})$, is that the constraint $\mbf{E}\cdot\mbf{B}=0$ is already satisfied to machine precision. However, regions of $E^2 - B^2 >0$ can be encountered during the evolution and should be monitored, either to trigger controlled termination or to issue fixes avoiding numerical breakdown.
In our scheme, we follow the reasoning of \cite{McKinney_2006MNRAS.367.1797M} and limit the drift velocity before it becomes superluminal. The physical motivation for this choice is that plasma inertia (and thus effects not modeled within FFE) must eventually become important and limit the achievable Lorentz factor of the plasma.

The conserved variables $\bm{U}$ and fluxes $\bm{F}$ are defined as  
\begin{align}
    \bm{U} = \left[ 
    \begin{array}{c}
        S_j  \\
        B^j 
    \end{array}
    \right],  
    &&
    \bm{F}^i = \left[ 
    \begin{array}{c}
        \alpha W^i_j - \beta^i S_j  \\
        \mathcal{V}^i B^j - B^i \mathcal{V}^j 
    \end{array}
    \right], 
\end{align}
where we define the transport velocity $\mathcal{V}^i = \alpha v^i - \beta^i$. The sources are defined as
\begin{equation}
    \bm{S} = \left[ 
    \begin{array}{c}
        \frac{1}{2}\alpha W^{ik} \partial_j \gamma_{ik} + S_i \partial_j \beta^i - U \partial_j \alpha  \\
        0^j
    \end{array}
    \right],
\end{equation}
with energy density
\begin{align}
    U & = \frac{1}{2}\left( E^2 + B^2 \right) = \frac{1}{2}\left( \frac{S^2}{B^2} + B^2 \right),
\end{align}
where we have used that $\mbf{E}\perp \mbf{B}$ in ideal magnetodynamics and hence $S^2/B^2 = E^2$. Similarly, because the velocity in the MD regime is the $\mbf{E}\times\mbf{B}$ drift velocity, $E^2 = v^2 B^2$. The three-energy momentum tensor $W^{ij}$ is defined as
\begin{align}
    W^{ij}
    & = \frac{S^i S^j}{B^2} - \frac{B^i B^j}{\Gamma^2} + p_{\rm{EM}} \gamma^{ij}
\end{align}
where we have introduced the total pressure $p_{\rm{EM}}$
\begin{equation}
    p_{\rm{EM}} = \frac{1}{2} \left( B^2 - E^2 \right) = \frac{B^2}{2\Gamma^2},
\end{equation}
and the Lorentz factor which can be written as
\begin{equation}
    \Gamma^2 = \frac{B^4}{B^4 - S^2}.
\end{equation}
Further details are presented in \autoref{sec:md}.

\subsection{Resistive Force-free Electrodynamics (FFE)}

The lack of an explicit resistivity in FFE is a potential issue when modeling any resistive layer, as detailed in \autoref{sec:intro}. Physical modeling of these layers requires either current prescriptions or sub-grid models which evolve the current parallel to the magnetic field \cite{Mahlmann2021}. An attempt to remedy this problem has been to introduce an explicit resistivity, which in the ideal limit, $\eta \rightarrow 0$, reduces to ideal FFE (e.g. \cite{Lyutikov2003, gruzinov2007dissipativestrongfieldelectrodynamics,Li2012,Parfrey2017, Alic2012_FFR}). We will refer to this method as resistive FFE.

The resistive FFE formulation in \verb+BHAC+ \cite{2021JPlPh..87f9014T, Ripperda2021_FFR} solves Faraday’s law and Amp\`{e}re's law to evolve the magnetic field, $\mbf{B}$, and the electric field, $\mbf{E}$.
The conserved variables $\bm{U}$ and fluxes $\bm{F}$ are defined as
\begin{align}
    & \bm{U} = \left[ 
    \begin{array}{c}
        B^j  \\
        E^j 
    \end{array}
    \right],  \\
    & \bm{F}^i = \left[ 
    \begin{array}{c}
       \beta^j B^i - \beta^i B^j + \gamma^{1/2} \epsilon^{jik} \alpha E_k  \\
        \beta^j E^i - \beta^i E^j - \gamma^{1/2} \epsilon^{jik} \alpha B_k 
    \end{array}
    \right], 
\end{align}
the sources are defined as
\begin{equation}
    \bm{S} = \left[ 
    \begin{array}{c}
        0^j  \\
        -\alpha J^j + \beta^j q
    \end{array}
    \right].
\end{equation}
The model and scheme use the current prescription presented in \cite{Alic2012_FFR},
\begin{equation}
\begin{aligned}
    J^i = & q  \frac{\gamma^{1/2} \epsilon^{ijk} E_j B_k}{B^l B_l} \\
    & + \frac{1}{\eta} \left[ E^l B_l \frac{B^i}{B^l B_l} + \Theta(E^l E_l - B^l B_l)\frac{E^i}{B^l B_l} \right]
    \end{aligned}
    \label{eq:JFF}
\end{equation}
where $\eta$ is an explicit resistivity and $\Theta$ is the Heaviside function which acts to damp the electric field in regions in which $E^2>B^2$ on a resistive timescale. Notice $J^i_\parallel$, the resistive current parallel to the magnetic field present in this formulation which acts to resistively decay violations to the FFE condition $\bm{E}\cdot\bm{B}=0$,
\begin{equation}
    J^i_\parallel  =\frac{1}{\eta} \left[ E^l B_l \frac{B^i}{B^l B_l} \right].
\end{equation}
Further details are presented in \autoref{sec:ffr}.

\subsection{Initial Conditions}\label{sec:initial conditions}

\begin{figure*}[htbp]
    \centering
    \includegraphics[width=\textwidth]{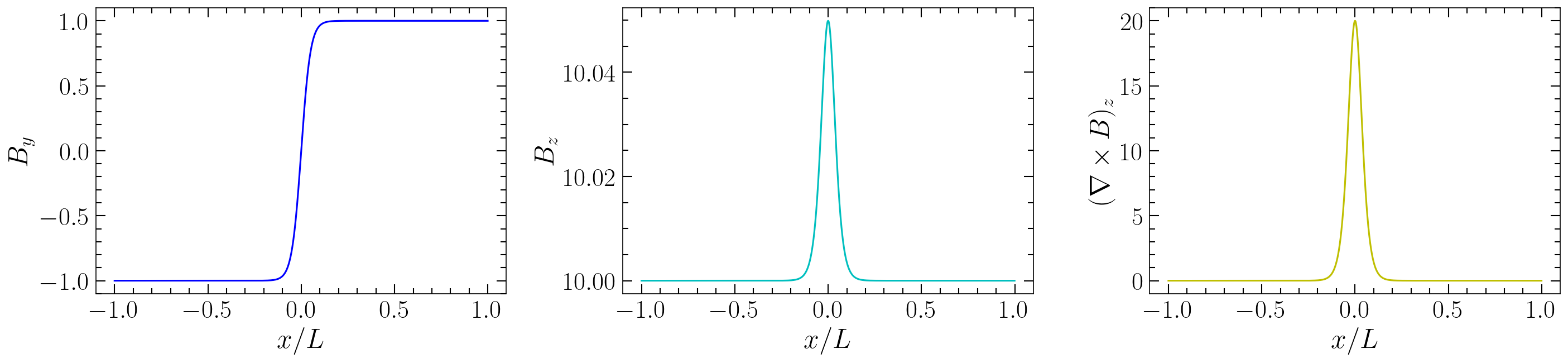}
    \caption{guide field-balanced Harris sheet: upstream field (left), $B_y$, the guide field (middle), $B_z$, the $z$ component (right) of $\bm{\nabla} \times \mbf{B}$, for a configuration with $B_0 =1$, $B_{\rm{guide}}=10$, and $a=0.05$. }
    \label{fig:magfields}
\end{figure*}

\begin{figure*}[htbp]
    \centering
    \includegraphics[width=\textwidth]{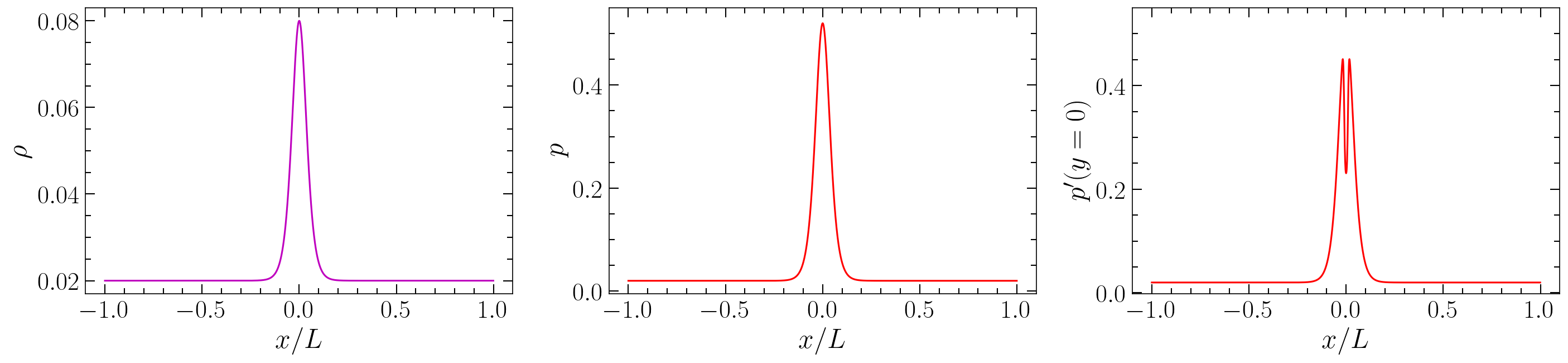}
    \caption{Pressure-balanced Harris sheet: density profile (left), $\rho$, the pressure profile (middle), $p$, and the perturbed pressure profile at $y=0$ (right), $p^\prime (y=0)$, for a configuration with $B_0 =1$, $a=0.05$, $\sigma =10$, and $T=0.001$. Note that the pressure pinch has a finite spatial extent in $y$. }
    \label{fig:pressure_profile}
\end{figure*}

    We consider a flat-spacetime metric (i.e. $\alpha =1$, $\beta^i = 0$ and $\gamma^{ij} = \delta^{ij}$ in \autoref{eq:metric_tensor}), and a plasma which is assumed to be a relativistic ideal gas with adiabatic index $\hat{\gamma}=4/3$. The 3-vector velocity components are initialized as zero in all directions, $\mbf{v}=(0,0,0)$. An initial upstream magnetization,
    \begin{equation}
        \sigma_{\text{hot}} = \frac{B_0^2 + B_{\text{guide}}^2 }{\rho \left(  1 + \displaystyle\frac{\hat{\gamma}}{\hat{\gamma} - 1} T \right)}, \label{eq:sigmahot}
    \end{equation}
    where $B_0$ is the upstream amplitude of the in-plane field and $B_{\rm{guide}}$ is the upstream value of the guide field is chosen, Similarly, the temperature, $T$, is chosen which in turn sets the initial upstream density and the upstream pressure, $p = \rho T$.

    The initial conditions for all the numerical simulations are that of a single thin current sheet with $B_y$ which reverses direction, in either a strong guide field balance in static force-free equilibrium (\autoref{sec:ffeHarris} or a pressure-balanced sheet (that we perturb out of equilibrium) with no $B_{\text{guide}}$ (\autoref{sec:pressureHarris}). When unperturbed, the guide field configuration decays Ohmically $\sim \eta\nabla^2\bm{B}$, and in \autoref{sec:Maximum Current Amplitude Test} we show that the relativistic equations reduce to the Newtonian resistive induction equation in this configuration.  When perturbed via a pinch (a localized removal at the origin) in $B_{\text{guide}}$ (\autoref{sec:guidefield_perturbed}), the current sheet undergoes reconnection. The pressure-balanced sheet is perturbed via a pressure pinch and undergoes reconnection. The exact expressions of the perturbations are given in the following subsections. 

    The Ohmic decay simulations are ``1.75D'' meaning $\partial_y=0$ and $\partial_z=0$ but vectors may have non-zero $y$- and $z$-components. All reconnection simulations are ``2.5D'', where $\partial_z=0$ but vectors may have a non-zero $z$-component. Both the pressure and guide field perturbation results in magnetic reconnection occurring in the sheet, breaking the symmetry in the $y$-direction.

\subsubsection{Guide Field-balanced Harris Sheet}\label{sec:ffeHarris}
     The $y$-component of the magnetic field is initialized through a vector potential,
    \begin{equation}
        A_{z} = - a B_0 \log\left[\cosh\left(\frac{x}{a}\right)\right],
    \end{equation}
    where $a$ is the initial thickness of the sheet. Initializing the in-plane field through a vector potential ensures that the magnetic field initially has zero divergence. Note that the current sheet is initialized at the origin. The magnetic field component corresponding to this vector potential is given by $\mbf{B} = \bm{\del} \times \mbf{A}$, 
    \begin{equation}
        B_y = B_0 \tanh \left( \frac{x}{a} \right). \label{eq:By0_sheet}
     \end{equation}
    The $z$-component of the magnetic field is initialized to begin the simulation in force-free equilibrium in the strong guide field regime
    \begin{equation}
        B_z = \sqrt{ B_{\text{guide}}^2 + B_0^2 - B_y^2  },
    \end{equation}
    where $B_{\text{guide}}>B_0$ is a constant and the value of $B_z$ upstream. A visual representation of the magnetic field configuration is shown in \autoref{fig:magfields}. For the guide field configuration, the density and pressure are uniform throughout the entire domain.
    
    We repeat the same initialization for the resistive setups, except we further initialize the electric field 3-vector, $\mbf{E}=(0,0,0)$, and the resistivity, $\eta$, which is constant and uniform. 
    
    Three resistivity values are chosen for our analysis of magnetic diffusion, $\eta = 10^{-3}, 10^{-4}$, and $10^{-5}$. The simulations with $\eta = 10^{-3}$ and $10^{-5}$ are for a fixed resolution (4096 grid cells)  to fully resolve the decay, while the simulations with $\eta = 10^{-4}$ are for a range of resolutions (16--4096 grid cells) to test how the magnetic decay is affected by being under-resolved. Similarly, six uniform resolutions are chosen, ranging from low resolution (128 grid cells) with strong numerical diffusion to high resolution (4096 grid cells) with weak numerical diffusion. All simulations have domains $-1\leq x/L \leq 1$, where $L=1$ in code units, $a/L=0.02$, $B_0=1$, $B_{\rm{guide}}=10$, $\sigma_{\rm{hot}}=10$, and $T=1.0$, with continuous boundaries, where quantities are extrapolated in the boundary cells.
    
    A complete list of the 1.75D simulations which use the initial conditions described in this section to model the Ohmic diffusion of a thin current sheet is presented in \autoref{tab:Diffusion_Sims}.

\subsubsection{Perturbed Guide Field-balanced Harris Sheet}\label{sec:guidefield_perturbed}

To investigate the reconnection rate in a guide field-balanced Harris sheet \cite{1962NCim...23..115H}, the initial force-free equilibrium is perturbed by pinching the guide field (a localized removal at the origin), violating equilibrium and initializing magnetic reconnection. The setup described in \autoref{sec:ffeHarris} is perturbed with
\begin{equation}
\begin{split}
    B_z & =  \sqrt{ B_{\text{guide}}^2 + B_0^2 - B_y^2  } \\
    & \times\left[ 1 + 0.25 \left( \tanh(200 y - 10) + \tanh(-10 -200 y) \right) \right. \\ & \left.  \times\left( \tanh(200 x +2) + \tanh(2 - 200 x) \right) \right].
\end{split}
\end{equation}

Nine resistivity values are chosen for our analysis of magnetic reconnection in a strong guide field, ranging from the low in-plane Lundquist number ($S_{\perp}\sim 10^2$) Sweet--Parker regime to the high in-plane Lundquist number ($S_{\perp}\sim 10^7$) asymptotic plasmoid regime. 
The in-plane Lundquist number is defined as
\begin{equation}
    S_\perp = \frac{4\pi v_{\rm{A},\perp} \Delta }{\eta c^2}, \label{eq:Sperp}
\end{equation}
where $v_{\rm{A},\perp}$ is the in-plane Alfv\'{e}n speed. We discuss reconnection in a guide field in more detail in  \autoref{sec:Guide Field-balanced Reconnection}.
Similarly, nine resolutions are chosen for the ideal simulations, ranging from the low resolution (32 grid cells per sheet half-length, analogous to Sweet--Parker) to the high resolution (8192 grid cells per sheet half-length, analogous with a resolution-independent reconnection rate). An additional high resolution MD simulation with 16384 cells per sheet half-length is run to confirm the asymptotic behavior of the reconnection rate. All simulations have domains  $-1 \leq x/L \leq 1$ and $-2 \leq y/L \leq 2$, where $L=1$ in  code units, $a/L = 0.02$, sheet half-length of $\Delta/L = 2$, $B_0 = 1$, $B_{\rm{guide}} = 2$, $\sigma_{\rm{hot}} = 10$, and $T = 1.0$ with open boundaries. Note that open boundaries are used in the reconnection setups to avoid the formation of large plasmoids, which allows reconnection to proceed for a longer time and for a cleaner measure of the reconnection rate. 

To minimize the computational cost of high resolution ideal MHD simulations of magnetic reconnection we use fixed mesh refinement levels in space, not adaptive or dynamic, over 8 initial sheet widths, or 2 initial sheet widths for the highest resolution MD simulations, ensuring that the reconnection layer is within the highly refined region. To test this method, we consider the three simulations in \autoref{sec:num_L}, two with and one without mesh refinement.

A complete list of the 2.5D simulations that use the initial conditions described in this section to analyze the reconnection rate of a guide field-balanced current sheet is presented in \autoref{tab:Vrec_sims_guide}.

\subsubsection{Perturbed Pressure-balanced Harris Sheet}\label{sec:pressureHarris}

Similar to the guide field-balanced Harris sheet, the pressure-balanced Harris sheet has the same in-plane magnetic field profile, \autoref{eq:By0_sheet}, but with zero guide field, $B_{\rm{guide}}=0$. The density profile has an amplitude of 
\begin{equation}
    \rho_1 = 3 \rho_0
\end{equation}
with profile
\begin{equation}
    \rho = \rho_0 + \rho_1 \sech^2\left( \frac{x}{a}\right).
\end{equation}
The initial temperature is constant and uniform in the upstream, and in turn sets the uniform upstream pressure outside the sheet, $p_0 = \rho_0 T$.
Force balance is achieved in the sheet via a pressure gradient. The pressure has an unperturbed profile,
\begin{equation}
    p = p_0 + \frac{1}{2} B_0^2 \sech^2\left(\frac{x}{a} \right),
\end{equation}
and the perturbed pressure profile is
\begin{equation}
\begin{split}
    p^\prime & = p_0 + \frac{1}{2} B_0^2 \sech^2\left(\frac{x}{a} \right) \\
    & \times \left[ 1 + 0.15 \left( \tanh(200 y - 10) + \tanh(-10 -200 y) \right) \right. \\ & \times\left.  \left( \tanh(200 x +2) + \tanh(2 - 200 x) \right) \right].
\end{split}
\end{equation}
A visual representation of the density and pressure profiles is shown in \autoref{fig:pressure_profile}. For the resistive setups, the electric field components are initialized as zero, $\mbf{E}=(0,0,0)$, and the resistivity, $\eta$, is constant and uniform.

Eight resistivity values are chosen for our analysis of magnetic reconnection in a pressure-balanced current sheet, ranging from the low Lundquist number ($S \sim 2\times 10^2$) Sweet--Parker regime to the high Lundquist number ($S\sim 2\times 10^7$) asymptotic plasmoid regime. The definition of $S$ is given in \autoref{eq:lundquist}. Similarly, six resolutions are chosen for the ideal simulations, ranging from the low resolution (64 cells per sheet half length) analogue to Sweet--Parker to the high resolution (2048 cells per sheet half length) analogue with a plateaued reconnection rate. All simulations have domains  $-1 \leq x/L \leq 1$ and $-2 \leq y/L \leq 2$, where $L=1$ in  code units, $a/L = 0.02$, sheet half length of $\Delta/L = 2$, $B_0 = 1$, $\sigma_{\rm{hot}} = 10$, and $T = 1.0$ with open boundaries. Note that open boundaries are used in the reconnection setups to avoid the formation of large plasmoids, which allows reconnection to proceed for a longer time and for a cleaner measure of the reconnection rate.


To reduce the computational cost of high resolution ideal MHD simulations of magnetic reconnection, we employ fixed spatial mesh refinement covering eight initial sheet widths, ensuring that the reconnection layer remains entirely within the highest-resolution region.
To test this method we consider the three simulations in \autoref{sec:num_L}, two with and one without mesh refinement. Resolution tests for the resistive MHD simulation \verb+vec_1e-5+ are shown in \autoref{app:resolution_testing}.

A complete list of the 2.5D simulations which use the initial conditions described in this section to analyze the reconnection rate of a pressure-balanced current sheet is presented in \autoref{tab:Vrec_sims}.

\begin{table*}[htbp]
\caption{\label{tab:Diffusion_Sims}%
1.75D Ohmic Diffusion Simulation Parameters. 
} 
\begin{ruledtabular}
\begin{tabular}{cllcc|llccc}

&\multicolumn{3}{c}{\textbf{Ideal Simulations}} &  &  \multicolumn{4}{c}{\textbf{Resistive Simulations 
}}  \\
\hline
&\multicolumn{1}{c}{\textrm{Sim. ID}}&
\multicolumn{1}{c}{\textrm{Sim. ID}}&
\textrm{Resolution}& &
\multicolumn{1}{c}{\textrm{Sim. ID}}&
\multicolumn{1}{c}{\textrm{Sim. ID}}&
\textrm{Resolution}& 
\textrm{$\eta\, [t_c]$}\\
&\multicolumn{1}{c}{(1)} & \multicolumn{1}{c}{(1)} & \multicolumn{1}{c}{(2)} & \multicolumn{1}{c|}{} & \multicolumn{1}{c}{(1)} &\multicolumn{1}{c}{(1)} & \multicolumn{1}{c}{(2)} & \multicolumn{1}{c}{(3)}  \\
\hline
&\Verb+1D_128+  & \Verb+MD_128+  & 128 & & \Verb+1D_1e-3_4096+ & \Verb+1D_FFR_1e-3_4096+ & 4096 & $10^{-3}$ \\
&\Verb+1D_256+ & \Verb+MD_256+  & 256 & & \Verb+1D_1e-4_16+ & \Verb+1D_FFR_1e-4_16+ & 16 & $10^{-4}$  \\
&\Verb+1D_512+ & \Verb+MD_512+  & 512 & & \Verb+1D_1e-4_64+ & \Verb+1D_FFR_1e-4_64+ & 64 & $10^{-4}$ \\
&\Verb+1D_1024+ & \Verb+MD_1024+ & 1024 & & \Verb+1D_1e-4_128+ & \Verb+1D_FFR_1e-4_128+ & 128 & $10^{-4}$ \\
&\Verb+1D_2048+ & \Verb+MD_2048+ & 2048 & & \Verb+1D_1e-4_256+ & \Verb+1D_FFR_1e-4_256+ & 256 & $10^{-4}$ \\
&\Verb+1D_4096+ & \Verb+MD_4096+ & 4096 &  & \Verb+1D_1e-4_4096+ & \Verb+1D_FFR_1e-4_4096+ & 4096 & $10^{-4}$  \\
&  & &  & & \Verb+1D_1e-5_4096+ & \Verb+1D_FFR_1e-5_4096+ & 4096 & $10^{-5}$ \\

\end{tabular}
\end{ruledtabular}
\begin{tablenotes}[para]
        \textit{\textbf{Notes.}} Simulation parameters for the setup described in \autoref{sec:ffeHarris}. All simulations have domains $-1\leq x/L \leq 1$, where $L=1$ in code units, $a/L=0.02$, $B_0=1$, $B_{\rm{guide}}=10$, $\sigma_{\rm{hot}}=10$, and $T=1.0$, with continuous boundaries, where quantities are extrapolated outside the domain. \textbf{Column (1):} the unique simulation ID. 
        \textbf{Column (2):} the uniform resolution.
        \textbf{Column (3):} resistivity in units of light crossing time, $t_c = L/c$. 
    \end{tablenotes}
\end{table*}

\begin{table*}[htbp]
\caption{\label{tab:Vrec_sims_guide}%
2.5D Guide field-balanced Harris Sheet Reconnection Simulation Parameters. 
} 
\begin{ruledtabular}
\begin{tabular}{clcc|llccc}

& \multicolumn{3}{c|}{\textbf{Ideal Simulations }}  & \multicolumn{4}{c}{\textbf{Resistive Simulations }} & \\
\hline
& \multicolumn{1}{c}{\textrm{Sim. ID}} & \multicolumn{1}{c}{\textrm{Sim. ID}} & \multicolumn{1}{c|}{\textrm{Resolution}} & \multicolumn{1}{c}{\textrm{Sim. ID}}&
\multicolumn{1}{c}{\textrm{Sim. ID}}&
\textrm{Resolution}& 
\textrm{$\eta\, [t_c]$}
&
\\
& \multicolumn{1}{c}{(1)} & \multicolumn{1}{c}{(1)} & \multicolumn{1}{c|}{(2)} & \multicolumn{1}{c}{(1)} & \multicolumn{1}{c}{(1)} &\multicolumn{1}{c}{(2)} & \multicolumn{1}{c}{(3)}  &\\
\hline
& \Verb+vrec_guide_32+  & \Verb+vrec_MD_32+ & 32  & \Verb+vrec_guide_1e-2+ & \Verb+vrec_FFR_1e-2+ & 2048 &$10^{-2}$      &\\
&  \Verb+vrec_guide_64+ & \Verb+vrec_MD_64+ & 64  & \Verb+vrec_guide_5e-3+ & \Verb+vrec_FFR_5e-3+ & 4096 & $5\times 10^{-3}$ &\\
& \Verb+vrec_guide_128+  & \Verb+vrec_MD_128+ & 128  & \Verb+vrec_guide_1e-3+ & \Verb+vrec_FFR_1e-3+ & 4096 & $10^{-3}$          &\\
& \Verb+vrec_guide_256+  & \Verb+vrec_MD_256+ & 256  & \Verb+vrec_guide_5e-4+ & \Verb+vrec_FFR_5e-4+ & 4096  & $5\times 10^{-4}$  &\\
& \Verb+vrec_guide_512+  & \Verb+vrec_MD_512+ & 512  & \Verb+vrec_guide_1e-4+ & \Verb+vrec_FFR_1e-4+ & 4096  & $10^{-4}$        &\\
&  \Verb+vrec_guide_1024+ & \Verb+vrec_MD_1024+ & 1024  & \Verb+vrec_guide_5e-5+ & \Verb+vrec_FFR_5e-5+ & 4096 & $5\times 10^{-5}$  &\\
 & \Verb+vrec_guide_2048+  & \Verb+vrec_MD_2048+ & 2048 & \Verb+vrec_guide_1e-5+ & \Verb+vrec_FFR_1e-5+ & 4096  & $10^{-5}$          & \\
& \Verb+vrec_guide_4096+  & \Verb+vrec_MD_4096+ & 4096  & \Verb+vrec_guide_1e-6+ & \Verb+vrec_FFR_1e-6+ & 4096 & $10^{-6}$          &\\
&  \Verb+vrec_guide_8192+ & \Verb+vrec_MD_8192+ & 8192  & \Verb+vrec_guide_1e-7+ & \Verb+vrec_FFR_1e-7+ & 4096 & $10^{-7}$          &\\
&   & \Verb+vrec_MD_16384+ & 16384 &  &  &  &     &\\
\end{tabular}
\end{ruledtabular}
\begin{tablenotes}[para]
        \textit{\textbf{Notes.}} Simulation parameters for the setup described in \autoref{sec:guidefield_perturbed}. All simulations have domains  $-1 \leq x/L \leq 1$ and $-2 \leq y/L \leq 2$, where $L=1$ in  code units, $a/L = 0.02$, sheet half length of $\Delta/L = 2$, $B_0 = 1$, $B_{\rm{guide}} = 2$, $\sigma_{\rm{hot}} = 10$, and $T = 1.0$ with open boundaries. \textbf{Column (1):} the unique simulation ID. 
        \textbf{Column (2):} the effective resolution per sheet half length.
        \textbf{Column (3):} resistivity in units of light crossing time, $t_c = L/c$.

       For the ideal simulations, only spatially static mesh refinement was used over 8 times the initial sheet thickness, or 2 times the initial sheet thickness for the two highest resolution MD runs.

    \end{tablenotes}
\end{table*}

\begin{table*}[htbp]
\caption{\label{tab:Vrec_sims}%
2.5D Pressure-balanced Harris Sheet Reconnection Simulation Parameters. 
} 
\begin{ruledtabular}
\begin{tabular}{clcc|lccc}

&\multicolumn{2}{c}{\textbf{Ideal Simulations}} & \multicolumn{1}{c|}{\textbf{}} & \multicolumn{3}{c}{\textbf{Resistive Simulations 
}}  \\
\hline
&\multicolumn{1}{c}{\textrm{Sim. ID}}&
\multicolumn{1}{c}{\textrm{Resolution}}&
\multicolumn{1}{c|}{\textrm{}}&

\multicolumn{1}{c}{\textrm{Sim. ID}}&
\multicolumn{1}{c}{\textrm{Resolution}}&
\textrm{$\eta \, [t_c]$}&

\\
&\multicolumn{1}{c}{(1)} & \multicolumn{1}{c}{(2)} & \multicolumn{1}{c|}{} & \multicolumn{1}{c}{(1)} & \multicolumn{1}{c}{(2)}  & \multicolumn{1}{c}{(3)} & \\
\hline
&\Verb+vrec_64+    & 64   && \Verb+vrec_5e-3+  & 4096 & $5\times 10^{-3}$  &\\
&\Verb+vrec_128+   & 128  && \Verb+vrec_1e-3+  & 4096 & $10^{-3}$          &\\
&\Verb+vrec_256+   & 256  && \Verb+vrec_5e-4+  & 4096 & $5\times 10^{-4}$  &\\
&\Verb+vrec_512+   & 512  && \Verb+vrec_1e-4+  & 4096 & $10^{-4}$          &\\
&\Verb+vrec_1024+  & 1024 && \Verb+vrec_5e-5+  & 4096 & $5\times 10^{-5}$  &\\
&\Verb+vrec_2048+  & 2048 && \Verb+vrec_1e-5+  & 4096 & $10^{-5}$          &\\
&  &     && \Verb+vrec_1e-6+  & 4096 & $10^{-6}$          &\\
&  &     && \Verb+vrec_1e-7+  & 4096  & $10^{-7}$          &\\

\end{tabular}
\end{ruledtabular}
\begin{tablenotes}[para]
        \textit{\textbf{Notes.}} Simulation parameters for the setup described in \autoref{sec:pressureHarris}. All simulations have domains  $-1 \leq x/L \leq 1$ and $-2 \leq y/L \leq 2$, where $L=1$ in  code units, $a/L = 0.02$, sheet half length of $\Delta/L = 2$, $B_0 = 1$, $\sigma_{\rm{hot}} = 10$, and $T = 1.0$ with open boundaries. \textbf{Column (1):} the unique simulation ID. 
        \textbf{Column (2):} the effective resolution per sheet half length.
        \textbf{Column (3):} resistivity in units of light crossing time, $t_c = L/c$.
        For the ideal simulations, only spatially static mesh refinement was used over 8 times the initial sheet thickness.
    \end{tablenotes}
\end{table*}

\section{Comparison of Magnetic Diffusion}\label{sec:Comparison of Magnetic Diffusion}

In our analysis of numerical resistivity and its relation to Ohmic diffusion, we employ two primary tests: firstly, examining the decay of current sheet amplitudes, and secondly, observing the evolution of the spatial distribution of the current. These tests are directly compared to  analytical results in resistive (non-relativistic) MHD which we derive in the following subsections. We demonstrate that analytical results align well with fully resolved explicit resistive simulations, both resistive MHD and resistive FFE, while ideal simulations produce subdiffusive decay of the current sheet. Subdiffuse decay is characterized by a sub-linear relationship between the second moment and time, $\langle x^2 \rangle \propto t^\alpha$, $0 \leq \alpha<1$ \cite{2017PhRvE..96d2153A}, while Ohmic diffusion for a uniform resistivity is expected to be linear. Note that in contrast to \autoref{sec:Numerical Simulations and Models}, we work in cgs units for the remainder of the paper.

\subsection{Maximum Current Amplitude Test}\label{sec:Maximum Current Amplitude Test}

\begin{figure}[htbp]
    \centering
    \includegraphics[width=0.455\textwidth]{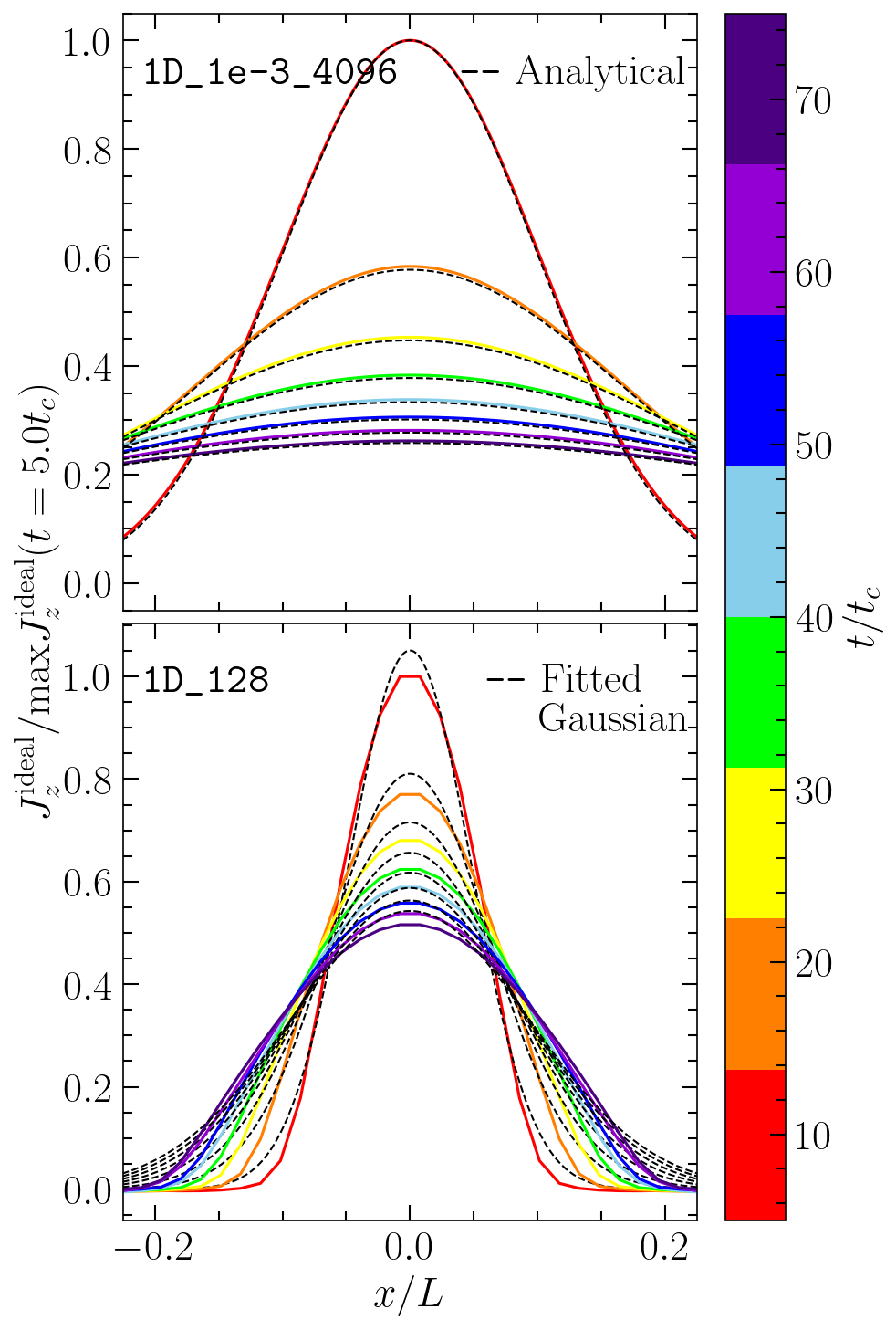}
    \cprotect\caption{Zoomed-in current profiles for resistive simulation \Verb+1D_1e-3_4096+ and ideal simulation \Verb+1D_128+. Note that the current is the ideal stationary current, $\mbf{J}^{\text{ideal}} = \frac{c}{4\pi} \bm{\nabla} \times \mbf{B}$. The analytical solutions plotted are given by \autoref{eq:J_guass}. The snapshots plotted occur from $t/t_c = 5$ to $t/t_c = 75$ and $L=c/t_c$ is the unit length which the simulation domain is normalized to.}
    \label{fig:current_profile}
\end{figure}

We initialize a Harris sheet that relaxes into the self-similar current sheet solution as found by \cite{2007MNRAS.382..995K}.
We study the evolution of the current in this configuration when embedded in a strong guide field.
Consider a thin current sheet embedded in a plasma with uniform Ohmic resistivity, such as the setup described in \autoref{sec:ffeHarris}. The evolution of the electric and magnetic fields are governed by Faraday's and Amp\`ere's law, and the current is determined by the relativistic Ohm's law \cite{1993PhRvL..71.3481B, 2003ApJ...589..893L, 2007MNRAS.382..995K},
\begin{align}
    & \partial_t \mbf{B} + c \bm{\del} \times \mbf{E} = \bm{0},\\
    & \partial_t \mbf{E} - c \bm{\del} \times \mbf{B} = - 4\pi \mbf{J},\\
    & \mbf{J} = q \mbf{v} + \frac{\Gamma}{\eta} \left[ \mbf{E} + \frac{\mbf{v}\times\mbf{B}}{c} - \frac{(\mbf{v}\cdot\mbf{E})}{c^2} \mbf{v} \right],
\end{align}
where $\Gamma$ is the Lorentz factor. The Ohmic decay of the thin current sheet is captured in the Newtonian limit, since the displacement current and space charge are insignificant because of the sub-relativistic speeds and large conductivity, 
\begin{align}
    & \frac{v^2}{c^2} \ll 1, \quad \Gamma \sim 1, \quad q \sim 0, \quad \left| \partial_t \mbf{E} \right| \sim \bm{0},
\end{align}
then,
\begin{align}
    & \partial_t \mbf{B} + c \bm{\del} \times \mbf{E} = \bm{0},\\
    &  \bm{\del} \times \mbf{B} -  \frac{4\pi}{c} \mbf{J} = \bm{0},\\
    & \mbf{J} = \frac{1}{\eta} \left[ \mbf{E} + \frac{\mbf{v}\times\mbf{B}}{c} \right]. 
\end{align}
These three equations can be rewritten as an evolution equation for the magnetic field by direct substitution into the $\bm{\del} \times \mbf{E}$ term in Faraday's law,
\begin{align}
    & \partial_t \mbf{B} + c  \mbf{\del} \times \left(  \frac{c \eta}{4\pi} \mbf{\del} \times \mbf{B} -  \frac{\mbf{v}\times\mbf{B}}{c}\right) = \bm{0}.
\end{align}
Then the non-ideal induction equation governing the evolution of the magnetic field in such a system is
\begin{equation}
    \frac{\partial \vecB}{\partial t} = \bm{\del}\times (\bm{v}\times\vecB) + \frac{c^2 \eta}{4\pi} \del^2 \vecB,\label{eq:induction_nonideal}
\end{equation}
which provides a relation between the diffusive length and time scales. Approximating  $|\bm{\del}| \sim \ell^{-1}$, $\partial_t \sim t^{-1}$, and assuming that the induction term is negligible yields a relation between the diffusive length and time scales
\begin{equation}
    \ell = \sqrt{\frac{c^2 \eta}{4\pi} t}.
\end{equation} 
Next, we seek solutions for the $\bm{v}=\bm{0}$ case for an effectively one-dimensional, thin current sheet, which is symmetric in $y$ and unchanging in $z$ with $\vecB = B_y \hat{y} + B_z \hat{z}$, which satisfy the diffusive partial differential equation 
\begin{equation}
    \frac{\partial B_y}{\partial t} = \tilde{\eta} \frac{\partial^2 B_y}{\partial x^2} \label{eq:diffusion_B}
\end{equation}
where $\tilde{\eta} = c^2 \eta/4\pi$.
The current sheet satisfies the boundary conditions $B_y\rightarrow \pm B_0$ as $x\rightarrow \pm \infty$. 
The $z$-component of the current is
\begin{equation}
    J_z = \frac{c}{4\pi} \frac{\partial B_y}{\partial x} \hat{z}.
\end{equation}
The boundary conditions for $J_z$ are $J_z\rightarrow 0$ as $x\rightarrow\pm\infty$. Since the current is initially locally peaked at $x=0$ we consider solutions of the form:
\begin{equation}
    J_z(x,t) = \frac{k}{\ell(t)} j \left( \frac{x}{\ell(t)} \right).
\end{equation}
where $k$ is a constant given by Amp\`{e}re's law. The current follows the same diffusion relation
\begin{equation}
    \frac{\partial J_z}{\partial t} = \tilde{\eta} \frac{\partial^2 J_z}{\partial x^2} \label{eq:diffusion_J}.
\end{equation}
From \autoref{eq:diffusion_J}, the differential equation for $j$ is
\begin{equation}
    j + \Tilde{x} j' + 2j'' =0, \label{eq:j_DE}
\end{equation}
where $\Tilde{x}=x/\ell(t)$, which has solution
\begin{equation}
    j = e^{-\tilde{x}^2/4}.
\end{equation}
Enforcing $J_z(x,t) = (c/4\pi) \partial_x B_y$ the complete solution is
\begin{align}
    J_z(x,t)
           & =   \sqrt{\frac{B_0^2}{4\pi^2 \eta t}} \, \exp \left(  - \frac{\pi x^2 }{c^2 \eta t}\right)\label{eq:J_guass},
\end{align}
in \autoref{fig:current_profile} we present visualizations of the current profiles of an ideal and resistive MHD simulation at various times (shown with different colors in the plot). We choose two simulations to demonstrate the evolution of resolved resistive MHD compared to low resolution ideal MHD which has significant diffusion over a similar timescale. Notice the excellent agreement between the explicit resistive simulations and the prediction from resistive theory in \autoref{eq:J_guass}. The current profiles from the ideal simulations have Gaussian fits performed which indicate that as the profile evolves, the tails of the profile spread too slowly compared to the Gaussian fit which is indicative of subdiffuse evolution. 

\begin{figure*}[htbp]
    \centering
    \begin{subfigure}{0.49\textwidth}
        \centering
        \includegraphics[width=\textwidth]{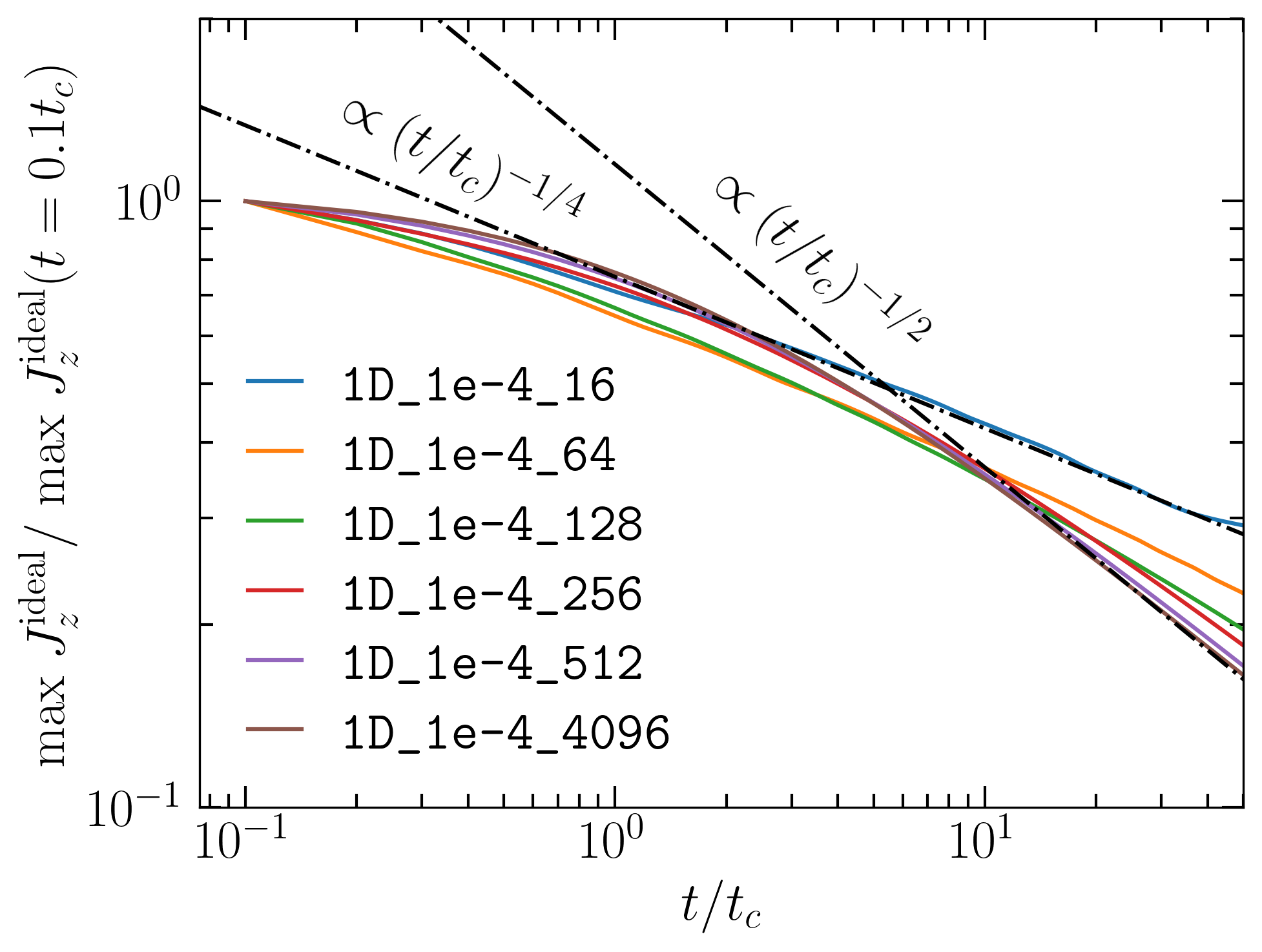}
        \caption{Resistive MHD.}
        \label{fig:rrmhd_current}
    \end{subfigure}
    \hfill
    \begin{subfigure}{0.49\textwidth}
        \centering
        \includegraphics[width=\textwidth]{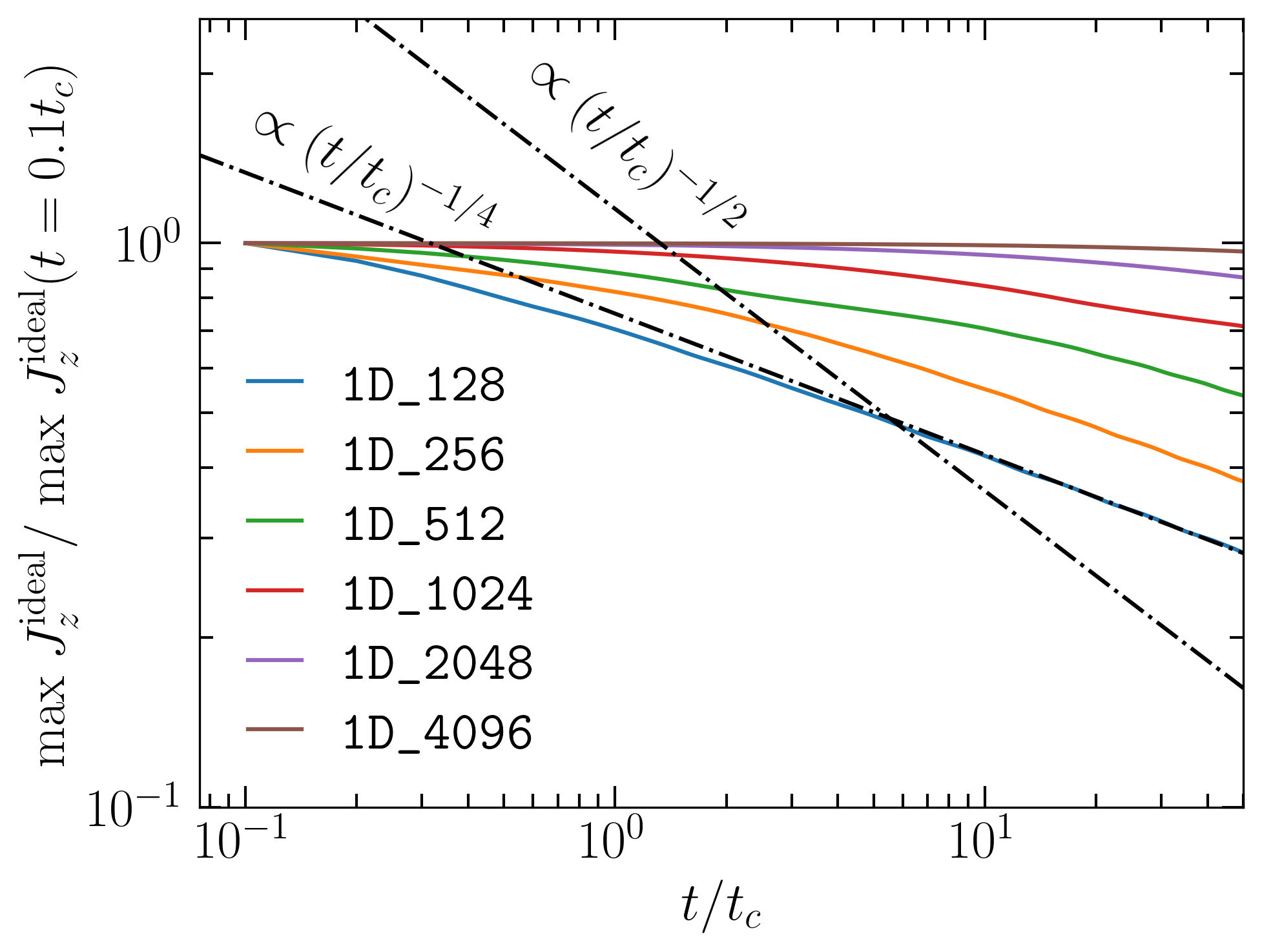}
        \caption{Ideal MHD.}
        \label{fig:ideal_current} 
    \end{subfigure}
    \hfill
    \begin{subfigure}{0.49\textwidth}
        \centering
        \includegraphics[width=\textwidth]{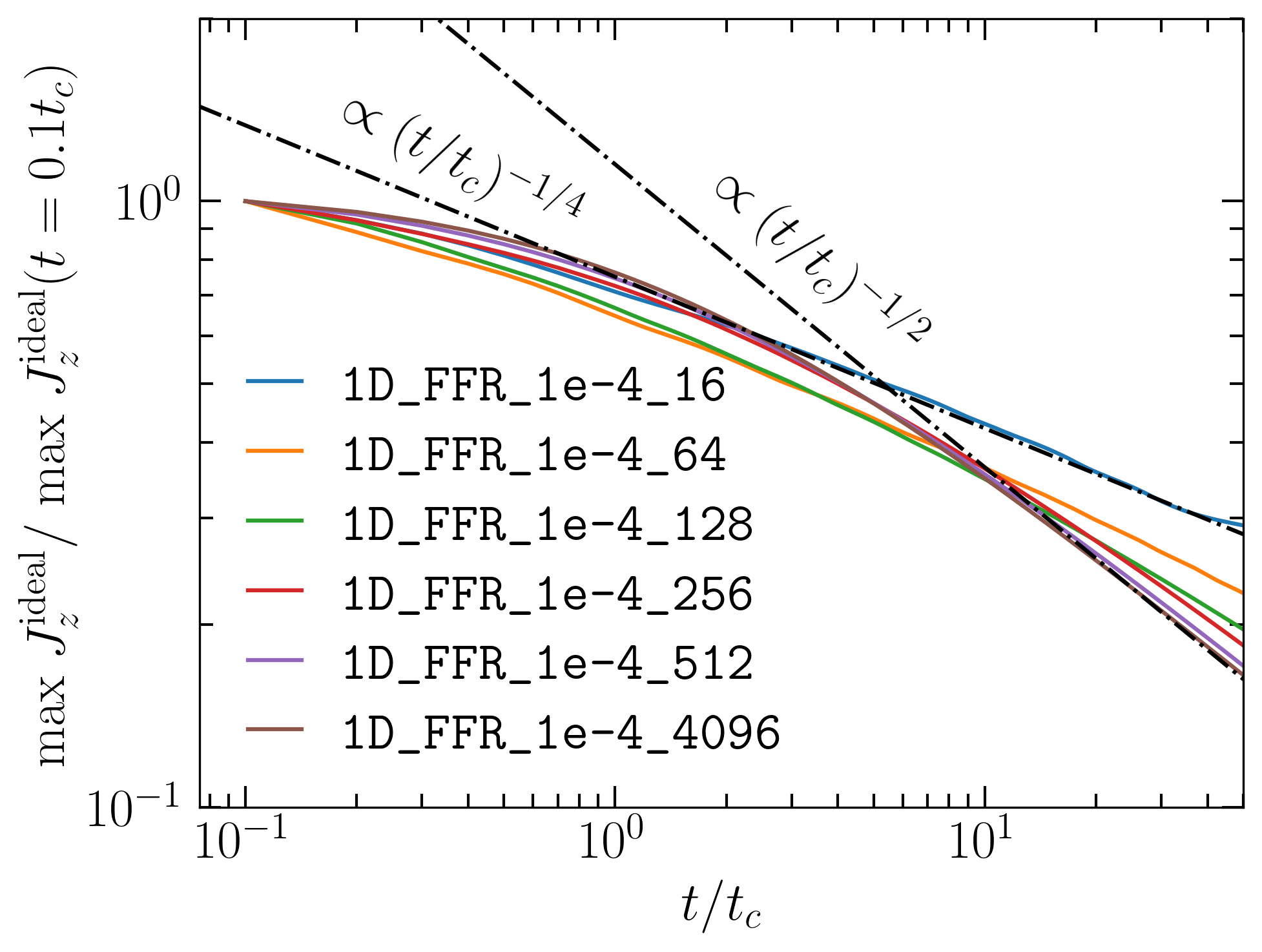}
        \caption{Resistive FFE.}
        \label{fig:FFR_current} 
    \end{subfigure}
    \hfill
    \begin{subfigure}{0.49\textwidth}
        \centering
        \includegraphics[width=\textwidth]{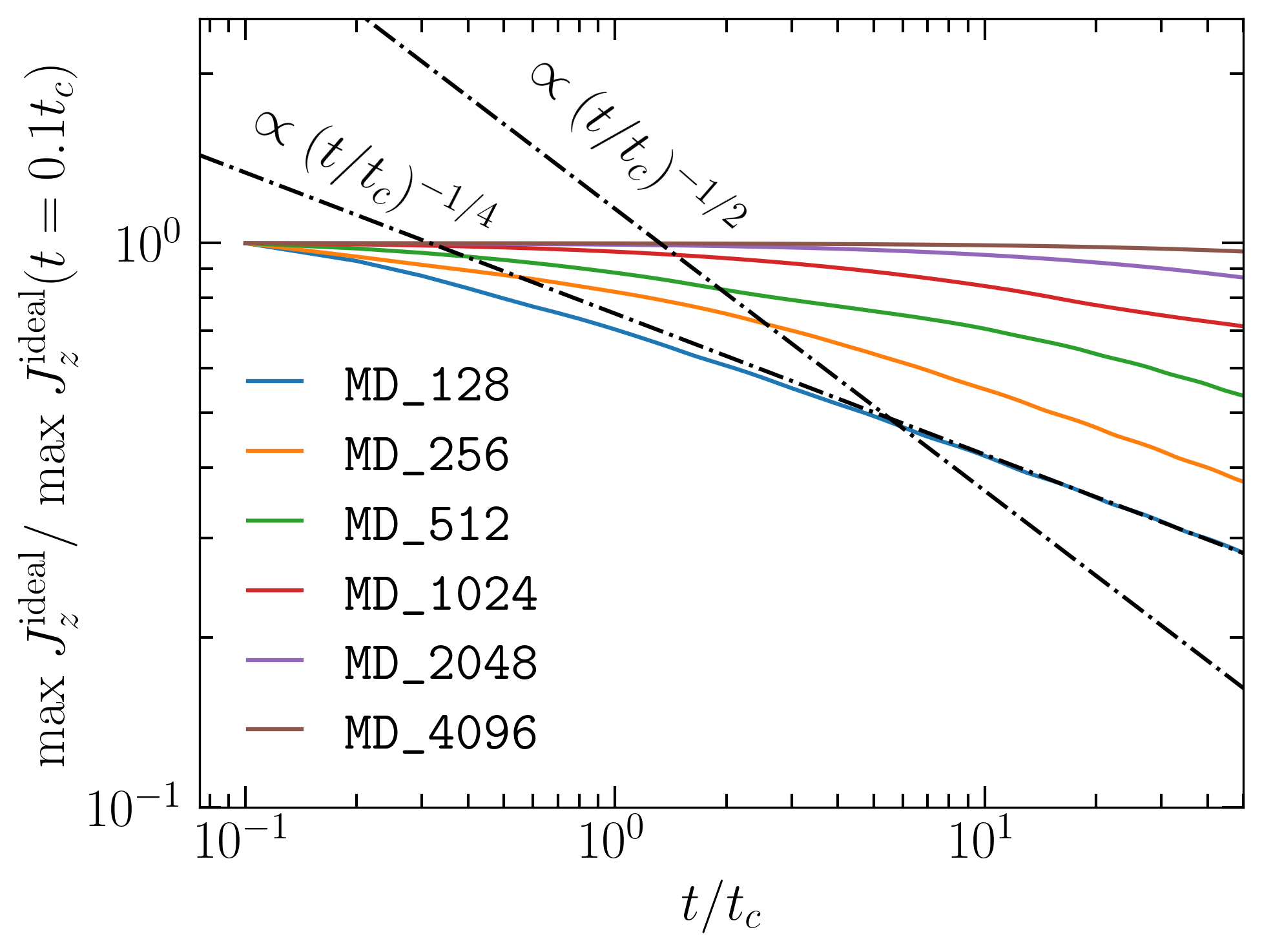}
        \caption{MD.}
        \label{fig:MD_current} 
    \end{subfigure}
    \hfill
    \begin{subfigure}{0.49\textwidth}
        \centering
        \includegraphics[width=\textwidth]{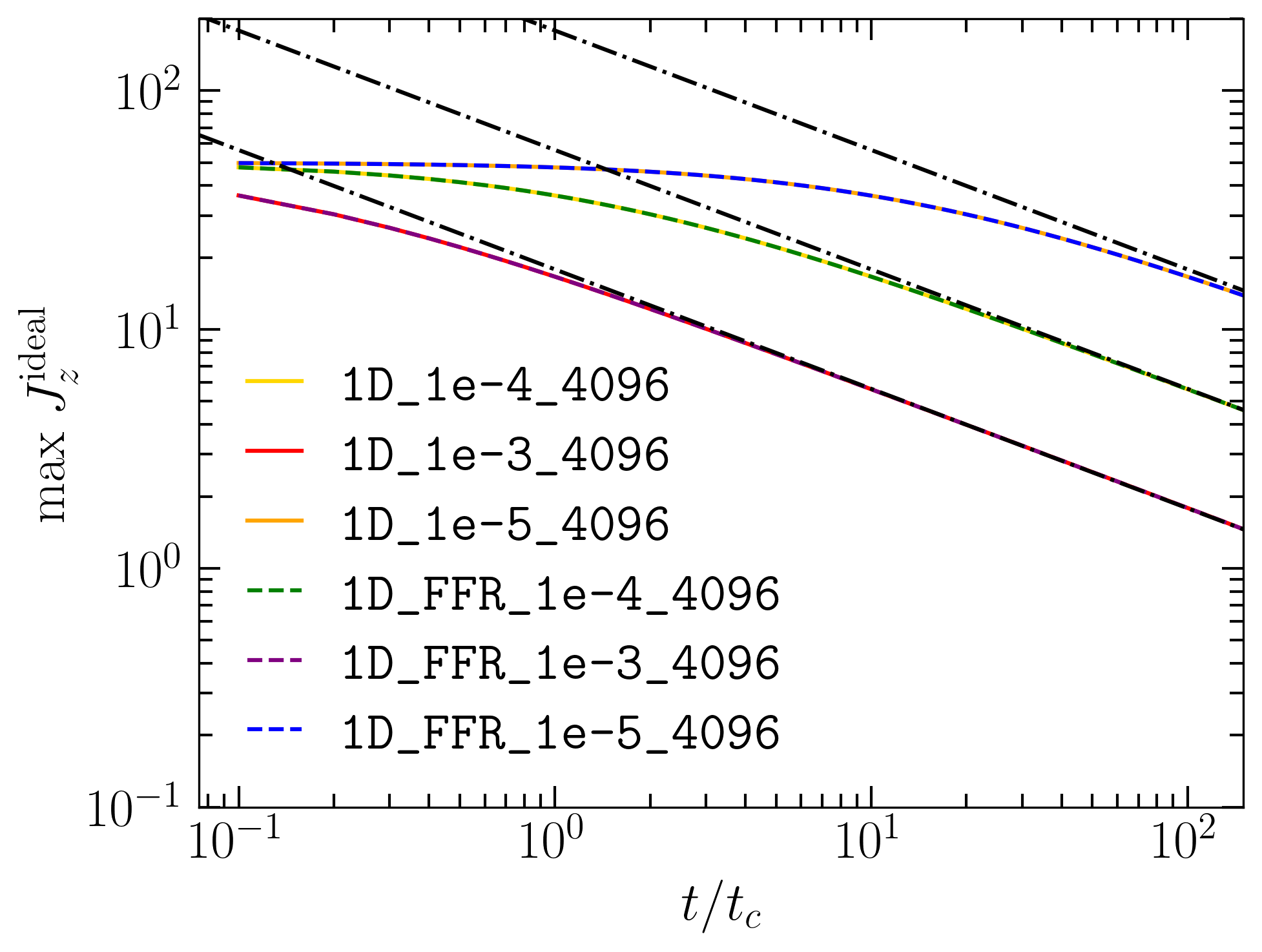}
        \caption{Resistive MHD and resistive FFE.}
        \label{fig:FFR_multieta} 
    \end{subfigure}
    \caption{
    The Ohmic decay of the current amplitude for a thin current sheet. The resistive simulations in \autoref{fig:rrmhd_current} and \autoref{fig:FFR_current} are for a fixed resistivity with increasing resolution, while the ideal simulations are ran with increasing resolution. \autoref{fig:FFR_multieta} demonstrates that the current amplitude decay perfectly matches between resistive FFE and resistive MHD, and that both schemes match onto the analytical expectation \autoref{eq:jmax_decay}.
    Note all currents here are the ideal stationary current $\bm{J}^{\rm{ideal}} = \frac{c}{4\pi}\bm{\del} \times \vecB$, and that the $(t/t_c)^{-1/4}$ scaling is purely for visualization purposes and is not an expected nor fitted result.  
    }
    \label{fig:current_decay}
\end{figure*}

From \autoref{eq:J_guass}, (when $t\neq 0$) taking the maximum removes the spatial dependence, leaving us with a test of the current amplitude's decay due to Ohmic resistivity,
\begin{equation}
    \max J_z(t) =  \sqrt{\frac{B_0^2}{4\pi^2 \eta t}}. \label{eq:jmax_decay}
\end{equation}
In \autoref{fig:current_decay} we show the current amplitude decay in ideal MHD, MD, resistive MHD, and resistive FFE simulations. We compare the analytic solution to the ideal stationary current, $\mbf{J}^{\text{ideal}} = \frac{c}{4\pi} \bm{\nabla} \times \mbf{B}$, because the relativistic resistive corrections due to the displacement current are negligible. As the resistive MHD and resistive FFE simulations sufficiently resolve the resistive scale, they follow the expected $t^{-1/2}$ scaling. This agreement is expected since both resistive MHD and the resistive FFE model \cite{Alic2012_FFR} analyzed in this paper reduce to the Newtonian resistive induction equation in the appropriate limit, as shown in \autoref{app:ff-induction} for the resistive FFE model.
The ideal MHD and MD results show that numerical diffusion leads to subdiffuse decay of the current amplitude. As the resolution of the ideal simulations increases, the decay becomes slower, approaching but never reaching the ideal limit of zero resistive decay, $t^{0}$.

\begin{figure*}[htbp]
    \centering
    \begin{subfigure}{0.49\textwidth}
        \centering
        \includegraphics[width=\textwidth]{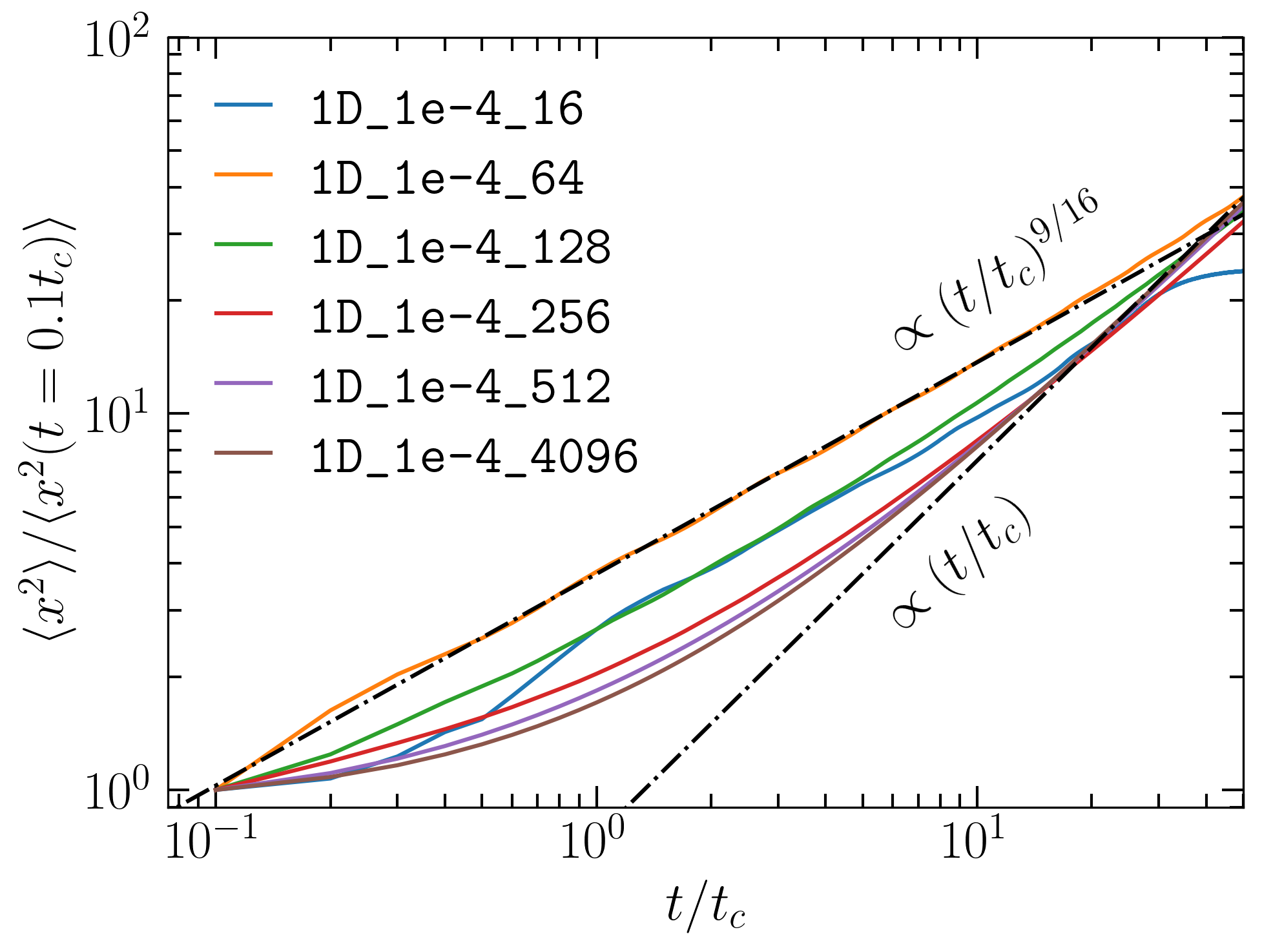}
        \caption{Resistive MHD.}
        \label{fig:rrmhd_x2}
    \end{subfigure}
    \hfill
    \begin{subfigure}{0.49\textwidth}
        \centering
        \includegraphics[width=\textwidth]{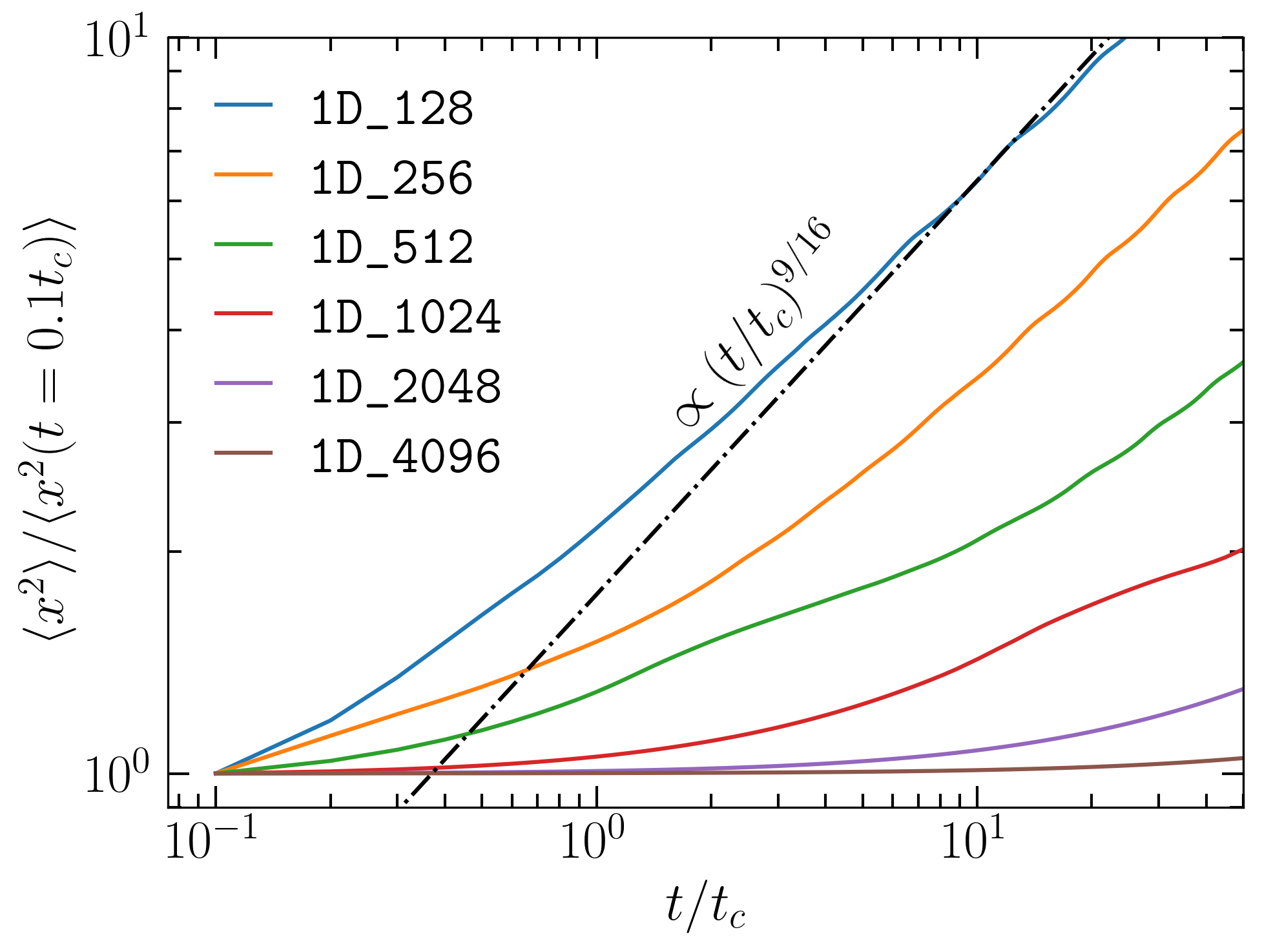}
        \caption{Ideal MHD.}
        \label{fig:ideal_x2} 
    \end{subfigure}
    \hfill
    \begin{subfigure}{0.49\textwidth}
        \centering
        \includegraphics[width=\textwidth]{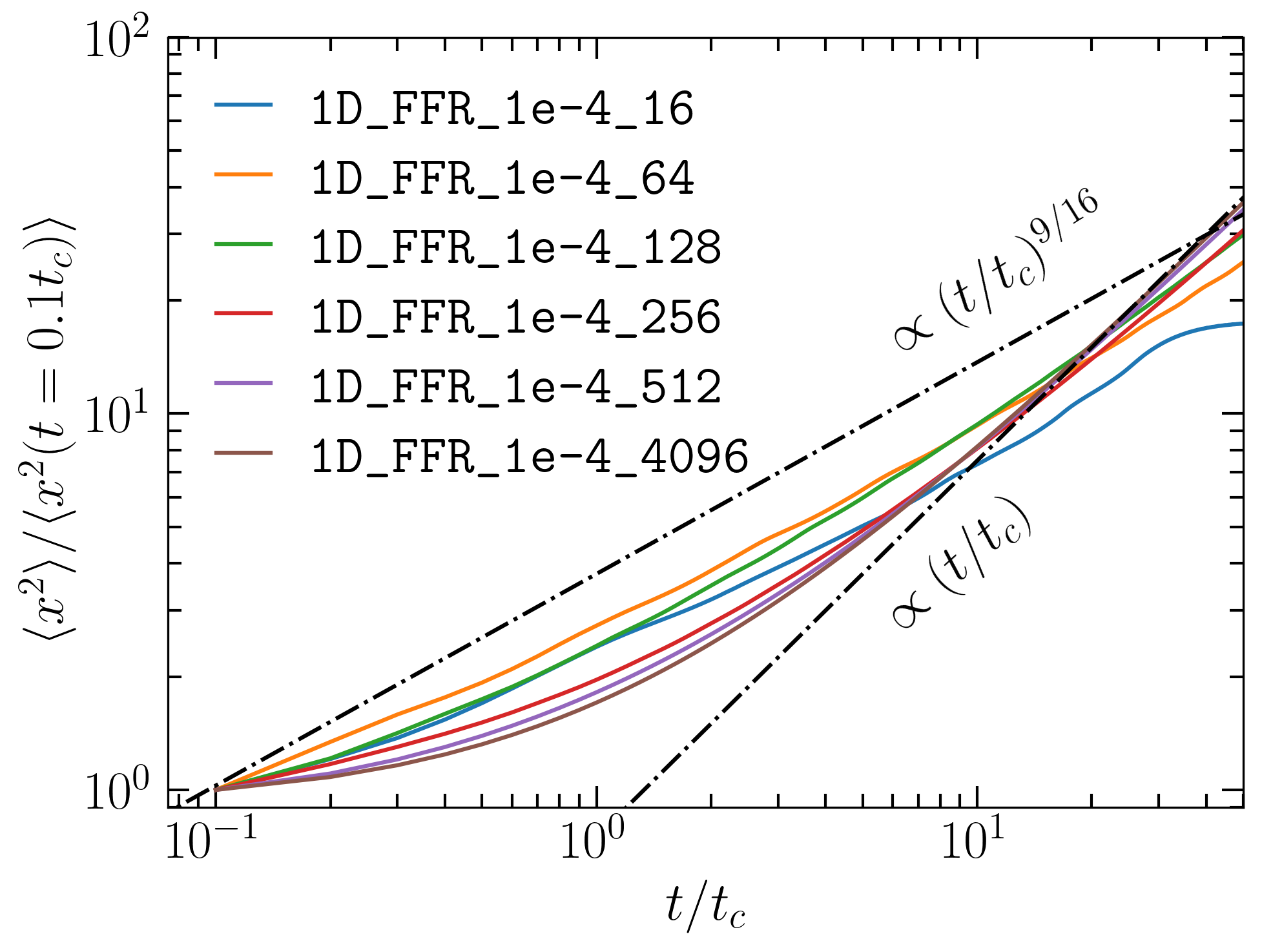}
        \caption{Resistive FFE.}
        \label{fig:FFR_x2} 
    \end{subfigure}
    \hfill
    \begin{subfigure}{0.49\textwidth}
        \centering
        \includegraphics[width=\textwidth]{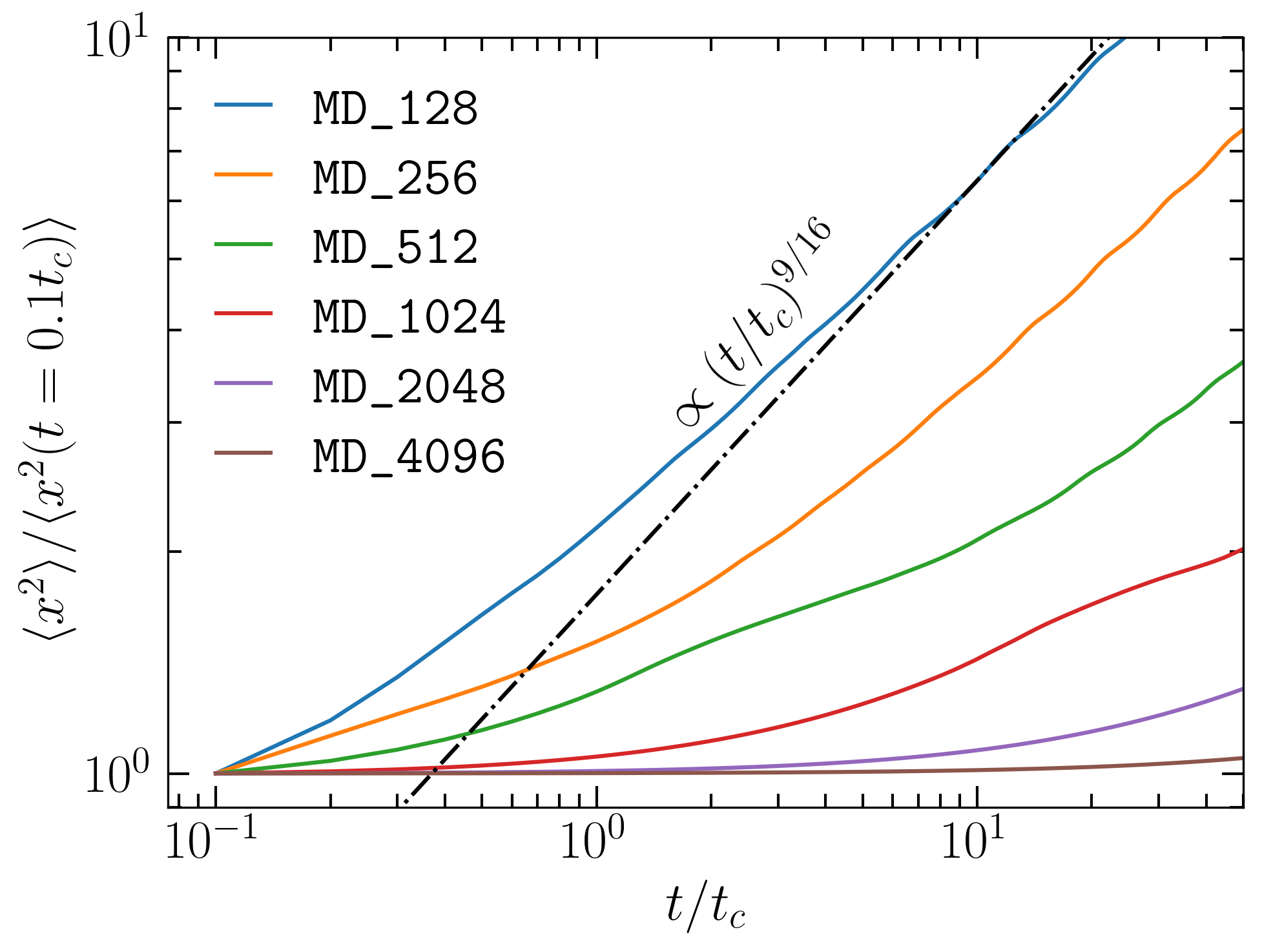}
        \caption{MD.}
        \label{fig:MD_x2} 
    \end{subfigure}
    \hfill
    \begin{subfigure}{0.49\textwidth}
        \centering
        \includegraphics[width=\textwidth]{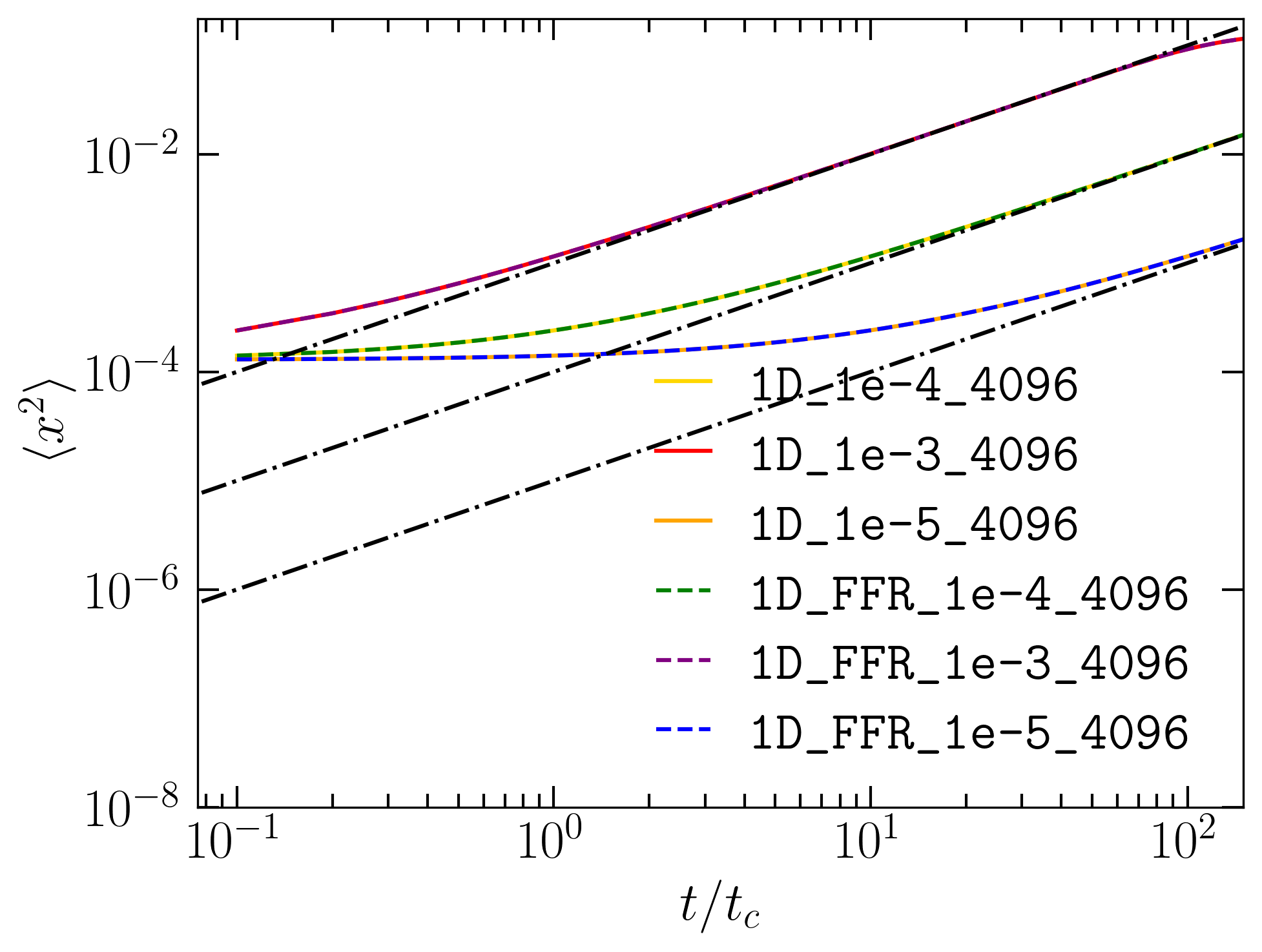}
        \caption{Resistive MHD and resistive FFE.}
        \label{fig:FFR_multi_x2} 
    \end{subfigure}
        \caption{
    The Ohmic evolution of the second moment, $\left\langle x^2 \right\rangle$, of the current distribution for a thin current sheet. The resistive simulations in \autoref{fig:rrmhd_x2} and \autoref{fig:FFR_x2} are for a fixed resistivity with increasing resolution, while the ideal simulations are ran with increasing resolution.
    \autoref{fig:FFR_multi_x2} demonstrates that the evolution of the second moment perfectly matches between resistive FFE and
resistive MHD, and that both schemes match onto the analytical expectation \autoref{eq:x2_decay}. 
    Note that the $(t/t_c)^{9/16}$ scaling is purely for visualization purposes and is not derived analytically nor a fitted result. 
    }
    \label{fig:x2_decay}
\end{figure*}

\subsection{Second Moment Test}\label{sec:Second Moment Test}
The second moment of the current distribution serves as a direct test to track the evolution of the spatial distribution of current as the current sheet decays due to Ohmic resistive (or numerical resistive) effects.
From \autoref{eq:J_guass}, the second moment in the distribution of $J(x,t)$ can also be derived. For standard Gaussian diffusion, this is expected to be a linear relation,
\begin{equation}
    \langle x^2 \rangle \propto  t.
\end{equation}
We normalize \autoref{eq:J_guass} to obtain a probability density function
\begin{equation}
    \mathbb{P}(x,t) = \frac{J_z^2(x,t)}{ \displaystyle\int_{-\infty}^{\infty} dx'\, J_z^2(x',t)   },
\end{equation}
the second moment is
\begin{align}
\langle x^2 \rangle & = \int_{-\infty}^{\infty} dx\, x^2 \mathbb{P}(x) = \frac{c^2 \eta t}{4 \pi }. 
\end{align}
If instead we consider a box of finite size $2L$ such that $-L \leq x \leq L$, then
\begin{equation}
    \mathbb{P}(x,t) = \frac{J_z^2(x,t)}{ \displaystyle\int_{-L}^{L} dx'\, J_z^2(x',t)   },
\end{equation}
and
\begin{align}
\langle x^2 \rangle & = \int_{-L}^{L} dx\, x^2 \mathbb{P}(x),\\ & = \frac{c^2 \eta t}{4 \pi } - \sqrt{\frac{4 c^2 \eta L^2 t}{8 \pi^2}} \frac{e^{- \frac{2 L^2 \pi }{ c^2 \eta t}}}{\rm{erf}\left( \sqrt{\frac{2 L^2 \pi}{ c^2 \eta t}} \right)}.
\end{align}
For $\frac{2 L^2 \pi }{ c^2 \eta t} \gg 1$, an appropriate approximation in these simulations (since $t\neq 0$, $c=1, \, L = 2$, and $\eta/4\pi \leq 10^{-3}$) is 
\begin{align}
\langle x^2 \rangle \simeq \frac{c^2 \eta t}{4 \pi } - \sqrt{\frac{4 c^2 \eta L^2 t}{8 \pi^2}} \exp\left(-\frac{ 4 L^2 \pi }{2 c^2 \eta t}  \right) \simeq \frac{c^2 \eta t}{4 \pi }.\label{eq:x2_decay}
\end{align}
 In \autoref{fig:x2_decay}, we show the time evolution of the second moment in ideal MHD, MD, resistive MHD, and resistive FFE simulations.  \autoref{fig:FFR_multieta} demonstrates that the evolution of the second moment perfectly matches between resistive FFE and resistive MHD, and that both schemes match onto the analytical expectation, \autoref{eq:x2_decay}. This agreement is expected, since the resistive FFE model \cite{Alic2012_FFR} analyzed in this paper reduces to the Newtonian resistive induction equation in the appropriate limit, as shown in \autoref{app:ff-induction}. As the resistive MHD and resistive FFE simulations resolve the explicit resistivity, they follow the analytical expectation, $\left\langle x^2 \right\rangle \propto t$. 
For increasing resolution in the ideal MHD and MD simulations, the evolution becomes increasingly subdiffuse, $\left\langle x^2 \right\rangle \propto t^{\alpha}$, $0 \leq \alpha < 1$, approaching but never reaching the ideal limit of zero resistive decay.

\section{Comparison of Magnetic Reconnection}\label{sec:Comparison of Magnetic Reconnection}

In our analysis of numerical resistivity and its impact on magnetic reconnection, we first probe the Sweet--Parker regime \cite{Sweet:1958IAUS....6..123S, Parker:1963ApJS....8..177P}, and second, the asymptotic plasmoid unstable regime \cite{Biskamp_1993, Lapenta.PhysRevLett.100.235001,2007PhPl...14j0703L, Bhattacharjee.2009PhPl...16k2102B, Cassak.PhysRevLett.95.235002, Huang_2010, Uzdensky_2010, Loureiro_2012, Ni2012, Huang.10.1063/1.4802941, Takamoto_2013, Comisso.10.1063/1.4918331, Loureiro_2016, Huang_2017, Comisso_2018} in ideal MHD, MD, resistive MHD, and resistive FFE. We describe the method for measuring the reconnection rate in \autoref{app:measure_vrec}. Our resistive MHD simulations reproduce the expected Sweet--Parker scaling of the reconnection rate with respect to the Lundquist number, $v_{\rm{rec}} = S^{-1/2}$, in both a pressure-balanced sheet and a guide field-balanced sheet. For large Lundquist number the reconnection rate plateaus, diverging from the Sweet--Parker scaling and entering the plasmoid-dominated asymptotic regime.

\subsection{Sweet--Parker Regime and the Asymptotic Reconnection Rate}\label{sec:SPandPlasmoid}

\begin{figure*}
    \centering
    \includegraphics[width=\textwidth]{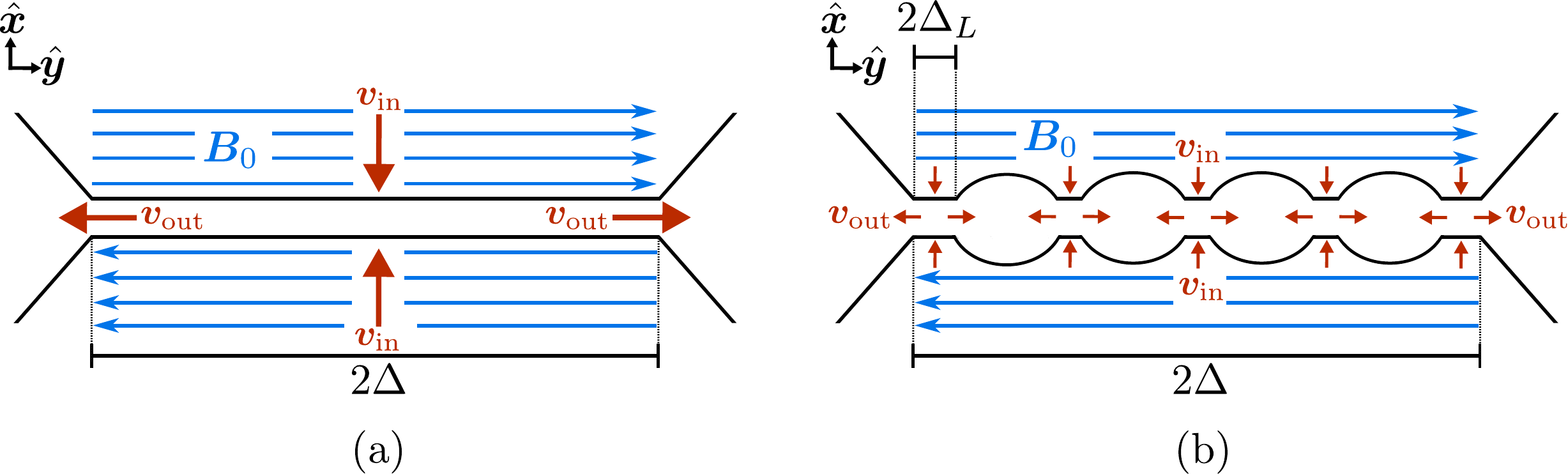}
    \caption{ A schematic diagram of a reconnecting Sweet--Parker sheet, panel (a), and a plasmoid unstable reconnection layer, panel (b), with upstream magnetic field amplitude $B_0$, half length $\Delta$, inflow speed $v_{\rm{in}}$ and outflow speed $v_{\rm{out}}$. Note in the plasmoid unstable sheet the system scale sheet is segmented into smaller sheets which have local half-length $\Delta_L$ and local Lundquist number, $S_L = 4\pi v_{\rm{A}} \Delta_L/(\eta c^2)$, of approximately the critical Lundquist number.
    }
    \label{fig:sheets}
\end{figure*}

In the Sweet--Parker model, a current layer of length equal to the system size undergoes steady state reconnection, where magnetic field lines flow into the layer at speed $v_{\rm{in}}$ and out along the outflow with speed $v_{\rm{out}}$, a visual example is given in \autoref{fig:sheets}. The reconnection rate has a direct dependence on the Lundquist number in the Sweet--Parker model,
\begin{equation}
    v_{\rm{rec}} = \sqrt{S^{-1}}.
\end{equation}
The analogue of the Sweet--Parker regime in numerical ideal MHD is probed by low-resolution simulations.
Current sheets with $S$ exceeding a critical threshold become plasmoid-unstable, fragmenting the sheet into regions where the local Lundquist number,
\begin{equation}
    S_{\rm{L}} = \frac{4\pi v_{\rm{A}} \Delta_{\rm{L}}}{\eta c^2},
\end{equation}
where $\Delta_{\rm{L}}$ is the local sheet half length, approaches the critical value, a simplified schematic is given in \autoref{fig:sheets}. In this regime, the reconnection rates measured in typical MHD simulations plateau at $v_{\rm{rec}} \sim \mathcal{O}(0.01)$ \cite{2007PhPl...14j0703L, Bhattacharjee.2009PhPl...16k2102B}. This finding has significant astrophysical implications, as many astrophysical systems possess resistivities so low that their Lundquist number far surpass the critical threshold. In the Sweet--Parker model the reconnection timescale depends directly on the Lundquist number
\begin{equation}
    t_{\rm{SP}} = \frac{L}{v_{\rm{A}}} \sqrt{S},
\end{equation}
while in the asymptotic regime the timescale becomes independent of $S$.
Without the fast reconnection facilitated by the plasmoid instability, the slower Sweet--Parker reconnection would be insufficient to explain the observed timescales in both solar and black hole flares. Therefore, it is crucial to produce the correct asymptotic behavior when performing global simulations to model these astrophysical environments. Despite the absence of an explicit resistivity, and hence Lundquist number, the asymptotic regime of ideal MHD simulations can be probed by high-resolution simulations where current sheets become plasmoid-unstable \cite{Nathanail_2020,Chashkina_2021,Ripperda_2022,Nathanail_2022,salas2024resolutionanalysismagneticallyarrested}. 

The maximum outflow is expected to be Alfv\'{e}nic; to see this, consider a pressure-balanced current sheet with no guide field and assume all the available energy is transferred into kinetic energy. From pressure balance in and out of the sheet \cite{Lyubarsky_2005MNRAS},
\begin{equation}
    p = \frac{B_0^2}{8\pi}. \label{eq:p_balance}
\end{equation}
Then conservation of energy from the reconnection layer to the upstream yields
\begin{equation}
    \rho h \Gamma^2 c^2 - p = \frac{B_0^2}{8\pi}
\end{equation}
where we have neglected the energy of the electric field in the reconnection layer, which is reasonable as $E \simeq \eta J$ and therefore $B_0^2 \gg E^2$ for a moderately conductive plasma. Substituting \autoref{eq:p_balance} we obtain
\begin{equation}
    \Gamma^2 = \frac{B^2}{4\pi \rho h c^2} = \sigma_{\rm{hot}},
\end{equation}
and hence $\Gamma^2 = \sigma_{\rm{hot}}$ (see \autoref{eq:sigmahot} for the definition of $\sigma_{\rm{hot}}$). Now consider the case where $\sigma_{\rm{hot}}\gg 1$ and suppose $v_{\rm{out}} \gg v_{\rm{in}}$, from which it follows that $\Gamma \simeq \Gamma_{\rm{out}}$ and
\begin{equation}
    \frac{1}{1-v_{\rm{out}}^2/c^2} = \sigma_{\rm{hot}} \simeq \sigma_{\rm{hot}} + 1,
\end{equation}
therefore
\begin{equation}
v_{\rm{out}} \simeq \sqrt{\frac{\sigma_{\rm{hot}}}{\sigma_{\rm{hot}} + 1}} \, c = v_{\rm{A}}.
\end{equation}
Interestingly, we arrive at the result $v_{\rm{out}} \simeq v_{\rm{A}}$ in the case of large $S$ and large $\sigma$ where $v_{\rm{rec}}\ll 1$ and $E^2 \ll B_0^2$. As well, this result holds for a reconnection layer in a strong guide field because the change in magnetic pressure in and out of the layer due to the guide field is negligible, in such a case the Alfv\'{e}nic outflow is along the total magnetic field, which is shown below.  

In \autoref{fig:rates_guide} we show the outflow and reconnection rate for a guide field-balanced sheet in ideal MHD, MD, resistive MHD, and resistive FFE.
In \autoref{fig:rates_pressure} we show the outflow and reconnection rate for a pressure-balanced sheet (algorithm shown in \autoref{app:measure_vrec}) in ideal MHD and resistive MHD. Notice that large $S$, or high-resolution ideal simulations, have outflow speeds that approach the in-plane Alfv\'{e}n speed as predicted above, and that the reconnection rate approaches the expected asymptotic value after steady state reconnection has begun.

\subsection{Guide Field-balanced Reconnection}\label{sec:Guide Field-balanced Reconnection}

For the guide field-balanced Harris sheet described in \autoref{sec:guidefield_perturbed}, the Alfv\'{e}n speed can be broken into two components \cite{Werner2017ApJ, Werner2021JPlPh}: along the guide field, $v_{\rm{A},\parallel}$; and, perpendicular to the guide field, $v_{\rm{A},\perp}$. The components are such that
\begin{equation}
    v_{\rm{A}}^2 = v_{\rm{A},\parallel}^2 + v_{\rm{A},\perp}^2,
\end{equation}
and
\begin{equation}
    v_{\rm{A}, \perp} = \sqrt{\frac{\sigma_{\rm{hot}, \perp}}{1 + \sigma_{\rm{hot}}}} \, c,  \label{eq:vAperp}
\end{equation}
where
\begin{equation}
    \sigma_{\rm{hot},\perp} = \sigma_{\rm{hot}} \left( \frac{B_0}{B_{\rm{guide}}} \right)^2 \left [ 1 + \left( \frac{B_0}{B_{\rm{guide}}} \right)^2  \right]^{-1},
\end{equation}
and
\begin{equation}
    v_{\rm{A}, \parallel} = \sqrt{\frac{\sigma_{\rm{hot}, \parallel}}{1 + \sigma_{\rm{hot}}}} \, c,    
\end{equation}
where
\begin{equation}
    \sigma_{\rm{hot},\parallel} = \sigma_{\rm{hot}} \left [ 1 + \left( \frac{B_0}{B_{\rm{guide}}} \right)^2  \right]^{-1}.
\end{equation}
For the choice of parameters in \autoref{tab:Vrec_sims_guide}, it follows that $\sigma_{\rm{hot},\perp} =2$, $\sigma_{\rm{hot}, \parallel} = 8$, and $v_{\rm{A},\perp} \sim 0.43 \, c$. 
In the force-free limit, where $\sigma_{\rm{hot}} \rightarrow \infty$, the total Alfv\'{e}n speed approaches the speed of light, but the in-plane and parallel components are limited by the guide field,
\begin{equation}
    \lim_{\sigma_{\rm{hot}} \rightarrow \infty }\frac{v_{\rm{A},\perp}}{c} =  \left( \frac{B_0}{B_{\rm{guide}}} \right) \left [ 1 + \left( \frac{B_0}{B_{\rm{guide}}} \right)^2  \right]^{-1/2} ,
\end{equation}
and
\begin{equation}
     \lim_{\sigma_{\rm{hot}} \rightarrow \infty }\frac{v_{\rm{A},\parallel}}{c} = \left [ 1 + \left( \frac{B_0}{B_{\rm{guide}}} \right)^2  \right]^{-1/2} .
\end{equation}
For the choice of parameters in \autoref{tab:Vrec_sims_guide}, it follows that in the force-free limit, $v_{\rm{A},\perp} \rightarrow c/\sqrt{5} \sim 0.45 \, c$ and $v_{\rm{A},\perp} \rightarrow 2c/\sqrt{5} \sim 0.89 \, c$. 

For this simulation setup, we define two Lundquist numbers, one in-plane
\begin{equation}
    S_\perp = \frac{4\pi v_{\rm{A},\perp} \Delta }{\eta c^2}, 
\end{equation}
and one out-of-plane
\begin{equation}
    S_\parallel = \frac{4\pi v_{\rm{A},\parallel} \Delta }{\eta c^2}.
\end{equation}
Note that because
\begin{equation}
v_{\rm{A}, \perp}< \sqrt{\frac{\sigma_{\rm{hot}, \perp}}{1 + \sigma_{\rm{hot}, \perp}}} \,c= v_{\rm{A}, \sigma_{\rm{hot}, \perp}},
\end{equation}
the $S_\perp$ and hence reconnection rate is suppressed due to the inertia of the strong guide field. This suppression has been found in Particle-in-Cell (PIC) studies to be of order $\sim \left( B_0/B_{\rm{guide}}\right)$ \cite{Werner2017ApJ} in the asymptotic regime. The expected resistive MHD scaling of reconnection rate with respect to $S_\perp$ is compared to the ideal MHD, MD, and resistive FFE result, a test to determine how well ideal MHD, MD, and resistive FFE with a resistive current prescription can mimic resolved magnetic reconnection in a force-free situation. 

\begin{figure*}[htbp]
    \centering
    \begin{subfigure}{\textwidth}
        \centering
        \includegraphics[width=\textwidth]{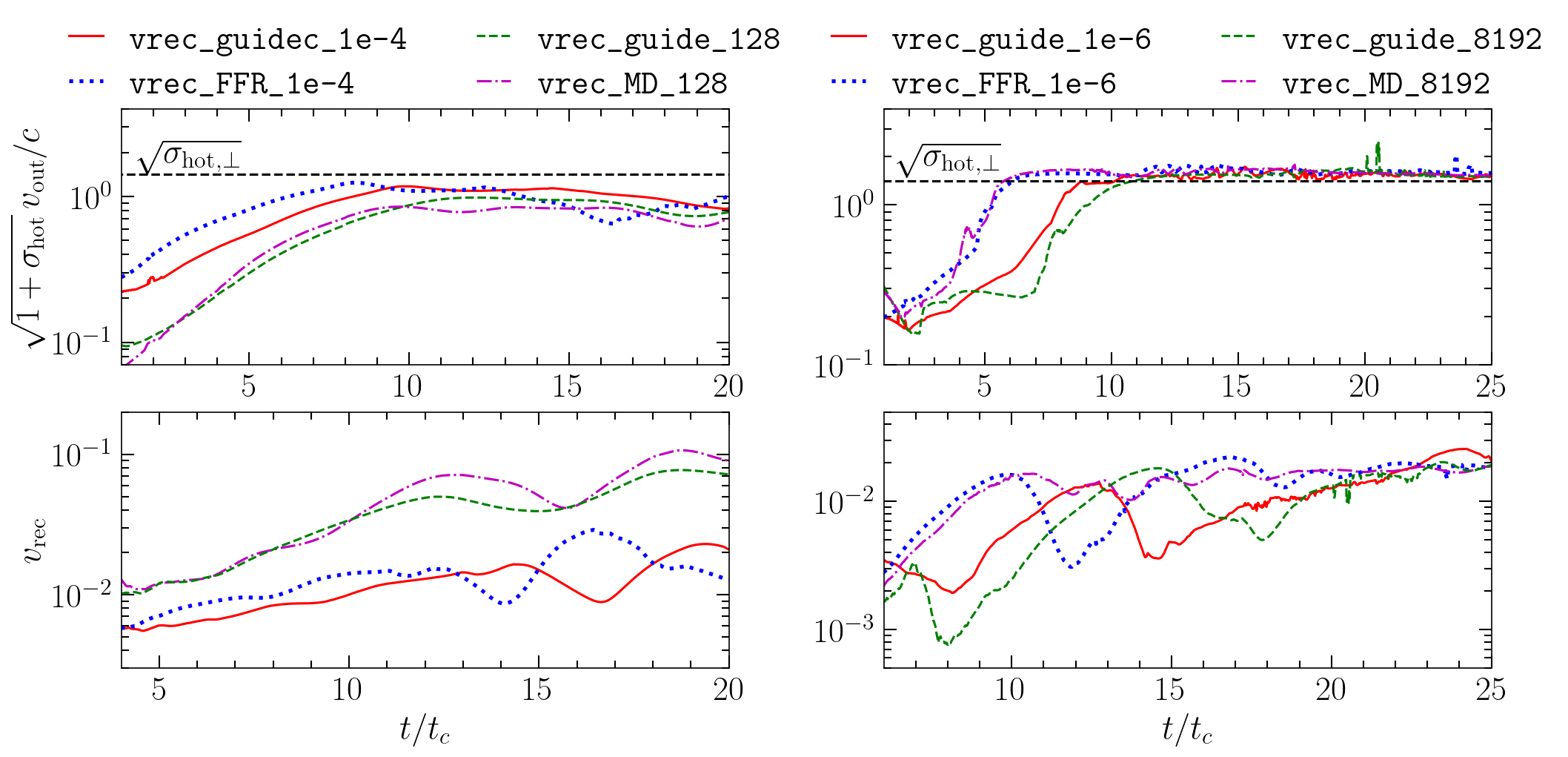}
        \caption{Guide field-balanced Harris sheet.}
        \label{fig:rates_guide}
    \end{subfigure}
    \hfill
    \begin{subfigure}{\textwidth}
        \centering
        \includegraphics[width=\textwidth]{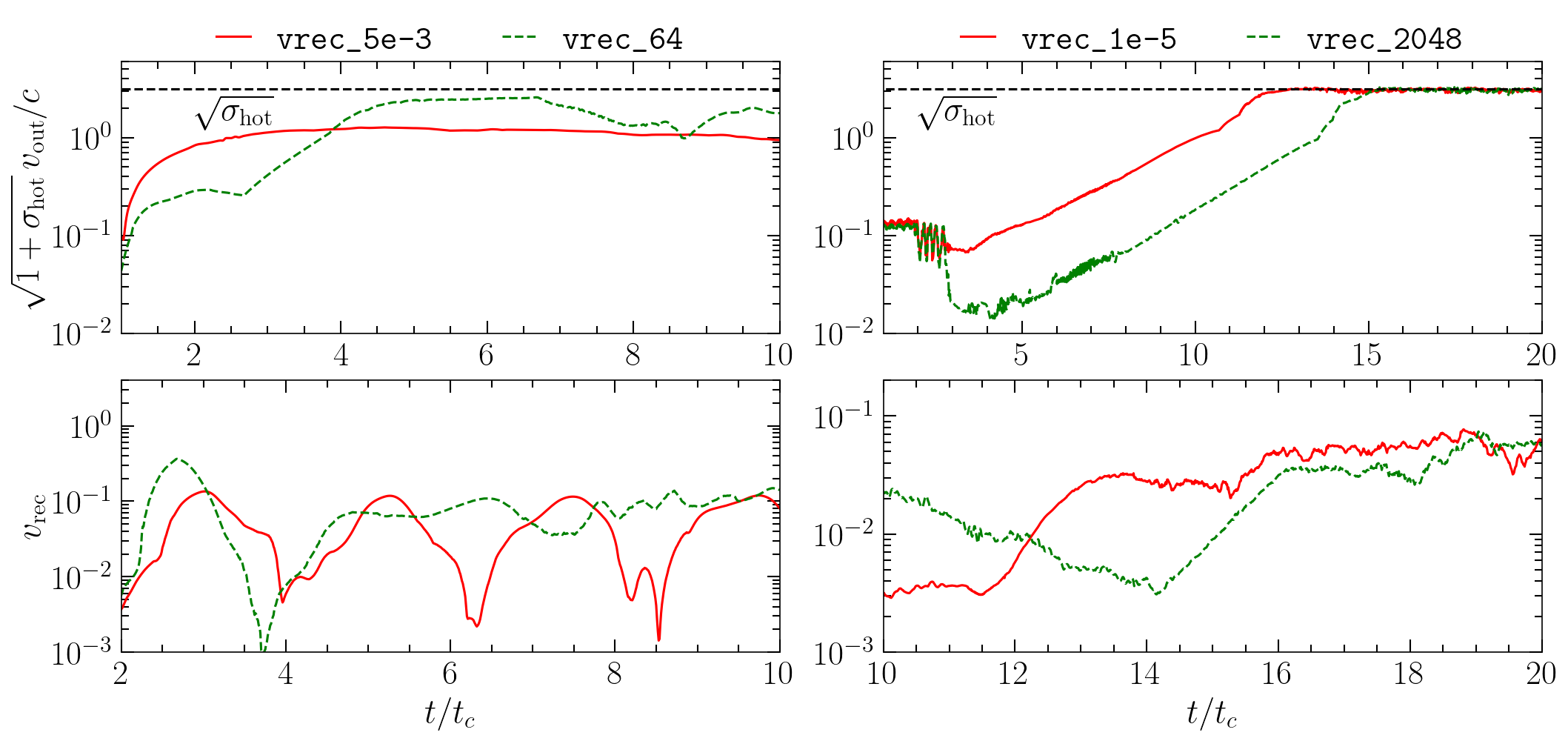}        \caption{Pressure-balanced Harris sheet.}
        \label{fig:rates_pressure} 
    \end{subfigure}
    \caption{ 
  The measured value of $\sqrt{1+\sigma_{\rm{hot}}}\, v_{\rm{out}}/c$  and $v_{\rm{rec}}$ as functions of time. As the outflow speed approaches the Alfv\'{e}n speed, $\sqrt{1+\sigma_{\rm{hot}}}\, v_{\rm{out}}/c$ will approach $\sqrt{\sigma_{\rm{hot}}}$. Note the simulations in the left column are low Lundquist number (low resolution) and in the Sweet--Parker regime, while the simulations in the right column are high Lundquist number (high resolution) and in the asymptotic regime. Note that for \autoref{fig:rates_guide}, the MD and resistive FFE simulations have the in-plane Alfv\'{e}n speed limited by the guide field to $\sim 0.45 c$, while in the ideal and resistive MHD simulations, with finite magnetization, the in-plane Alfv\'{e}n speed is $\sim 0.43 c$.
    }
    \label{fig:vout_and_vrec}
\end{figure*}

\begin{figure*}[htbp]
    \centering
    \includegraphics[width=\textwidth]{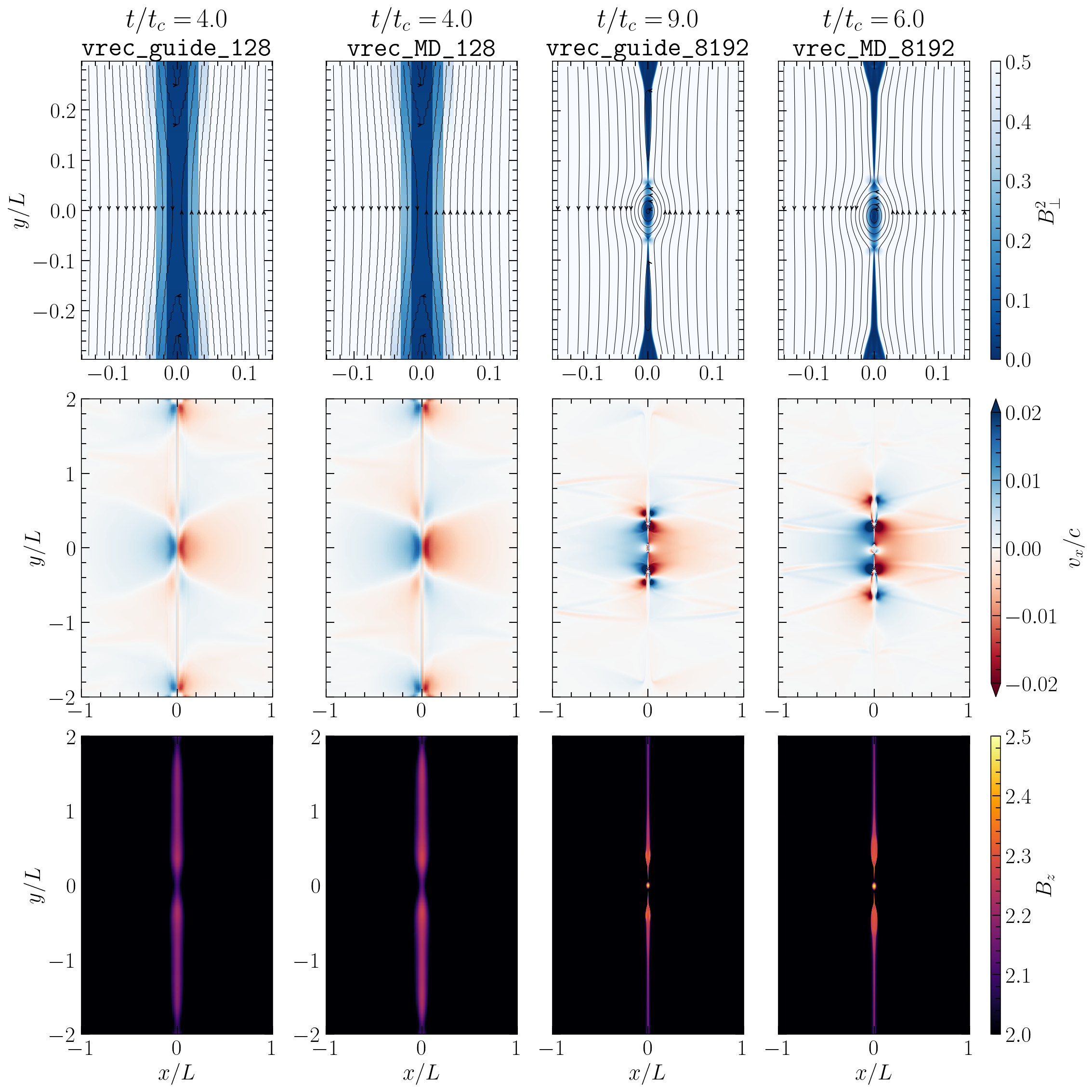}
    \caption{Guide field-balanced Harris sheet. A zoom-in and visualization of the in-plane magnetic field strength, $B_\perp^2$, and magnetic field lines, the speed in the $x$-direction, $v_x/c$, and the guide magnetic field, $B_z$, as the current sheet is undergoing reconnection. Note that for the MD simulation the velocity is the $\bm{E}\times\bm{B}$ drift. 
    Left Columns: Both the ideal MHD and MD simulations are shown at $t/t_c=4$. 
    Right Columns: The ideal MHD simulation is shown at $t/t_c=9$, while the MD simulation is shown at $t/t_c = 6$. 
    Time realizations were chosen such that both simulations were at a similar point in their reconnection evolution, having just plateaued in both the $v_{\rm{out}}$ and $v_{\rm{rec}}$, as shown in \autoref{fig:rates_guide}.
    The simulations in the left column are low resolution and in the Sweet--Parker regime, while the simulations in the right column are high resolution and in the asymptotic regime (\autoref{fig:vrec} shows the exact resolution criteria for the transition, based on the simulation parameters we use in this study). 
    }
    \label{fig:MD_beta_and_vx_and_B2}
\end{figure*}

\begin{figure*}[htbp]
    \centering
    \includegraphics[width=\textwidth]{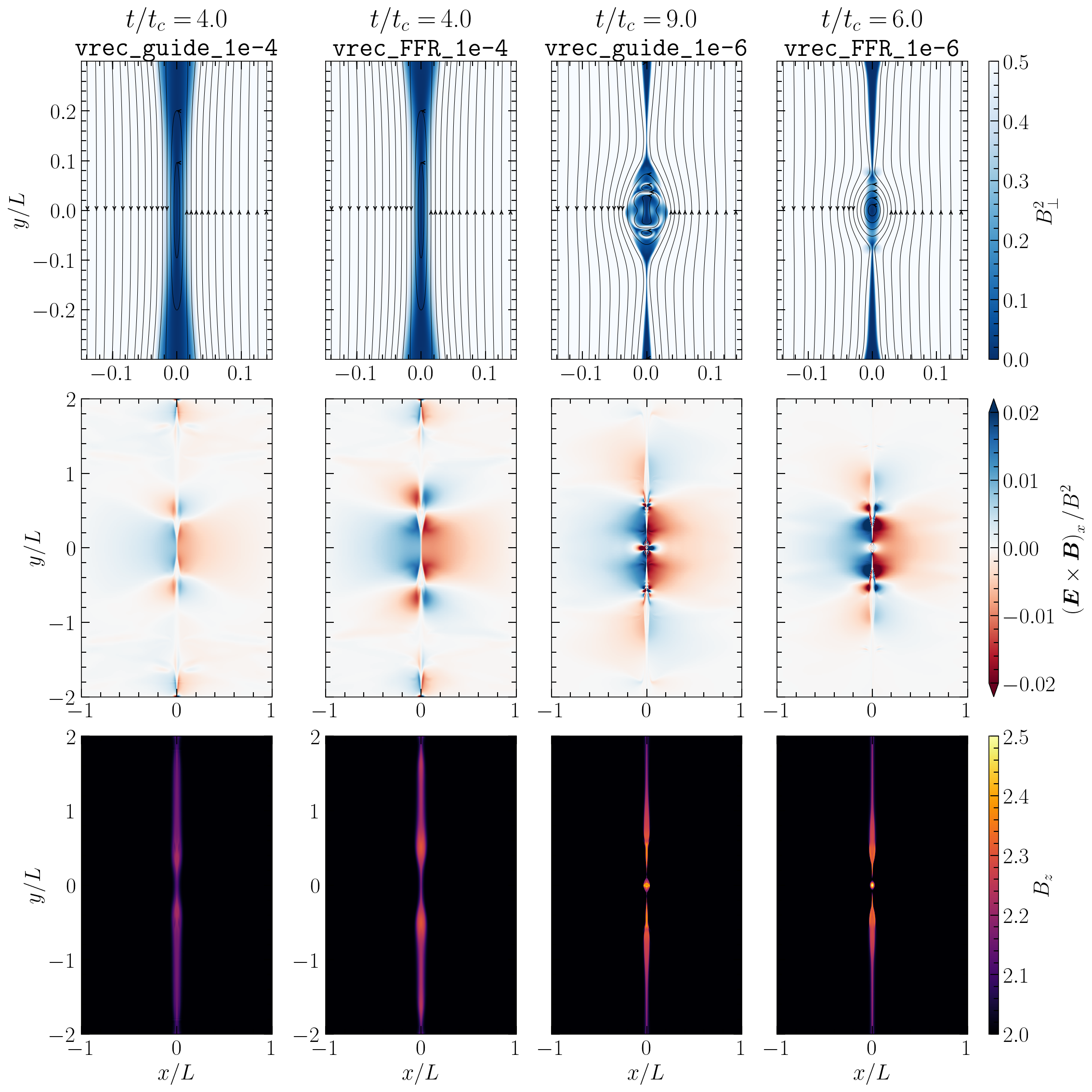}
    \caption{Guide field-balanced Harris sheet. A zoom-in and visualization of the in-plane magnetic field strength, $B_\perp^2$, and magnetic field lines, the drift speed in the $x$-direction, $\left(\vecE\times\vecB\right)_x/B^2$, and the guide field, $B_z$, as the current sheet is undergoing reconnection. 
    Left Columns: Both the resistive MHD and FFE simulations are shown at $t/t_c=8$. 
    Right Columns: The resistive MHD simulation is shown at $t/t_c=9$, while the resistive FFE simulation is shown at $t/t_c = 6$. 
    Time realizations were chosen such that both simulations were at a similar point in their reconnection evolution, having just plateaued in both the $v_{\rm{out}}$ and $v_{\rm{rec}}$, as shown in \autoref{fig:rates_guide}. The simulations in the left column are low Lundquist number and in the Sweet--Parker regime, whilst the simulations in the right column are high Lundquist number and in the asymptotic regime. 
    }
    \label{fig:rrmhdffr_Bz_ExB_B2perp}
\end{figure*}

\begin{figure*}[htbp]
    \centering
    \begin{subfigure}{0.49\textwidth}
        \centering
        \includegraphics[width=\textwidth]{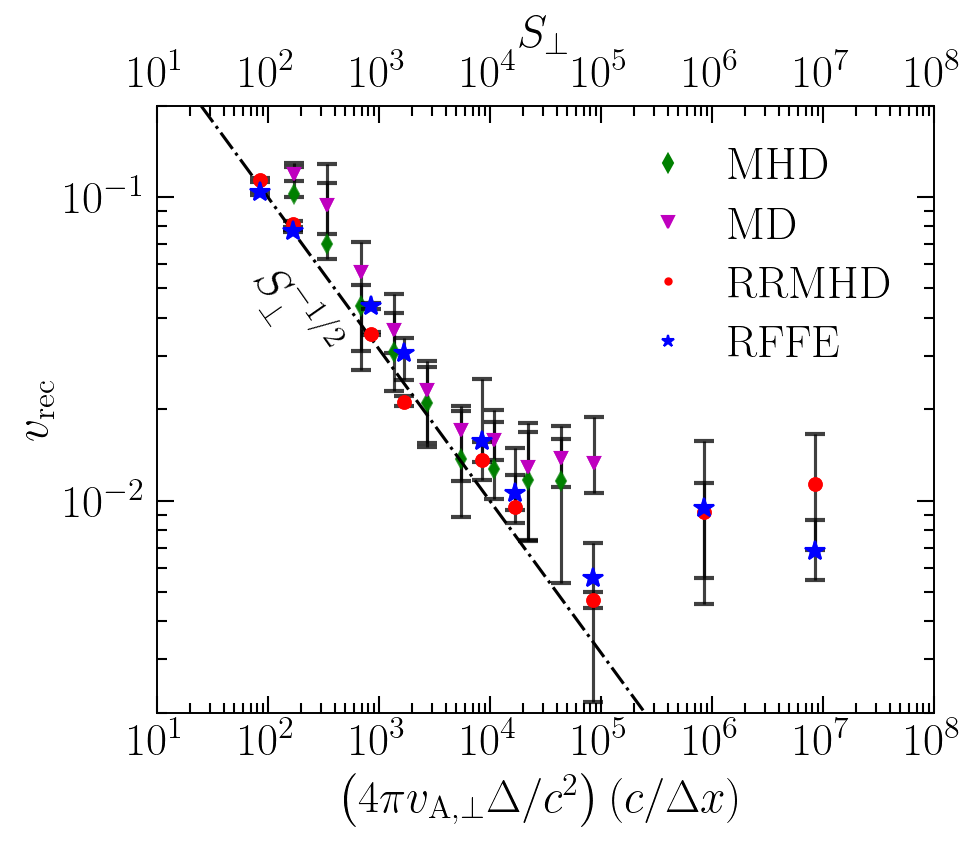}
        \caption{Guide field-balanced Harris sheet.}
        \label{fig:vrec_guide}
    \end{subfigure}
    \hfill
    \begin{subfigure}{0.49\textwidth}
        \centering
        \includegraphics[width=\textwidth]{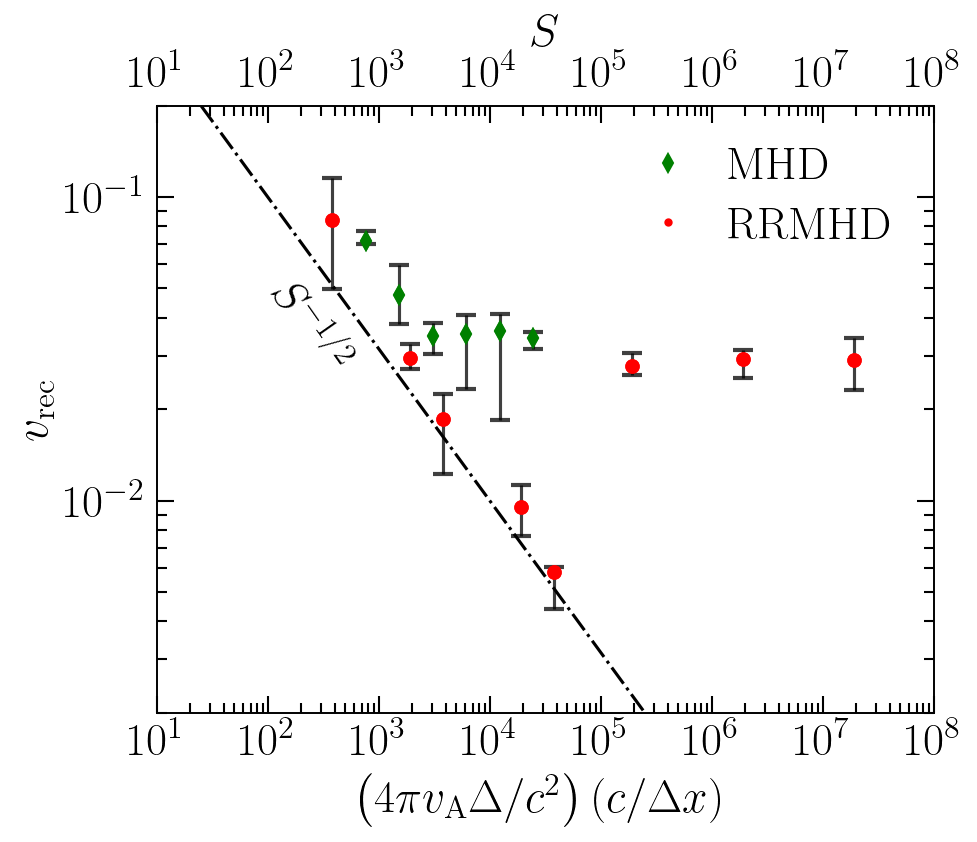}        \caption{Pressure-balanced Harris sheet.}
        \label{fig:vrec_pressure} 
    \end{subfigure}
    \caption{The time averaged reconnection rate for the simulations in \autoref{tab:Vrec_sims_guide} and \autoref{tab:Vrec_sims}. Details of how  the reconnection rate, $v_{\rm rec}$, is measured are presented in \autoref{app:measure_vrec}. The relativistic resistive MHD (RRMHD) and resistive FFE (RFFE) results are plotted as a function of Lundquist number, $S$, in the pressure-balanced sheet simulations, and as a function of in-plane Lundquist number, $S_\perp$, for the guide field-balanced simulations. The ideal MHD and MD results are plotted as a function of the inverse of the grid scale size, $(1/\Delta x)$. The effective Lundquist number is approximate by substituting $\eta_{\rm{num}} = c/\Delta x$ which approximately collapse all $v_{\rm rec}$ functions to a single curve, providing an approximate map between $1/\Delta x$ and $S$. Note the Sweet--Parker scaling, $v_{\rm rec} \propto S^{-1/2}$, holds relatively well for all low-resolution and low-$S$ simulations. }
    \label{fig:vrec}
\end{figure*}

At both low and high resolutions, ideal MHD and MD macroscopically appear similar to resistive MHD, as shown in \autoref{fig:MD_beta_and_vx_and_B2} and \autoref{fig:rrmhdffr_Bz_ExB_B2perp}.  The resistive FFE simulations match the resistive MHD simulations at low $S$ both qualitatively, as shown in \autoref{fig:rrmhdffr_Bz_ExB_B2perp}, and in terms of the outflow speeds and $v_{\rm rec}$, as shown in \autoref{fig:rates_guide}. At large $S$, the resistive FFE simulations agree qualitatively, as shown in \autoref{fig:rrmhdffr_Bz_ExB_B2perp}, and in terms of the outflow speeds and $v_{\rm rec}$ but with a difference in the onset time for reconnection, as shown in \autoref{fig:rates_guide}. Interestingly, in \autoref{fig:rates_guide}, at small $S$ or low resolution, the ideal schemes and resistive schemes agree best, while at large $S$ and high resolution, the force-free and MHD schemes agree best.
In \autoref{fig:vrec_guide}, we show that the ideal MHD and MD guide field-balanced reconnecting current sheet simulations have an analogue of the Sweet--Parker regime at low resolutions ($4\pi v_{\rm{A},\perp}\Delta/c\Delta x \lesssim 5\times 10^4$), and that resistive FFE agrees with the resistive MHD reconnection rate at low and high $S$. At high resolutions ($4\pi v_{\rm{A},\perp}\Delta/c\Delta x \gtrsim 5 \times 10^4$), the ideal MHD and MD simulations have an asymptotic $v_{\rm rec}$ similar to resistive MHD and resistive FFE has an asymptotic $v_{\rm rec}$, close to that of resistive MHD. 

It is important to note that magnetically supported current sheets in a strong guide field for which the force-free models were tested in this paper differ greatly from current sheets which have a significant plasma pressure. For example, in the equatorial current sheet about a rotating neutron star, the plasma pressure and inertia become important, which is not accounted for in force-free models\cite{2022ARA&A..60..495P}. As well, the plasma is unable to be heated in the FFE models which therefore miss a significant portion of the physical dissipation taking place during magnetic reconnection.

\begin{figure*}[htbp]
    \centering
    \includegraphics[width=\textwidth]{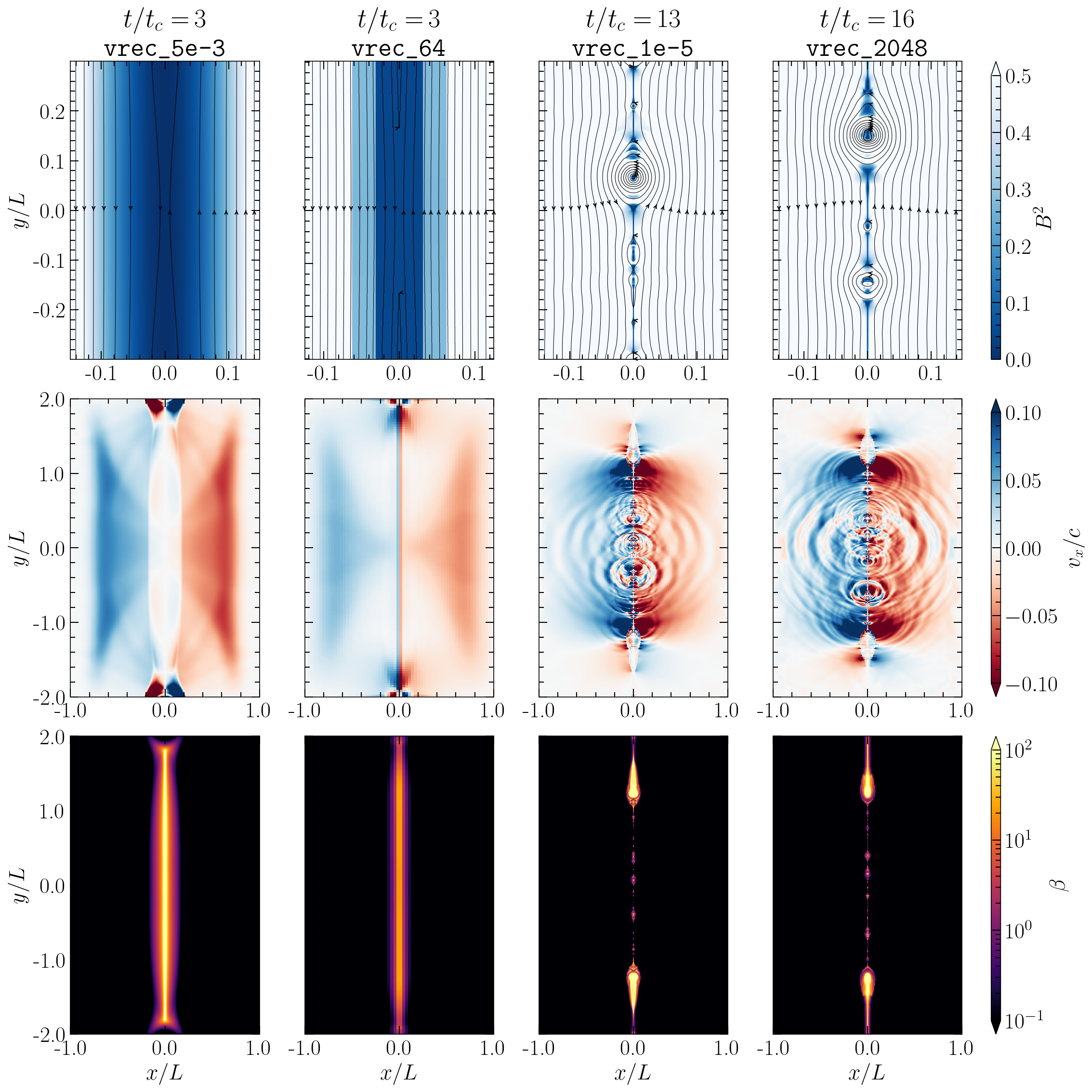}
    \caption{Pressure-balanced Harris sheet. A zoom-in and visualization of the magnetic field strength, $B^2$, and magnetic field lines, the velocity in the $x$-direction, $v_x$, and the gas ratio, $\beta = 8\pi p/B^2$, as the current sheet is undergoing reconnection. 
    Left Columns: Both the ideal MHD and resistive MHD simulations are shown at $t/t_c=3$. 
    Right Columns: The resistive MHD simulation is shown at $t/t_c=13$, while the ideal MHD simulation is shown at $t/t_c = 16$. 
    Time realizations were chosen such that both simulations were at a similar point in their reconnection evolution, having just plateaued in both the $v_{\rm{out}}$ and $v_{\rm{rec}}$, as shown in \autoref{fig:rates_pressure}.
     The simulations in the left column are low Lundquist number, or, low resolution, and in the Sweet--Parker regime, whilst the simulations in the right column are high Lundquist number, or, high resolution, and in the asymptotic regime. 
    }
    \label{fig:idealrrmhd_beta_and_vx_and_B2}
\end{figure*}

\subsection{Pressure-balanced Reconnection}

For the pressure-balanced Harris sheet described in \autoref{sec:pressureHarris}, there is no magnetic field component perpendicular to the plane of reconnection and therefore there is no velocity perpendicular to the plane of reconnection, hence the Alfv\'{e}n speed is 
\begin{equation}
    v_{\rm{A}} = \sqrt{\frac{\sigma_{\rm{hot}}}{1 + \sigma_{\rm{hot}}}} \, c,
\end{equation}
and the Lundquist number is 
\begin{equation}
    S = \frac{4\pi v_{\rm{A}} \Delta}{\eta c^2}.
\end{equation}
The reconnection is expected to proceed as described in \autoref{sec:SPandPlasmoid}, beginning in the Sweet--Parker regime until the Lundquist number becomes sufficiently large and enters the plasmoid-dominated regime where the rate becomes asymptotic. We compare the expected resistive MHD scaling of the reconnection rate with respect to $S$ to the ideal MHD scaling with respect to resolution -- a test to determine how well ideal MHD with $\sim$ grid scale diffusion can faithfully resemble resistively resolved magnetic reconnection.

\autoref{fig:vrec_pressure} shows that the pressure-balanced reconnecting current sheet agrees with the guide field simulations demonstrating that ideal MHD has a low resolution ($4\pi v_{\rm{A}}\Delta/c\Delta x \lesssim 5 \times 10^3$) analogue of the Sweet--Parker regime, and that the resistive MHD results reproduce the Sweet--Parker scaling in the expected $S$ regime. Crucially, both ideal and resistive MHD plateau at a similar value for $v_{\rm rec}$ in the fast reconnection, plasmoid dominated regime at high Lundquist or high resolution ($4\pi v_{\rm{A}}\Delta/c\Delta x \gtrsim 5\times 10^3$), similar to the results seen in the guide field-balanced configuration. Additionally, both low $S$, resistive MHD simulations and low resolution ideal MHD simulations, and large $S$, resistive MHD simulations and high resolution ideal MHD simulations appear qualitatively similar macroscopically, i.e. in overall current sheet structure, plasmoid structure, magnetic field structure and inflow velocities as shown in \autoref{fig:idealrrmhd_beta_and_vx_and_B2}. 
Additionally, more quantitatively, the outflow and $v_{\rm rec}$ agree, as shown in \autoref{fig:rates_pressure}, where the high $S$, resistive simulation and high resolution simulation demonstrate similar structure in their time dependence, albeit with an offset in time. In \autoref{fig:rates_pressure}, the low $S$ simulation and low resolution simulation do not have as similar structure in the outflow and $v_{\rm rec}$.

\section{Discussion and Conclusions}\label{sec:Discussion}

We have compared the diffusion and reconnection properties of four physical models commonly used in high energy astrophysical simulations: ideal MHD, resistive MHD, MD, and resistive FFE. The diffusion properties of the two ideal models, ideal MHD and MD, agree. Similarly, the diffusion properties of the two resistive models, resistive MHD and resistive FFE, agree. We tested diffusion in the strong guide field, high magnetization regime where the FFE and MHD models are expected to agree best. The reconnection properties of all four models agree remarkably well, each having some Sweet--Parker like regime and an asymptotic regime with similar asymptotic reconnection rates. We discuss the diffusion and reconnection properties of these models in detail below.

Our reconnection experiments with a pressure-balanced sheet show that it is computationally less expensive to obtain a reconnection rate of $\mathcal{O}(0.01)$ with ideal MHD. Reaching the asymptotic rate in the case of a current sheet with a strong guide field in ideal MHD is more expensive by approximately a factor of ten in resolution for a guide field strength of $B_{\rm{guide}}/B_0 = 2$, compared to ideal MHD with a pressure-balanced sheet, matching the change seen in the required in-plane Lundquist number to asymptote. The resolutions required for asymptotic reconnection for a current sheet with $B_{\rm guide} = 0$ in ideal MHD agree with what previous global simulations have found \cite{Ripperda_2022, 2024MNRAS.533..254S}. 

By choosing a proportionality constant for $S_{\rm{num}} \propto 4\pi \Delta v_{\rm{A}}/c\Delta x$ of 1,  in \autoref{fig:vrec} we found that all of the different simulations, regardless of underlying model, collapse to a single curve that goes like $S_{\rm{num}}^{-1/2}$ in the Sweet--Parker regime, and becomes independent of $S$ for $S_{\rm{num}} \gtrsim 10^4$. This means that the effective (numerical) Lundquist number is roughly the Lundquist number with $\eta \rightarrow \Delta x/c$, for both a pressure-balanced sheet 
\begin{equation}
    S_{\rm{num}} \simeq \frac{4\pi v_{\rm{A}} \Delta}{c \Delta x}, \label{eq:S_num_pressure}
\end{equation}
and a guide field-balanced sheet,
\begin{equation}
    S_{\rm{num}, \perp} \simeq \frac{4\pi v_{\rm{A},\perp} \Delta}{c \Delta x}, \label{eq:S_num_guide}
\end{equation}
where $\Delta x$ is the size of an individual grid cell. 
From \autoref{eq:S_num_pressure} and \autoref{eq:S_num_guide} the approximate numerical resistivity, $\eta_{\rm{num}}$, for a reconnecting sheet can be approximated as
\begin{equation}
    \eta_{\rm{num}} \simeq \frac{\Delta x}{c}. \label{eq:eta_num}
\end{equation}
This suggests that if one runs an ideal simulation, regardless of the underlying physical model, one can use \autoref{eq:eta_num} to determine an effective $S$ for a given resolution when using an overall second order method. Importantly, the order of the spatial reconstruction method does not have a significant impact on this result. Furthermore, it allows one to estimate the resolution required to resolve the resistive scale features of a reconnecting current sheet in a resistive MHD simulation. To resolve the resistive scale features of a current sheet with Lundquist number $S$, the resolution must be chosen such that $S_{\rm{num}} \gtrsim S$. Essentially the maximum $S$ which can be resolved is given by $S_{\rm{num}}$.

We show that the Ohmic decay of current structures is not properly captured by numerical resistivity in both ideal MHD and MD. Instead, the diffusion of the current sheet is subdiffusive $\left\langle x^2 \right\rangle \propto t^{\alpha}$, $\alpha < 1$, deviating from the well-established Ohmic decay laws predicted by resistive MHD theory, e.g., $\left\langle x^2 \right\rangle \propto t$. The exponent depends on both the resolution and the order of the spatial reconstruction scheme. As expected, a lower order more diffusive spatial reconstruction scheme has a larger exponent than a higher order less diffusive scheme.
Astrophysical plasmas are typically highly conductive, leading to slow Ohmic heating rates. Thus, using high resolutions with correspondingly slow numerical Ohmic dissipation provides an appropriate approximation, provided that the details of Ohmic heating are not the focus of the study.
Importantly, if the deviation from $\langle x^2 \rangle \propto t$ was solely due to the numerical resistivity being a function of time, decreasing as the sheet becomes more well resolved as it expands, then the current profile would remain a Gaussian which it is clearly not as shown in \autoref{fig:current_profile}. Resistive FFE using the prescription in \cite{Alic2012_FFR} was shown to perfectly match resistive MHD for the Ohmic decay of a current sheet in a strong guide field.
In non-uniform grids, e.g. when employing AMR schemes, the effective numerical resistivity varies with resolution—stronger dissipation in coarse regions, weaker in refined ones. This can create uneven dynamics: asymptotic reconnection where resolution is high, Sweet--Parker like where it is low. In global compact-object simulations of plasma flows, such as those using modified Kerr--Schild coordinates \cite{2004ApJ...611..977M, Porth:2016rfi}, this imbalance can shape the system’s evolution, altering reconnection rates and Ohmic decay.

It is important to state that converging on the reconnection rate 
does not mean that the smallest scale features in the current sheet are fully resolved. 
Typical astrophysical $S$ greatly exceed the critical $S$ such that there are many orders of magnitude between the largest and smallest current sheet structures in a reconnection layer. Due to numerical limitations it has not been possible to resolve more than a couple of order of magnitude in separation and it is therefore difficult to conclude whether the physics is indeed asymptotic.
Resolving the substructure of reconnecting current sheets is necessary for careful studies, e.g. local studies of the initial kick of test particles in $E>B$ regions \cite{2025arXiv250100979G,2003ApJ...586...72L}. Or for global configurations where acceleration in reconnection layers is important, e.g. particle energization in the jet boundary reconnection layer \cite{2021ApJ...907L..44S, 2022PhRvL.128n5101M}. Crucially, in the two ideal models $E>B$ regions do not exist which are essential for particle acceleration.
Some configurations or measurements are not sensitive to the substructure of reconnecting current sheets, e.g. the flux decay of a balding black hole depends directly on the reconnection rate \cite{2021PhRvL.127e5101B}. 
Reconnection in a turbulent plasma, or 3D reconnection which can result in a turbulent current sheet, is still an unsolved problem due to the high resolution required \cite{2021ApJ...923L..13C,2022SciA....8N7627D}. It is therefore unclear whether the 2D understanding of asymptotic reconnection, and hence our conclusions, translate to 3D systems which should be the subject of future studies.
As well, throughout this paper we have neglected viscous effects. For highly magnetized MHD plasmas, the ratio between the resistive and viscous dissipation rates scales as $\sqrt{\sigma} \gg 1$ \cite{2021ApJ...923L..13C}.
Viscosity can nevertheless play an important role (e.g., in accretion disks; e.g., \cite{2023PhRvL.130k5201G, 2015ApJ...810..162C}). It is therefore important to distinguish between physical and numerical viscosity, as well as to consider the role of the magnetic Prandtl number in reconnection \cite{2016PhPl...23c2111C}; in this context, a similar comparison in the non-relativistic regime would be valuable for assessing the generality of our results.

\section*{Acknowledgments}

The authors thank Amitava Bhattacharjee, Alexander Philippov, Lorenzo Sironi, and Christopher Thompson for productive conversations and guidance. 

We acknowledge the support of the Natural Sciences and Engineering Research Council of Canada (NSERC), [funding reference number 568580]
Cette recherche a \'et\'e financ\'ee par le Conseil de recherches en sciences naturelles et en g\'enie du Canada (CRSNG), [num\'ero de r\'ef\'erence 568580].

We acknowledge the support of the Natural Sciences and Engineering Research Council of Canada (NSERC), [CGS D - 588952 - 2024]. Cette recherche a \'{e}t\'{e} financ\'{e}e par le Conseil de recherches en sciences naturelles et en g\'{e}nie du Canada (CRSNG), [CGS D - 588952 - 2024].

M.~P.~G., T.~G. and B.~R. are supported by the Natural Sciences \& Engineering Research Council of Canada (NSERC), the Canadian Space Agency (23JWGO2A01), and by a grant from the Simons Foundation (MP-SCMPS-00001470). B.~R. acknowledges a guest researcher position at the Flatiron Institute, supported by the Simons Foundation.
F.~B.\ acknowledges support from the FED-tWIN programme (profile Prf-2020-004, project ``ENERGY'') issued by BELSPO, and from the FWO Junior Research Project G020224N granted by the Research Foundation -- Flanders (FWO).

J.~R.~B. further acknowledges the support from NSF Award 2206756, as well as high-performance computing resources provided by the Leibniz Rechenzentrum and the Gauss Center for Supercomputing grant~pn76gi~pr73fi and pn76ga.

The computational resources and services used in this work were partially provided by facilities supported by the VSC (Flemish Supercomputer Center), funded by the Research Foundation Flanders (FWO) and the Flemish Government – department EWI, by the Scientific Computing Core at the Flatiron Institute, a division of the
Simons Foundation, and by Compute Ontario and the Digital Research Alliance of Canada (alliancecan.ca) compute allocation rrg-ripperda.

\appendix

\section{Magnetodynamic Formulation}\label{sec:md}

The magnetodynamic formulation in \verb+BHAC+ solves Faraday’s law and the momentum equation for the Poynting flux to evolve the magnetic field, $\mbf{B}$, and the Poynting flux, $\mbf{S} = \mbf{E}\times \mbf{B}$ \cite{Komissarov2002, Komissarov2004c, 2007A&A...473...11D}. The constrained transport method is used to evolve the magnetic flux and to preserve $\del \cdot \mbf{B}=0$ to machine precision \cite{Olivares2019}. Note throughout this section we work in units with $c=k_B=1$, $B/\sqrt{4\pi} \rightarrow B$, $E/\sqrt{4\pi} \rightarrow E$, $\sqrt{4\pi} J \rightarrow J$, $\sqrt{4\pi} q \rightarrow q$, and $\eta/4\pi \rightarrow \eta$. As well, we work in the 3+1 ADM formalism for completeness
\begin{equation}
    g^{\mu \nu} = \left(  
    \begin{array}{cc}
        -1/\alpha^2 & \beta^j/\alpha^2 \\
        \beta^j/\alpha^2 & \gamma^{ij} - \beta^i \beta^j/\alpha^2 
    \end{array}
    \right),
\end{equation}
while all simulations present in this paper are in flat Minkowski spacetime with $\alpha =1$, $\beta^i = 0$ and $\gamma^{ij} = \delta^{ij}$. The evolution equations are written in conservative form as
\begin{equation}
    \partial_t \left( \sqrt{\gamma} \bm{U} \right) + \partial_i \left( \sqrt{\gamma} \bm{F}^i \right) = \sqrt{\gamma} \bm{S},
\end{equation}
with conserved variables $\bm{U}$ and fluxes $\bm{F}$ defined as
\begin{align}
    \bm{U} = \left[ 
    \begin{array}{c}
        S_j  \\
        B^j 
    \end{array}
    \right],  
    &&
    \bm{F}^i = \left[ 
    \begin{array}{c}
        \alpha W^i_j - \beta^i S_j  \\
        \mathcal{V}^i B^j - B^i \mathcal{V}^j 
    \end{array}
    \right], 
\end{align}
where we define the transport velocity $\mathcal{V}^i = \alpha v^i - \beta^i$. The sources are defined as
\begin{equation}
    \bm{S} = \left[ 
    \begin{array}{c}
        \frac{1}{2}\alpha W^{ik} \partial_j \gamma_{ik} + S_i \partial_j \beta^i - U \partial_j \alpha  \\
        0
    \end{array}
    \right],
\end{equation}
with energy density
\begin{align}
    U & = \frac{1}{2}\left( E^2 + B^2 \right) = \frac{1}{2}\left( \frac{S^2}{B^2} + B^2 \right) = \frac{B^2}{2} \left( 1 + v^2 \right),
\end{align}
where we have used that $\mbf{E}\perp \mbf{B}$ in ideal magnetodynamics and hence $S^2/B^2 = E^2$, and similarly that because the velocity will be the $\mbf{E}\times\mbf{B}$ drift velocity that $E^2 = v^2 B^2$. The three-energy momentum tensor $W^{ij}$ is defined as
\begin{align}
    W^{ij} & = \gamma^i_\mu \gamma^j_\nu T^{\mu\nu}\\ & = - E^i E^j - B^i B^j + \left[ \frac{1}{2} (E^2 + B^2) \right] \gamma^{ij}\\
    & = \frac{S^i S^j}{B^2} - \frac{B^i B^j}{\Gamma^2} + p_{\rm{EM}} \gamma^{ij}
\end{align}
where we have introduced the total pressure $p_{\rm{EM}}$
\begin{equation}
    p_{\rm{EM}} = \frac{1}{2} \left( B^2 - E^2 \right) = \frac{B^2}{2\Gamma^2},
\end{equation}
and the Lorentz factor which can be written as
\begin{equation}
    \Gamma^2 = \frac{B^4}{B^4 - S^2}.
\end{equation}

In principle, the system can be solved entirely in conserved variables. However, for reconstruction, it might be advised to work on a robust variable that does not lead to violations of e.g. $v^2 > 1$. We hence introduce Lorentz-factor times (contravariant) 3-drift velocity $\tilde{u}^i = \Gamma v^i$ in our set of primitive variables
\begin{equation}
    \bm{P} = [\tilde{u}^i, B^i].
\end{equation}
We chose not to introduce an auxiliary variable, but to re-compute $\Gamma$ from either primitives or conserved variables as
\begin{equation}
    \Gamma^2 = 1 + \tilde{u}^2 = \frac{B^4}{B^4 - S^2}.
\end{equation}
The $j$-th component of the momentum-flux is
\begin{align}
    F^i(S_j) & = \alpha W_j^i - \beta^i S_j\\
             & = \mathcal{V}^i S_j + \alpha p_{\rm{EM}} \delta^i_j - \alpha \frac{B^i  B_j}{\Gamma^2}
\end{align}
where the latter form is amenable to split-off the transport flux $\mathcal{V}^i S_j$. Faraday's law is already in a form to split-off the transport flux
\begin{equation}
    F^i(B^j) = \mathcal{V}^i B^j - B^i \mathcal{V}^j.
\end{equation}
The source term of the momentum equation can be written as
\begin{equation}
\begin{split}
    S(S_j) = &\ \frac{1}{2} \alpha \left( \frac{S^i S^j}{B^2} - \frac{B^i B^j}{\Gamma^2} + p_{\rm{EM}} \gamma^{ij}  \right) \partial_j \gamma_{ik} \\
    &+ S_i \partial_j \beta^i - \frac{1}{2} \left( \frac{S^2}{B^2} + B^2 \right) \partial_j \alpha.
\end{split}
\end{equation}
The characteristic waves are all moving with the speed of light and hence we set
\begin{equation}
    \lambda_{\pm}^i = \pm \alpha \sqrt{\gamma^{ii}} - \beta^i.
\end{equation}

Ideal force-free electrodynamics implies the Lorentz invariant constraints $\mbf{E}\cdot\mbf{B} =0$ and $E^2 - B^2 <0$. These constraints are non-evolutionary. A numerical scheme should be designed to ensure violations at least remain bounded. An advantage of the $(\mbf{S}, \mbf{B})$ formulation presented here, over the one in terms of $(\mbf{E}, \mbf{B})$ (e.g. \autoref{sec:ffr}) is that the constraint $\mbf{E}\cdot\mbf{B}=0$ is already satisfied to machine precision. Thus in the MD scheme, additional enforcement steps like removal of the parallel electric field component are not strictly required. In a similar vein, the force-free constraint $F_{\mu \nu} J^\mu =0$ is satisfied to machine precision by virtue of the conservative scheme solving the equivalent conservation law $\del_{\mu} T^{\mu\nu} =0$.

However, regions of $E^2 - B^2 >0$ can be encountered during the evolution and should be monitored, either to trigger controlled termination or to issue fixes avoiding numerical breakdown. In $(\mbf{E}, \mbf{B})$ schemes following \cite{Komissarov_2006MNRAS.367...19K} this is achieved by an increased cross-field conductivity, which exponentially dampens the electric field. Other schemes, as presented for example \cite{Alic2012_FFR}, modify Ohm's law to drive the solution towards the FFE compliant state.

In our scheme, we follow the reasoning of \cite{McKinney_2006MNRAS.367.1797M} and limit the drift velocity before it becomes superluminal. The physical motivation for this choice is that plasma inertia (and thus effects not modeled within FFE) must eventually become important and limit the achievable Lorentz factor of the plasma. Thus we introduce as an additional parameter, the maximum (squared) plasma 3-velocity, $v^2_{\rm{max}}$, or equivalently the maximum Lorentz factor $\Gamma_{\rm{max}}$, which we default to the large value of $\Gamma_{\rm{max}} = 10^3$. We thus introduce the algebraic constraint on the drift velocity
\begin{equation}
    v^i \rightarrow v^i\, {\rm{min}} \left( 1, \sqrt{\frac{v^2_{\rm{max}}}{v^2}} \right)
\end{equation}
which is enforced after every sub-step of the time-integration.

\section{Resistive Force Free Electrodynamic Formulation}\label{sec:ffr}

The resistive force free electrodynamic formulation in \verb+BHAC+ solves Faraday’s law and Amp\`{e}re's law to evolve the magnetic field, $\mbf{B}$, and the electric field, $\mbf{E}$.
The constrained transport method is used to evolve the magnetic flux and to preserve $\del \cdot \mbf{B}=0$ to machine precision \cite{Olivares2019}. Note throughout this section we work in units with $c=k_B=1$, $B/\sqrt{4\pi} \rightarrow B$, $E/\sqrt{4\pi} \rightarrow E$, $\sqrt{4\pi} J \rightarrow J$, $\sqrt{4\pi} q \rightarrow q$, and $\eta/4\pi \rightarrow \eta$.
As well, we work in the 3+1 ADM formalism for completeness
\begin{equation}
    g^{\mu \nu} = \left(  
    \begin{array}{cc}
        -1/\alpha^2 & \beta^j/\alpha^2 \\
        \beta^j/\alpha^2 & \gamma^{ij} - \beta^i \beta^j/\alpha^2 
    \end{array}
    \right),
\end{equation}
while all simulations presented in this paper are in flat Minkowski spacetime with $\alpha =1$, $\beta^i = 0$, and $\gamma^{ij} = \delta^{ij}$. The evolution equations are written in conservative form as
\begin{equation}
    \partial_t \left( \sqrt{\gamma} \bm{U} \right) + \partial_i \left( \sqrt{\gamma} \bm{F}^i \right) = \sqrt{\gamma} \bm{S},
\end{equation}
with conserved variables $\bm{U}$ and fluxes $\bm{F}$ defined as
\begin{align}
    & \bm{U} = \left[ 
    \begin{array}{c}
        B^j  \\
        E^j 
    \end{array}
    \right],  \\
    & \bm{F}^i = \left[ 
    \begin{array}{c}
       \beta^j B^i - \beta^i B^j + \gamma^{1/2} \epsilon^{jik} \alpha E_k  \\
        \beta^j E^i - \beta^i E^j - \gamma^{1/2} \epsilon^{jik} \alpha B_k 
    \end{array}
    \right], 
\end{align}
with $\epsilon^{ijk}$ the spatial Levi-Civita antisymmetric symbol, the sources are defined as
\begin{equation}
    \bm{S} = \left[ 
    \begin{array}{c}
        0  \\
        -\alpha J^j + \beta^j q
    \end{array}
    \right].
\end{equation}
The scheme uses the current prescription presented in \cite{Alic2012_FFR}
\begin{equation}
\begin{aligned}
    J^i = & q  \frac{\gamma^{1/2} \epsilon^{ijk} E_j B_k}{B^l B_l} \\
    & + \frac{1}{\eta} \left[ E^l B_l \frac{B^i}{B^l B_l} + \Theta(E^l E_l - B^l B_l)\frac{E^i}{B^l B_l} \right]
    \end{aligned}
\end{equation}
where $\eta$ is an explicit resistivity and $\Theta$ is the Heaviside function which acts to damp the electric field in regions in which $E^2>B^2$ on a resistive timescale. Notice $J^i_\parallel$, the resistive current parallel to the magnetic field present in this formulation which acts to resistively decay violations to the FFE condition $\bm{E}\cdot\bm{B}=0$,
\begin{equation}
    J^i_\parallel  =\frac{1}{\eta} \left[ E^l B_l \frac{B^i}{B^l B_l} \right].
\end{equation}

\subsection{GRFFE solution with ImEx schemes}
\label{sec:imex}
The time evolution of $E^i$ in the GRFFE paradigm can be written explicitly as
\begin{equation}
    \partial_t (\sqrt{\gamma}\vecE) = \vecQ_{\vecE}(\sqrt{\gamma}\vecB,\sqrt{\gamma}\vecE) +\frac{1}{\eta}\vecR_{\vecE}(\sqrt{\gamma}\vecB,\sqrt{\gamma}\vecE),
\end{equation}
where we distinguish a contribution $\vecQ$ related to nonstiff terms and a stiff part of the update $\vecR$ (whose stiffness is determined by $\eta$). In particular, Eq.~\eqref{eq:JFF} shows that the current contributes to the nonstiff part with the term $q\gamma^{1/2} \epsilon^{ijk} E_j B_k/(B^l B_l)$, and to the stiff part with
\begin{equation}
    \left[ E^l B_l \frac{B^i}{B^l B_l} + \Theta(E^l E_l - B^l B_l)\frac{E^i}{B^l B_l} \right]
\end{equation}
\verb+BHAC+ handles the presence of stiff terms by evolving the electric field with an ``ImEx12'' implicit-explicit Runge-Kutta scheme \cite{Ripperda_2019}, involving a second-order time discretization of the nonstiff variables and a first-order time discretization of the stiff variables. At any given substep of ImEx12 (after the magnetic field has been updated) the electric-field update therefore reads
\begin{equation}
    \vecE^\mathrm{new} = \vecE^* + \frac{a_{jj}\Delta t}{\eta} \vecR_{\vecE}(\vecB^\mathrm{new},\vecE^\mathrm{new}),
\end{equation}
where we have simplified the notation as $\sqrt{\gamma}\vecE\to\vecE$ and $\sqrt{\gamma}\vecB\to\vecB$. Here, $\vecE^{*} = \vecE^\mathrm{old} + a_{jj}\Delta t \vecQ_{\vecE}(\vecB^\mathrm{old},\vecE^\mathrm{old})$ (i.e.\ the explicitly updated electric field), and $a_{jj}$ is the coefficient from the Butcher tableau for the current ImEx12 substep \cite{2010arXiv1009.2757P,2009MNRAS.394.1727P}.

We thus observe that an expression for $\vecE^\mathrm{new}$ cannot typically be written in closed form, since the equation above is (in general) nonlinearly implicit in $\vecE^\mathrm{new}$ itself. For the FF current in Eq.~\eqref{eq:JFF}, the time-update equation reads
\begin{equation}
    E^{i} = E^{i,*} + \frac{a_{jj}\Delta t}{\eta} \left[ E^l B_l \frac{B^i}{B^l B_l} + \Theta(E^l E_l - B^l B_l)\frac{E^i}{B^l B_l} \right],
    \label{eq:Eimplupdate}
\end{equation}
where we have further simplified the notation as $E^\mathrm{new}\to E$ and $B^\mathrm{new}\to B$. This equation appears linear, but the Heaviside damping term prevents in fact the use of an explicit solution. We conclude that the implicit part of the electric-field update, at each ImEx12 substep, cannot be carried out unless an iterative method is applied to solve the nonlinear equation \eqref{eq:Eimplupdate}. This nonlinear iterative solution is the central engine of the GRFFE method.

To perform the electric-field update, at each ImEx12 substep we apply an iterative procedure to minimize the residual functions $\bm{f}(\vecE)=\bm{0}$ in the unknown electric field $\vecE$, where at the $k$-th iteration we have
\begin{equation}
\begin{split}
    \bm{f}(\vecE^k) =  & \vecE^k - \vecE^* - \frac{a_{jj}\Delta t}{\eta} \left[ (\vecE^k \cdot \vecB) \frac{\vecB}{B^2}  \right. \\ & \left. + \Theta((\vecE^k)^2 - B^2)\frac{\vecE^k}{B^2} \right].
\end{split}
\end{equation}
The nonlinear system is solved in \verb+BHAC+ with a Jacobian-free Newton-Krylov iteration (e.g. \cite{Kelley1995}), only requiring the residual functions as input. The nonlinear iteration is not guaranteed to converge in a finite number of steps; in such cases, we simply apply the explicit (linear) solution of Eq.~\eqref{eq:Eimplupdate} without the Heaviside term, and adopt the result as the updated electric field. We have empirically found that the rate of inversion failures, however, is extremely low.

\section{Dependence on Spatial Reconstruction and Time Integration Order}\label{app:order}

\begin{figure*}[htbp]
    \centering
    \begin{subfigure}{0.49\textwidth}
        \centering        \includegraphics[width=\textwidth]{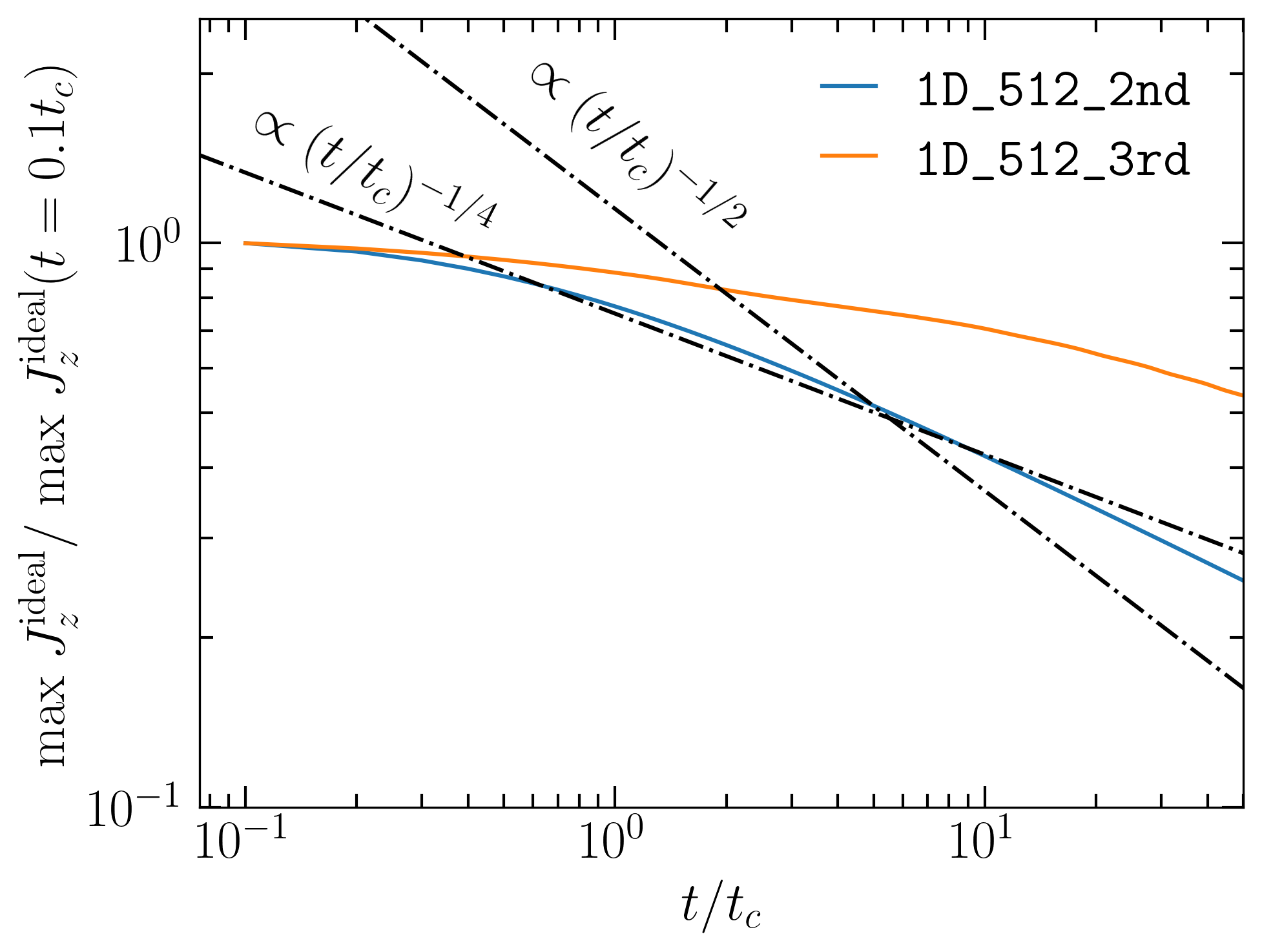} 
        \caption{}
        \label{fig:xjmax_spaceorder} 
    \end{subfigure}
    \hfill
    \begin{subfigure}{0.49\textwidth}
        \centering
    \includegraphics[width=\textwidth]{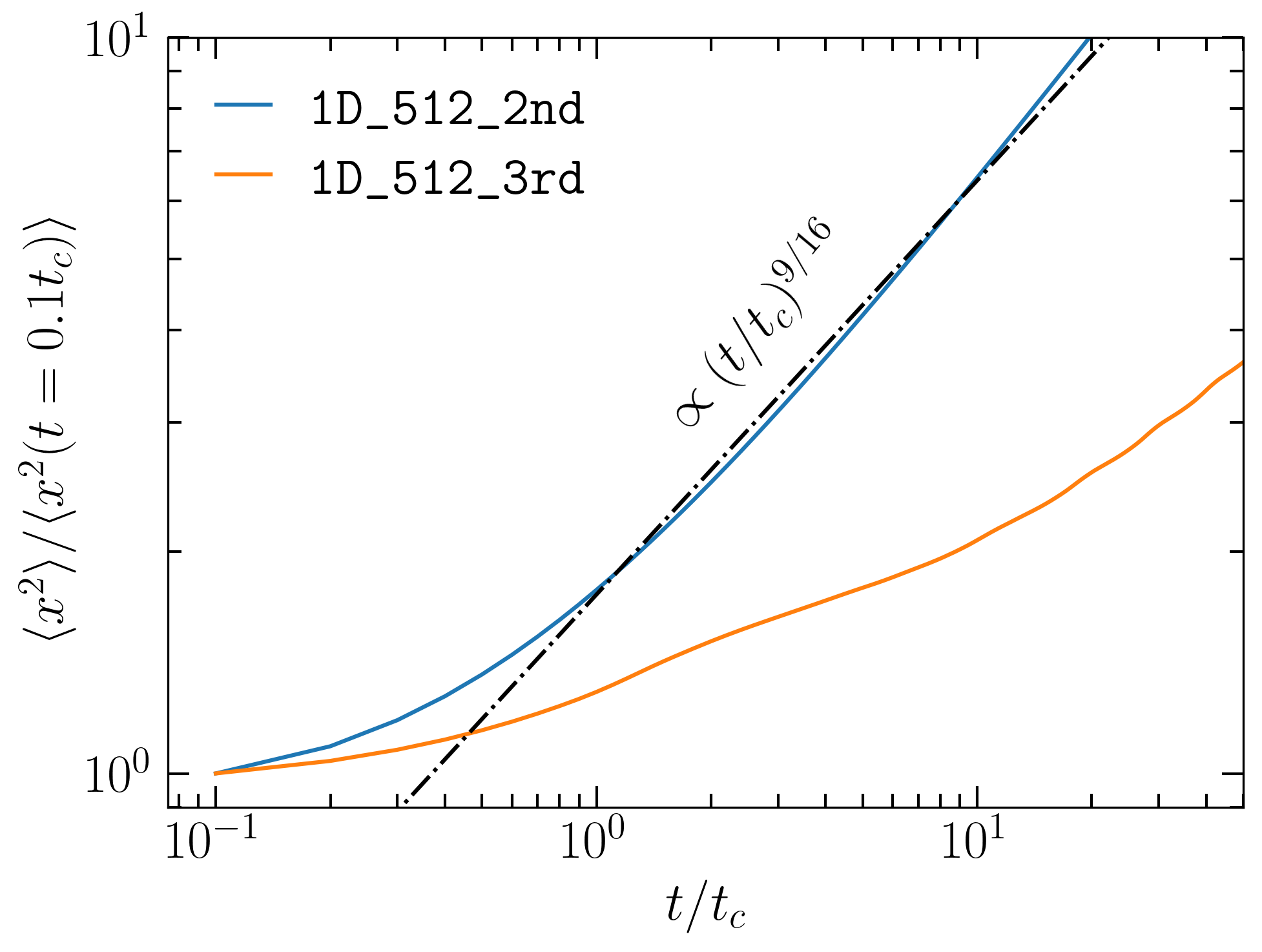}
            \caption{}
        \label{fig:x2_spaceorder}
    \end{subfigure}
    \begin{subfigure}{0.49\textwidth}
        \centering
    \includegraphics[width=\textwidth]{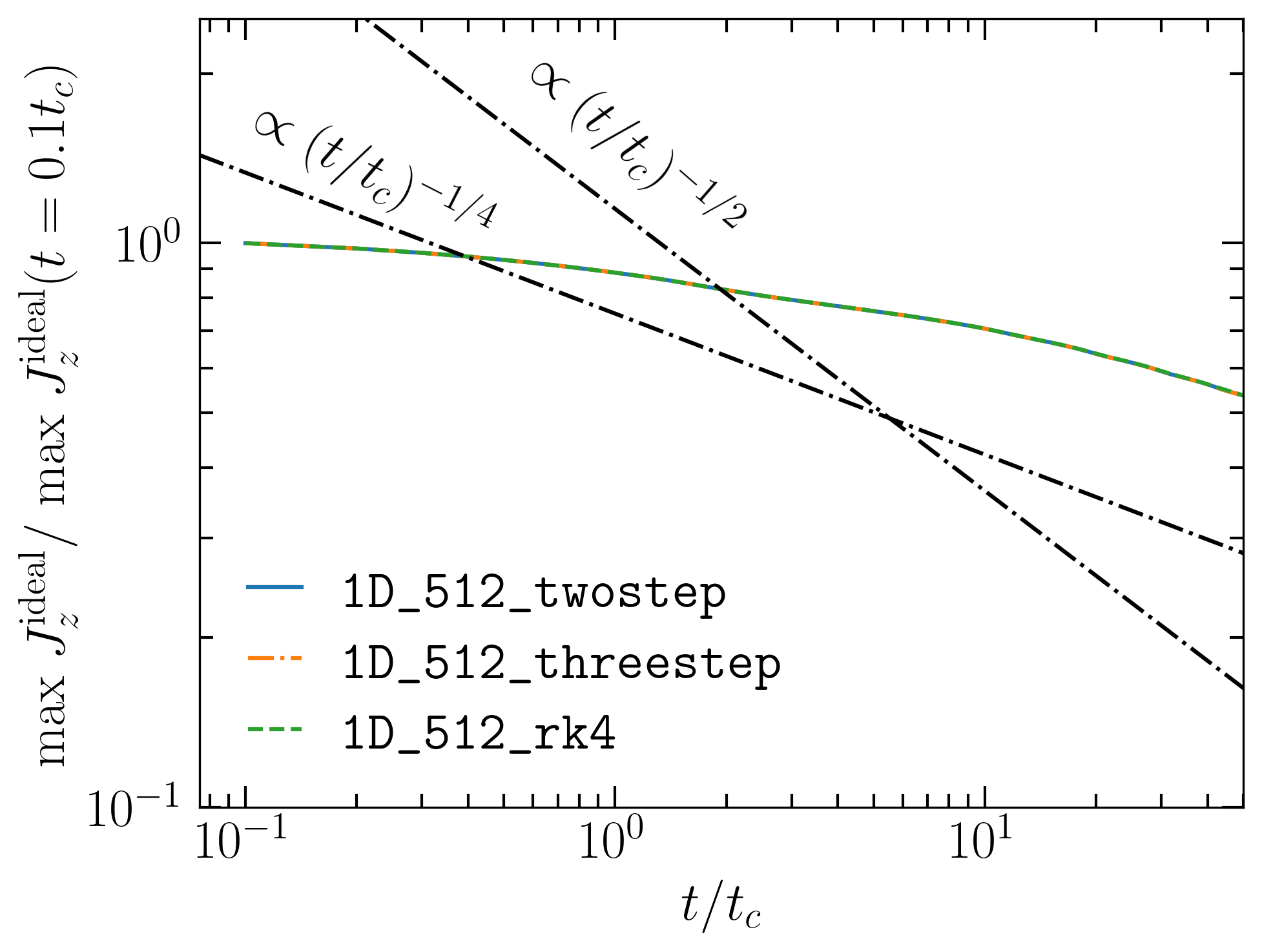}  
            \caption{}
        \label{fig:xjmax_timeorder} 
    \end{subfigure}
    \hfill
    \begin{subfigure}{0.49\textwidth}
        \centering
    \includegraphics[width=\textwidth]{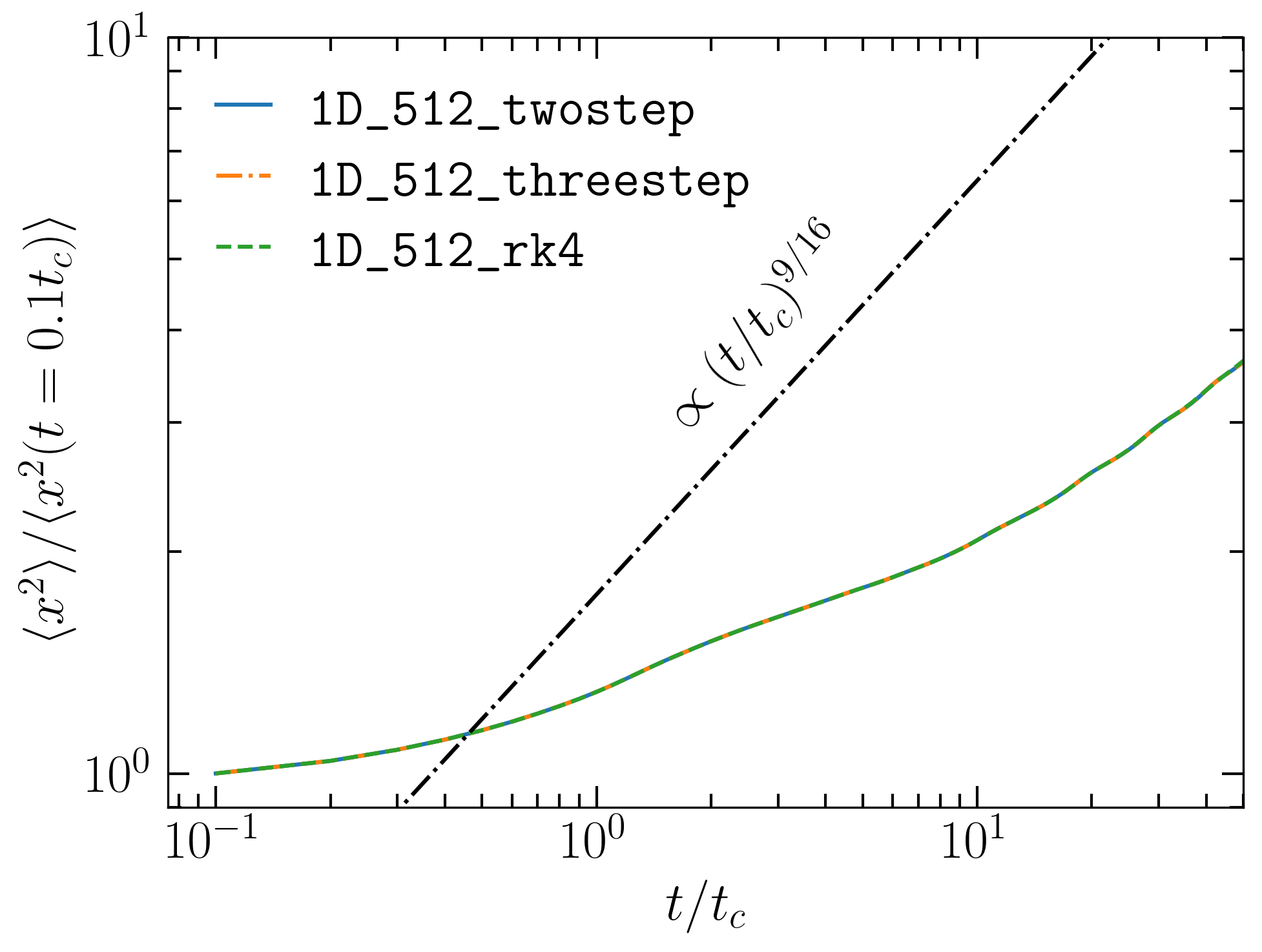}
            \caption{}
        \label{fig:x2_timeorder}
    \end{subfigure}
    
    \caption{
        The Ohmic decay of the current amplitude and evolution of the second moment, $\langle x^2 \rangle$, of the current distribution for a thin current sheet. The top row corresponds to simulations with a fixed grid resolution but varying between 2nd and 3rd order spatial reconstruction schemes, as indicated in the legends. The bottom row shows results for a fixed spatial reconstruction scheme while varying the time integration order, which has no effect on $\langle x^2 \rangle$ or max $J_z^{\rm{ideal}}$. 
    }
    \label{fig:space_time_order}
\end{figure*}

To determine the dependence of numerical diffusion on the order of the spatial reconstruction scheme and the time integration method, we consider the setup in \autoref{sec:ffeHarris}. We vary the orders of both the spatial reconstruction scheme and the time integration scheme for a fixed resolution, ideal MHD simulation. For a complete list of the diffusion simulations considered in this section see \autoref{tab:order}. We show in \autoref{fig:xjmax_spaceorder} and \autoref{fig:x2_spaceorder} that a lower order, more diffusive, spatial reconstruction scheme results in stronger numerical diffusion for a fixed resolution, as expected. Interestingly, the numerical diffusion is still subdiffuse and resembles the evolution of the 3rd order method at one quarter the resolution as shown in \autoref{fig:ideal_x2}. We show in \autoref{fig:xjmax_timeorder} and \autoref{fig:x2_timeorder}  that the order of the time integration method has no effect on the numerical diffusion, with methods of 2nd to 4th order demonstrating identical diffusive evolution. Hence, as expected, the nature of numerical diffusion is contained within the reconstruction method for the spatial fluxes.

\begin{table}[htbp]
\caption{\label{tab:order}%
Spatial Reconstruction and Time Integration Order Tests of Numerical Diffusion
} 
\begin{ruledtabular}
\begin{tabular}{clcccc}

& \multicolumn{1}{c}{\textrm{Sim. ID}}&
\textrm{Resolution}& 
\textrm{Spatial}&
\textrm{Temporal}
&
\\
&&
& 
\textrm{Order}&
\textrm{Order}
&
\\
 & \multicolumn{1}{c}{(1)}  &\multicolumn{1}{c}{(2)}  &\multicolumn{1}{c}{(3)} & \multicolumn{1}{c}{(4)} &\\
\hline
& \Verb+1D_512_2nd+  & 512 & 2 & 2 &\\
& \Verb+1D_512_3rd+  & 512 & 3 & 2&\\
& \Verb+1D_512_twostep+  & 512 & 3 & 2&\\
& \Verb+1D_512_threestep+  & 512 & 3 & 3&\\
& \Verb+1D_512_rk4+  & 512 & 3 & 4&\\

\end{tabular}
\end{ruledtabular}
\begin{tablenotes}[para]
        \textit{\textbf{Notes.}} Simulation parameters for the setup described in \autoref{sec:ffeHarris}. All simulations are ideal MHD. All simulations have domains $-1\leq x/L \leq 1$, where $L=1$ in code units, $a/L=0.02$, $B_0=1$, $B_{\rm{guide}}=10$, $\sigma_{\rm{hot}}=10$, and $T=1.0$, with continuous boundaries, where quantities are extrapolated outside the domain. \textbf{Column (1):} the unique simulation ID. 
        \textbf{Column (2):} the uniform resolution.
        \textbf{Column (3):} the order of the spatial reconstruction scheme. 
        \textbf{Column (4):} the order of the time integration scheme. 
    \end{tablenotes}
\end{table}

\begin{figure}
    \centering
\includegraphics[width=0.5\textwidth]{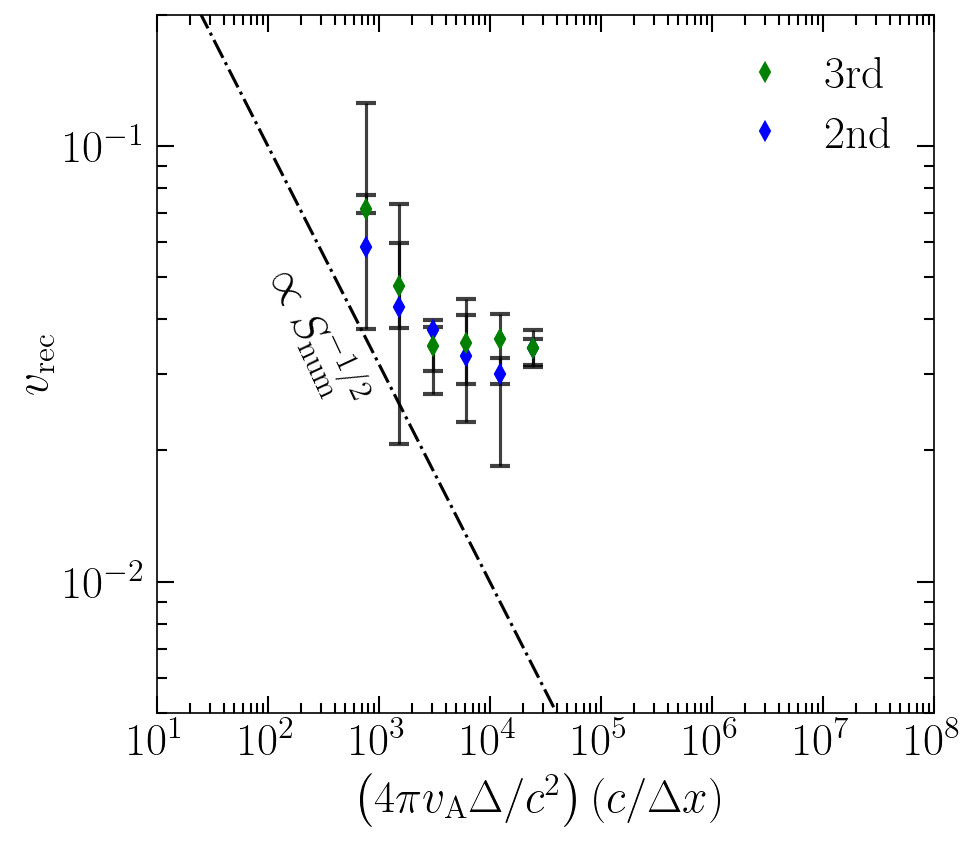}
    \caption{
    The time averaged reconnection rate for the simulations in \autoref{tab:vrec_order}. Details of how  the reconnection rate, $v_{\rm rec}$, is measured are presented in \autoref{app:measure_vrec}. The results are plotted as a function of the inverse of the grid scale size, $(1/\Delta x)$. The effective Lundquist number is approximate by substituting $\eta_{\rm{num}} = \Delta x/c$. 
    }
    \label{fig:vrec_order}
\end{figure}

\begin{figure*}[htbp]
    \centering
    \begin{subfigure}{\textwidth}
        \centering
        \includegraphics[width=\textwidth]{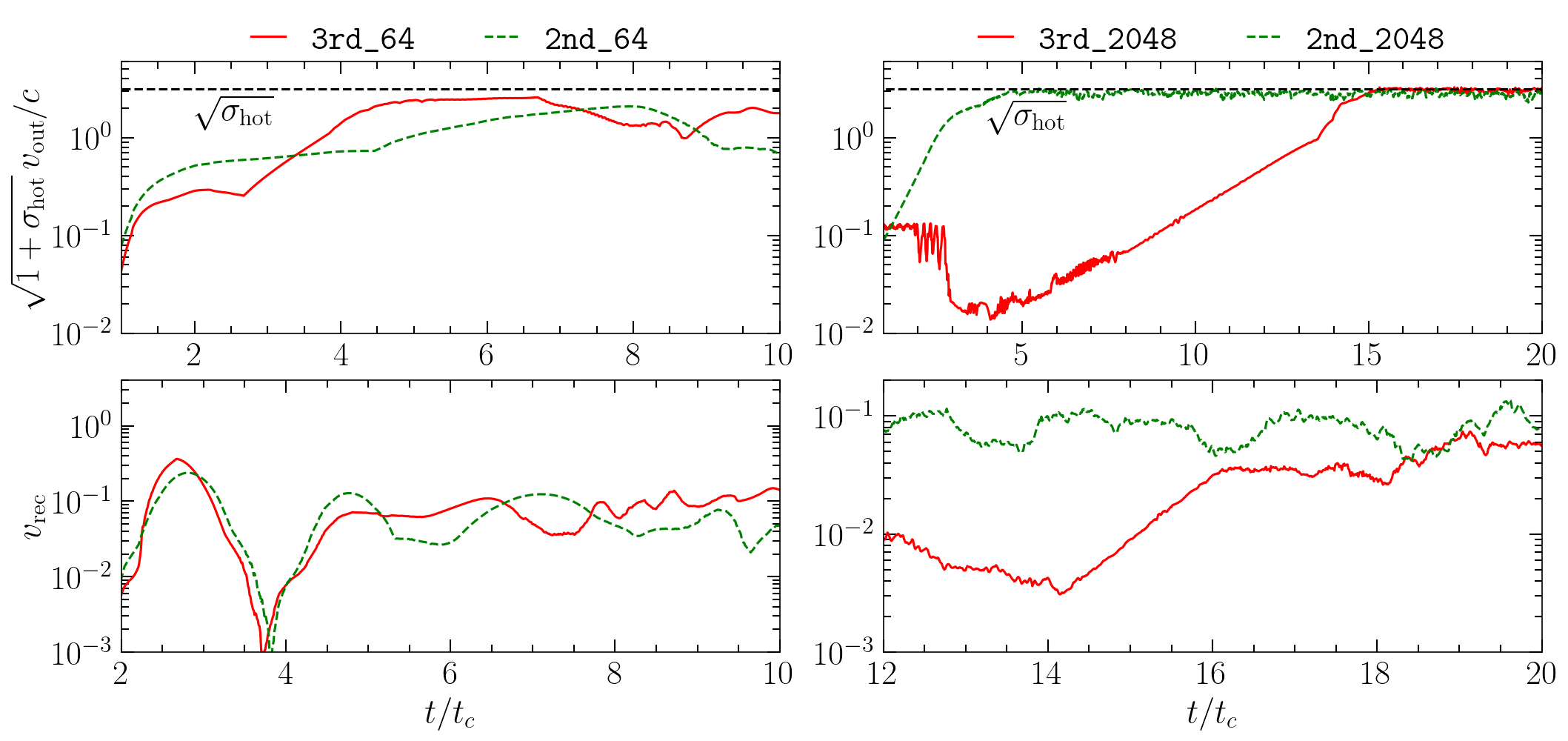}        \caption{Pressure-balanced Harris sheet.}
    \end{subfigure}
    \caption{ 
  The measured value of $\sqrt{1+\sigma_{\rm{hot}}}\, v_{\rm{out}}/c$  and $v_{\rm{rec}}$ as functions of time. As the outflow speed approaches the Alfv\'{e}n speed, $\sqrt{1+\sigma_{\rm{hot}}}\, v_{\rm{out}}/c$ will approach $\sqrt{\sigma_{\rm{hot}}}$. Note the simulations in the left column are low resolution, while the simulations in the right column are high resolution and in the asymptotic regime.
    }
    \label{fig:vrec_vout_order}
\end{figure*}

To determine the dependence of numerical reconnection on the order of the spatial reconstruction scheme, we consider the pressure balanced Harris sheet setup in \autoref{sec:pressureHarris}. We vary the orders of the spatial reconstruction scheme for a range of resolutions, from low resolution (64 cells per sheet half length) to high resolution (2048 cells per sheet half length). For a complete list of the reconnection simulations considered in this section see \autoref{tab:vrec_order}. We show in \autoref{fig:vrec_order} that the order of the spatial reconstruction scheme has an insignificant effect on the time-averaged reconnection rate. This is potentially because the reconnection itself is sensitive to the aspect ratio of the sheet (captured by the standard Lundquist number dependence, \autoref{eq:lundquist}), which in the ideal models, regardless of reconstruction order, is being set by the grid scale. Hence, as we show in \autoref{fig:vrec_order}, both the second- and third-order methods follow a Sweet--Parker-like scaling at low resolutions and plateau at similar values in the asymptotic Lundquist number regime. However, even though the time-averaged reconnection rates have similar values and asymptotic properties, \autoref{fig:vrec_vout_order} shows that there is a difference in onset time of reconnection. This is consistent with the lower-order reconstruction scheme being more diffuse, as shown in \autoref{fig:xjmax_spaceorder} and \autoref{fig:x2_spaceorder}, since the onset is associated with the diffusion timescale, which controls how a thick current sheet thins before reconnection ignites.

\begin{table}[htbp]
\caption{\label{tab:vrec_order}%
Spatial Reconstruction Tests of Numerical Reconnection
} 
\begin{ruledtabular}
\begin{tabular}{clccc}

& \multicolumn{1}{c}{\textrm{Sim. ID}}&
\textrm{Resolution}& 
\textrm{Spatial}&
\\
&&
& 
\textrm{Order}&
\\
 & \multicolumn{1}{c}{(1)}  &\multicolumn{1}{c}{(2)}  &\multicolumn{1}{c}{(3)} & \\
\hline
& \Verb+2nd_64+  & 64 & 2 &\\
& \Verb+2nd_128+  & 128 & 2 &\\
& \Verb+2nd_256+  & 256 & 2 &\\
& \Verb+2nd_512+  & 512 & 2 &\\
& \Verb+2nd_1024+  & 1024 & 2 &\\
& \Verb+2nd_2048+  & 2048 & 2 &\\

& \Verb+3rd_64+  & 64 & 3 &\\
& \Verb+3rd_128+  & 128 & 3 &\\
& \Verb+3rd_256+  & 256 & 3 &\\
& \Verb+3rd_512+  & 512 & 3 &\\
& \Verb+3rd_1024+  & 1024 & 3 &\\
& \Verb+3rd_2048+  & 2048 & 3 &\\

\end{tabular}
\end{ruledtabular}
\begin{tablenotes}[para]
        \textit{\textbf{Notes.}} Simulation parameters for the setup described in \autoref{sec:pressureHarris}. All simulations have domains  $-1 \leq x/L \leq 1$ and $-2 \leq y/L \leq 2$, where $L=1$ in  code units, $a/L = 0.02$, sheet half length of $\Delta/L = 2$, $B_0 = 1$, $\sigma_{\rm{hot}} = 10$, and $T = 1.0$ with open boundaries. \textbf{Column (1):} the unique simulation ID. 
        \textbf{Column (2):} the effective resolution per sheet half length. 
        \textbf{Column (3):} the order of the spatial reconstruction scheme. 
    \end{tablenotes}
\end{table}

\section{Local Resolution Dependence}\label{sec:num_L}

\begin{figure}[htbp]
    \centering
    \includegraphics[width=0.49\textwidth]{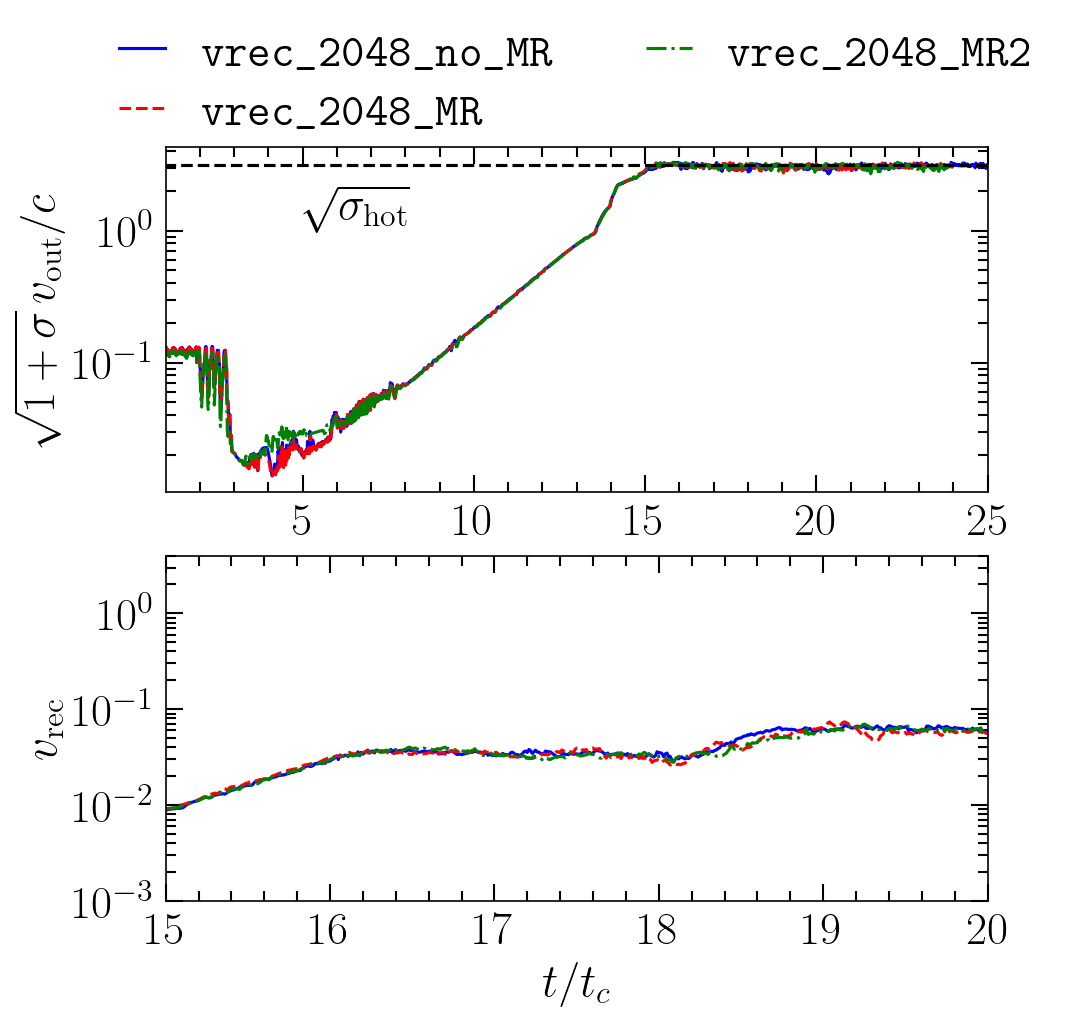}
    \caption{ 
     For a pressure-balanced sheet, the measured value of $\sqrt{1+\sigma}\, v_{\rm{out}}/c$,  and $v_{\rm{rec}}$ as functions of time. For three ideal MHD simulations, one without and two with block refinement over eight times the initial sheet thickness and two times the initial sheet thickness. 
    }
    \label{fig:MR_test}
\end{figure}

To determine the effect which mesh refinement has on the ideal simulations we consider the three simulations in \autoref{tab:MR_test}: no mesh refinement; mesh refinement over eight initial sheet widths; and, mesh refinement over two initial sheet widths.
In  \autoref{fig:MR_test}, the agreement between the outflow speed and the reconnection rate of the pressure-balanced ideal MHD simulations with varying size of fixed refinement levels in space and time demonstrates that the use of block refinement is sufficient. This test has the additional benefit of demonstrating that it is the local grid resolution of the current sheet which determines the numerical Lundquist number and that it is independent of the upstream resolution.

\begin{table}[htbp]
\caption{\label{tab:MR_test}%
Ideal pressure-balanced Harris Sheet Mesh Refinement Testing
} 
\begin{ruledtabular}
\begin{tabular}{lcccc}

 \multicolumn{1}{c}{\textrm{Sim. ID}}&
\textrm{Base}& 
\textrm{MR}&
\textrm{Width}&
\textrm{Resolution}
\\
  \multicolumn{1}{c}{(1)}  &\multicolumn{1}{c}{(2)}  &\multicolumn{1}{c}{(3)} & \multicolumn{1}{c}{(4)} & \multicolumn{1}{c}{(5)}\\
\hline
 \Verb+vrec_2048_MR+  & 512 & 2 & 8 & 2048\\
 \Verb+vrec_2048_MR2+  & 512 & 2 & 2& 2048\\
 \Verb+vrec_2048_no_MR+  & 2048 & 0 & & 2048\\

\end{tabular}
\end{ruledtabular}
\begin{tablenotes}[para]
        \textit{\textbf{Notes.}} Simulation parameters for the setup described in \autoref{sec:pressureHarris}. All simulations have a domain of width 2 and length 4 in code units, initial sheet thickness of $a = 0.02$, sheet half length of $\Delta = 2$, $B_0=1$, $\sigma_{\rm{hot}}=10$, and $T=1.0$. \textbf{Column (1):} the unique simulation ID. 
        \textbf{Column (2):} the base number of cells per sheet half length.
        \textbf{Column (3):} employed level of mesh refinement. 
        \textbf{Column (4):} width of mesh refinement region relative to initial sheet thickness. 
        \textbf{Column (5):} the effective resolution per sheet half length.
    \end{tablenotes}
\end{table}

\section{Resolution Testing}\label{app:resolution_testing}

\begin{figure}
    \centering
    \includegraphics[width=0.49\textwidth]{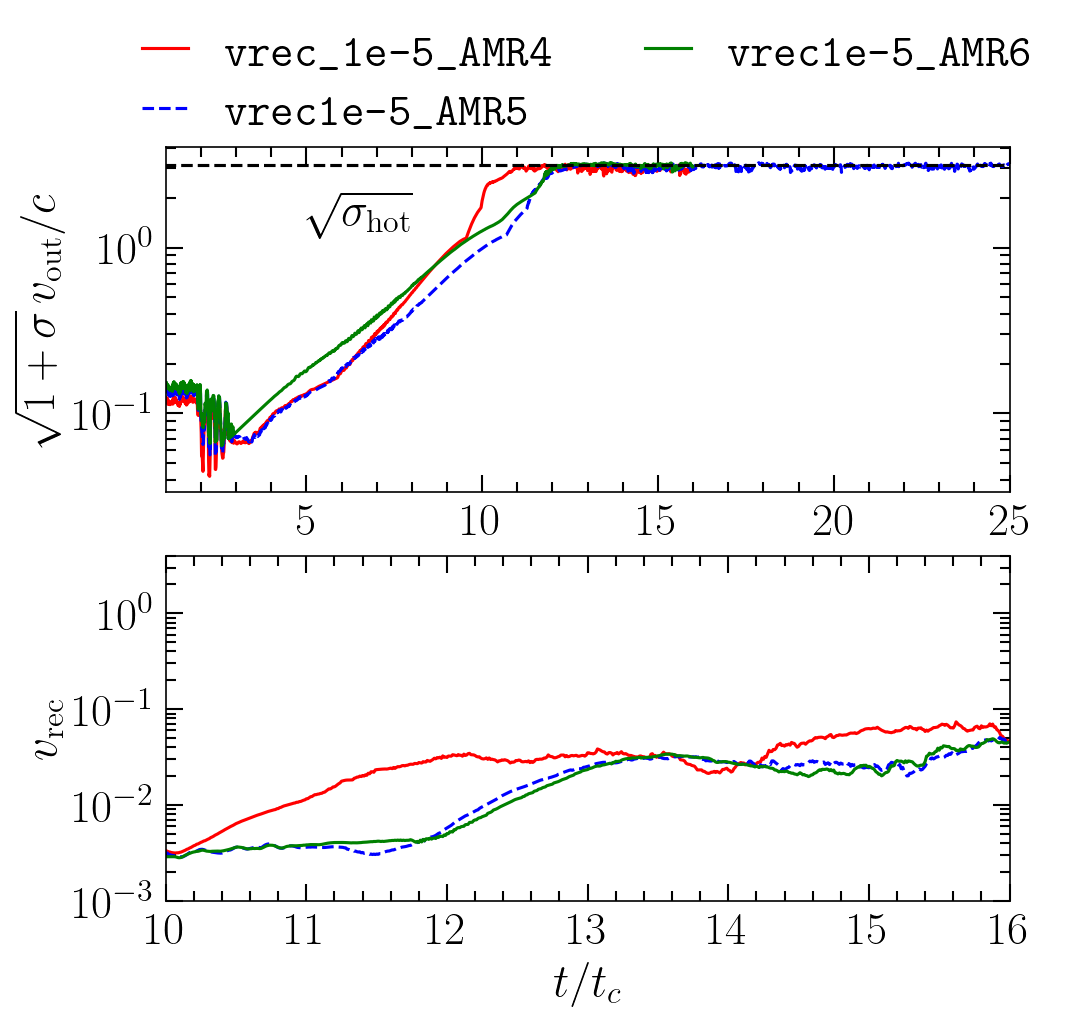}
    \caption{
For a pressure-balanced sheet, the measured value of $\sqrt{1+\sigma}\, v_{\rm{out}}/c$,  and $v_{\rm{rec}}$ as functions of time. For the first three resistive MHD simulations in \autoref{tab:resolution_test}, with various levels of adaptive mesh refinement. 
    }
    \label{fig:1e-5_res_test}
\end{figure}

To ensure that numerical effects are minimized for the reconnection rate measurement we chose to resolution test simulation \verb+eta_1e-5+, which is a pressure-balanced Harris sheet with large Lundquist number in the asymptotic regime. Note the reason lower resistivity simulations were not chosen to resolution test is because the simulations become prohibitively expensive as the Lundquist number is increased at very high resolutions. For the interests of this paper, the reconnection rate and macroscopic structure of the current sheet, extremely high resolutions are not necessary. The plasmoid unstable regime requires significantly more resolution than the Sweet--Parker regime, this is exactly why a plasmoid unstable run was tested. In \autoref{tab:resolution_test} we perform resolution testing by running \verb+eta_1e-5+ at multiple levels of AMR, all with a base grid of 128 cells per sheet half length. Each of these runs were analyzed and the reconnection rate and outflow speed compared between the runs. The simulation was determined to be resolved when adding an additional level of AMR did not significantly change either quantity. \autoref{fig:1e-5_res_test}, demonstrates that between 4 and 5 levels of AMR that there is a noticeable difference in the reconnection rate, between 5 and 6 levels the rate is similar, hence we conclude that \verb+vrec_1e-5+ is sufficiently resolved at 5 levels of AMR.

\begin{table}[htbp]
\caption{\label{tab:resolution_test}%
Pressure-balanced Harris Sheet Resolution Testing  
} 
\begin{ruledtabular}
\begin{tabular}{lccc}

\multicolumn{1}{c}{\textrm{Sim. ID}}&
\textrm{Base}& 
\textrm{AMR}&
\textrm{Resolution}

\\
\multicolumn{1}{c}{(1)}  &\multicolumn{1}{c}{(2)}  &\multicolumn{1}{c}{(3)}  &\multicolumn{1}{c}{(4)}\\
\hline
\Verb+vrec_1e-5_AMR4+  & 128 & 4 & 2048\\
\Verb+vrec_1e-5_AMR5+  & 128 & 5 & 4096\\
\Verb+vrec_1e-5_AMR6+  & 128 & 6 & 8192\\

\end{tabular}
\end{ruledtabular}
\begin{tablenotes}[para]
        \textit{\textbf{Notes.}} Simulation parameters for the setup described in \autoref{sec:pressureHarris}. All simulations have a domain of width 2 and length 4 in code units, initial sheet thickness of $a = 0.02$, sheet half length of $\Delta = 2$, $B_0=1$, $\sigma_{\rm{hot}}=10$, and $T=1.0$. \textbf{Column (1):} the unique simulation ID. 
        \textbf{Column (2):} the base number of cells per sheet half length.
        \textbf{Column (3):} employed levels of adaptive mesh refinement (AMR).
        \textbf{Column (4):} the effective resolution per sheet half length.

    \end{tablenotes}
\end{table}

\section{Resistive Induction Equation in Resistive Force-Free}\label{app:ff-induction}

Consider the Newtonian limit in which $\del \cdot \vecE =  4\pi q \sim 0$, $\partial_t \vecE \sim 0$ and $E^2 < B^2$. Then Ampere's law takes the form
\begin{equation}
    \del \times \vecB = \frac{4\pi}{c}\bm{J}, 
\end{equation}
and the resistive force-free current is \cite{Alic2012_FFR}
\begin{equation}
    \bm{J} = \frac{1}{\eta}\left[ (\vecE\cdot\vecB) \frac{\vecB}{B^2}\right].
\end{equation}
Combining these expressions for the current, taking the curl of either side, using $\del \cdot \vecB =0$, and making use of the vector identity $\vecB \times (\vecE \times \vecB) = B^2 \vecE - (\vecB \cdot \vecE)\vecB$, 
\begin{equation}
    - \del^2 \vecB = \frac{4\pi}{\eta c} \left[  \del \times \vecE   + \del \times \left( \left( \frac{\vecE \times \vecB}{B^2}  \right)  \times \vecB \right)   \right],
\end{equation}
inserting Faraday's law yields a resistive induction equation with a velocity $\bm{v}/c = \vecE\times\vecB/B^2$,
\begin{equation}
    \frac{\partial \vecB}{\partial t} = \bm{\del} \times \left[ \left( \frac{\vecE\times\vecB}{B^2} \right) \times \vecB  \right] c + \frac{\eta c^2}{4\pi} \del^2 \vecB.
\end{equation}

\section{Measuring the Reconnection Rate}\label{app:measure_vrec}

\begin{figure}[htbp]
    \centering
    \includegraphics[width=0.49\textwidth]{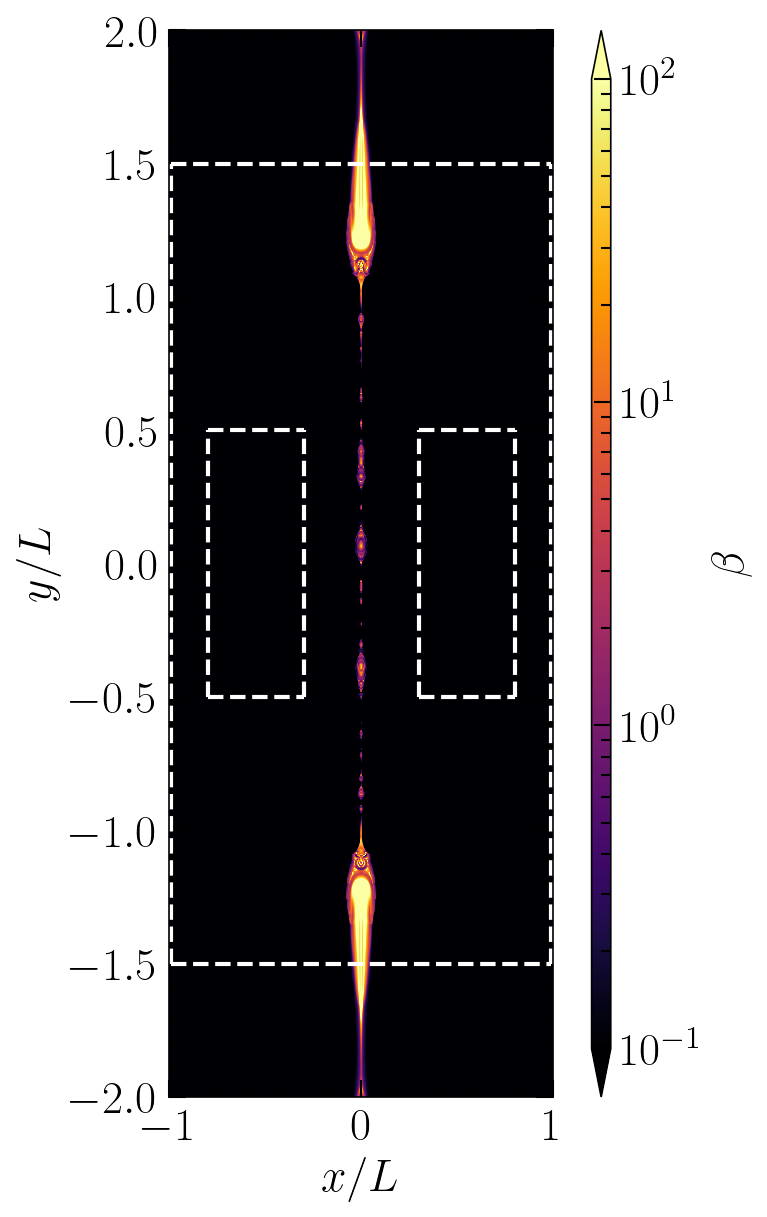}
    \cprotect\caption{ 
     For a pressure-balanced sheet, simulation \Verb+vrec_1e-5+, visualization of the gas ratio, $\beta= 8\pi p/B^2$ as the current sheet is undergoing reconnection. The small regions outlined in white indicate the sub region in the upstream over which in flowing $v_x$ values are averaged to obtain $v_{\rm{in}}$, the large outlined region is the sub region over which the maximum of $|v_y|$ is taken to determine $v_{\rm{out}}$. Note $L = c/t_c$ is the unit length which the simulation is normalized to.
    }
    \label{fig:vrec_vis}
\end{figure}

 The dimensionless number corresponding to the ratio of the inflow and outflow speeds of field lines into the current layer is the reconnection rate,
\begin{equation}
    v_{\rm{rec}} = \frac{v_{\rm{in}} }{v_{\rm{out}}}.
\end{equation}
To measure \( v_{\rm{rec}} \) in a simulation, the computational domain is sliced into 1D horizontal cuts perpendicular to the initial current sheet at intervals of 0.01 $c/t_c$ and with an output frequency of 0.01 $t_c$. 
The inflow speed is calculated by taking the average of in-flowing $v_x$ values (e.g. $v_x>0$ to the left of the sheet) over a spatial region in the upstream which avoids the computational boundary. The outflow speed is calculated over a larger region, excluding the edge of the domain, and is found as the maximum value of $|v_y|$. The reconnection rate is then calculated at each instance in time as $\langle v_x \rangle / v_{\rm{out}}$, taking only the inflow velocities into account for $\langle v_x \rangle$,  this is performed on both sides of the sheet.
To obtain a time-averaged reconnection rate, a start time is chosen based on visual inspection of the simulation, a clear indicator of reconnection occurring is the outflow velocity plateauing, and in the case of large Lundquist simulations, approaching the Alfv\'{e}n speed. The rate is then computed over various time intervals by varying both the start and end times over the entire time interval. The median of the reconnection rates calculated across these intervals is taken as the time-averaged reconnection rate, and the 16th and 84th percentile of the reconnection rate between the chosen min and max time are taken as the error below and above.
In particular, for our measurements of the reconnection rate, the inflow is determined in a subregion of size $-0.8 \leq x \leq -0.3$, to the left of the sheet, and $0.3 \leq x \leq 0.8$, to the right of the sheet, while for both regions $-0.5 \leq y \leq 0.5$. The outflow is calculated over the region $-1.5 \leq y \leq 1.5$. We show an example of these subregions in \autoref{fig:vrec_vis}.

For the guide field-balanced simulations, the reconnection rate is calculated with an identical procedure for the resistive and ideal MHD simulations, with the outflow being only the in-plane outflow in the $y$-direction. For the resistive FFE simulations, the velocity used is not the fluid velocity but instead the $\vecE\times\vecB/B^2$ drift speed. Hence, the $x$-component of $\vecE\times\vecB/B^2$ determines the inflow and the $y$-component of $\vecE\times\vecB/B^2$ the outflow. Identical spatial regions are used in the calculations. Note that the outflow speed is compared to the in-plane Alfv\'{e}n speed which accounts for the guide field inertia \autoref{eq:vAperp}. Tests were performed to ensure that the reconnection rate in the resistive MHD simulations was identical when using the $\vecE\times\vecB/B^2$ drift velocity as opposed to the fluid velocity $\bm{v}$.

\clearpage
\bibliography{ref}

\begin{thebibliography}{120}%
\makeatletter
\providecommand \@ifxundefined [1]{%
 \@ifx{#1\undefined}
}%
\providecommand \@ifnum [1]{%
 \ifnum #1\expandafter \@firstoftwo
 \else \expandafter \@secondoftwo
 \fi
}%
\providecommand \@ifx [1]{%
 \ifx #1\expandafter \@firstoftwo
 \else \expandafter \@secondoftwo
 \fi
}%
\providecommand \natexlab [1]{#1}%
\providecommand \enquote  [1]{``#1''}%
\providecommand \bibnamefont  [1]{#1}%
\providecommand \bibfnamefont [1]{#1}%
\providecommand \citenamefont [1]{#1}%
\providecommand \href@noop [0]{\@secondoftwo}%
\providecommand \href [0]{\begingroup \@sanitize@url \@href}%
\providecommand \@href[1]{\@@startlink{#1}\@@href}%
\providecommand \@@href[1]{\endgroup#1\@@endlink}%
\providecommand \@sanitize@url [0]{\catcode `\\12\catcode `\$12\catcode `\&12\catcode `\#12\catcode `\^12\catcode `\_12\catcode `\%12\relax}%
\providecommand \@@startlink[1]{}%
\providecommand \@@endlink[0]{}%
\providecommand \url  [0]{\begingroup\@sanitize@url \@url }%
\providecommand \@url [1]{\endgroup\@href {#1}{\urlprefix }}%
\providecommand \urlprefix  [0]{URL }%
\providecommand \Eprint [0]{\href }%
\providecommand \doibase [0]{https://doi.org/}%
\providecommand \selectlanguage [0]{\@gobble}%
\providecommand \bibinfo  [0]{\@secondoftwo}%
\providecommand \bibfield  [0]{\@secondoftwo}%
\providecommand \translation [1]{[#1]}%
\providecommand \BibitemOpen [0]{}%
\providecommand \bibitemStop [0]{}%
\providecommand \bibitemNoStop [0]{.\EOS\space}%
\providecommand \EOS [0]{\spacefactor3000\relax}%
\providecommand \BibitemShut  [1]{\csname bibitem#1\endcsname}%
\let\auto@bib@innerbib\@empty
\bibitem [{\citenamefont {{O. Porth et al.}}(2019)}]{Porth-short:2019ApJS..243...26P}%
  \BibitemOpen
  \bibfield  {author} {\bibinfo {author} {\bibnamefont {{O. Porth et al.}}},\ }\bibfield  {title} {\bibinfo {title} {{The Event Horizon General Relativistic Magnetohydrodynamic Code Comparison Project}},\ }\href {https://doi.org/10.3847/1538-4365/ab29fd} {\bibfield  {journal} {\bibinfo  {journal} {\apjs}\ }\textbf {\bibinfo {volume} {243}},\ \bibinfo {eid} {26} (\bibinfo {year} {2019})},\ \Eprint {https://arxiv.org/abs/1904.04923} {arXiv:1904.04923 [astro-ph.HE]} \BibitemShut {NoStop}%
\bibitem [{\citenamefont {{White}}\ \emph {et~al.}(2019)\citenamefont {{White}}, \citenamefont {{Stone}},\ and\ \citenamefont {{Quataert}}}]{white:2019ApJ...874..168W}%
  \BibitemOpen
  \bibfield  {author} {\bibinfo {author} {\bibfnamefont {C.~J.}\ \bibnamefont {{White}}}, \bibinfo {author} {\bibfnamefont {J.~M.}\ \bibnamefont {{Stone}}},\ and\ \bibinfo {author} {\bibfnamefont {E.}~\bibnamefont {{Quataert}}},\ }\bibfield  {title} {\bibinfo {title} {{A Resolution Study of Magnetically Arrested Disks}},\ }\href {https://doi.org/10.3847/1538-4357/ab0c0c} {\bibfield  {journal} {\bibinfo  {journal} {\apj}\ }\textbf {\bibinfo {volume} {874}},\ \bibinfo {eid} {168} (\bibinfo {year} {2019})},\ \Eprint {https://arxiv.org/abs/1903.01509} {arXiv:1903.01509 [astro-ph.HE]} \BibitemShut {NoStop}%
\bibitem [{\citenamefont {{Ripperda}}\ \emph {et~al.}(2020)\citenamefont {{Ripperda}}, \citenamefont {{Bacchini}},\ and\ \citenamefont {{Philippov}}}]{Ripperda:2020ApJ...900..100R}%
  \BibitemOpen
  \bibfield  {author} {\bibinfo {author} {\bibfnamefont {B.}~\bibnamefont {{Ripperda}}}, \bibinfo {author} {\bibfnamefont {F.}~\bibnamefont {{Bacchini}}},\ and\ \bibinfo {author} {\bibfnamefont {A.~A.}\ \bibnamefont {{Philippov}}},\ }\bibfield  {title} {\bibinfo {title} {{Magnetic Reconnection and Hot Spot Formation in Black Hole Accretion Disks}},\ }\href {https://doi.org/10.3847/1538-4357/ababab} {\bibfield  {journal} {\bibinfo  {journal} {\apj}\ }\textbf {\bibinfo {volume} {900}},\ \bibinfo {eid} {100} (\bibinfo {year} {2020})},\ \Eprint {https://arxiv.org/abs/2003.04330} {arXiv:2003.04330 [astro-ph.HE]} \BibitemShut {NoStop}%
\bibitem [{\citenamefont {Ripperda}\ \emph {et~al.}(2022)\citenamefont {Ripperda}, \citenamefont {Liska}, \citenamefont {Chatterjee}, \citenamefont {Musoke}, \citenamefont {Philippov}, \citenamefont {Markoff}, \citenamefont {Tchekhovskoy},\ and\ \citenamefont {Younsi}}]{Ripperda_2022}%
  \BibitemOpen
  \bibfield  {author} {\bibinfo {author} {\bibfnamefont {B.}~\bibnamefont {Ripperda}}, \bibinfo {author} {\bibfnamefont {M.}~\bibnamefont {Liska}}, \bibinfo {author} {\bibfnamefont {K.}~\bibnamefont {Chatterjee}}, \bibinfo {author} {\bibfnamefont {G.}~\bibnamefont {Musoke}}, \bibinfo {author} {\bibfnamefont {A.~A.}\ \bibnamefont {Philippov}}, \bibinfo {author} {\bibfnamefont {S.~B.}\ \bibnamefont {Markoff}}, \bibinfo {author} {\bibfnamefont {A.}~\bibnamefont {Tchekhovskoy}},\ and\ \bibinfo {author} {\bibfnamefont {Z.}~\bibnamefont {Younsi}},\ }\bibfield  {title} {\bibinfo {title} {Black hole flares: Ejection of accreted magnetic flux through 3d plasmoid-mediated reconnection},\ }\href {https://doi.org/10.3847/2041-8213/ac46a1} {\bibfield  {journal} {\bibinfo  {journal} {The Astrophysical Journal Letters}\ }\textbf {\bibinfo {volume} {924}},\ \bibinfo {pages} {L32} (\bibinfo {year} {2022})}\BibitemShut {NoStop}%
\bibitem [{\citenamefont {{Porth}}\ and\ \citenamefont {{Fendt}}(2010)}]{2010ApJ...709.1100P}%
  \BibitemOpen
  \bibfield  {author} {\bibinfo {author} {\bibfnamefont {O.}~\bibnamefont {{Porth}}}\ and\ \bibinfo {author} {\bibfnamefont {C.}~\bibnamefont {{Fendt}}},\ }\bibfield  {title} {\bibinfo {title} {{Acceleration and Collimation of Relativistic Magnetohydrodynamic Disk Winds}},\ }\href {https://doi.org/10.1088/0004-637X/709/2/1100} {\bibfield  {journal} {\bibinfo  {journal} {\apj}\ }\textbf {\bibinfo {volume} {709}},\ \bibinfo {pages} {1100} (\bibinfo {year} {2010})},\ \Eprint {https://arxiv.org/abs/0911.3001} {arXiv:0911.3001 [astro-ph.HE]} \BibitemShut {NoStop}%
\bibitem [{\citenamefont {{Bromberg}}\ and\ \citenamefont {{Tchekhovskoy}}(2016)}]{2016MNRAS.456.1739B}%
  \BibitemOpen
  \bibfield  {author} {\bibinfo {author} {\bibfnamefont {O.}~\bibnamefont {{Bromberg}}}\ and\ \bibinfo {author} {\bibfnamefont {A.}~\bibnamefont {{Tchekhovskoy}}},\ }\bibfield  {title} {\bibinfo {title} {{Relativistic MHD simulations of core-collapse GRB jets: 3D instabilities and magnetic dissipation}},\ }\href {https://doi.org/10.1093/mnras/stv2591} {\bibfield  {journal} {\bibinfo  {journal} {\mnras}\ }\textbf {\bibinfo {volume} {456}},\ \bibinfo {pages} {1739} (\bibinfo {year} {2016})},\ \Eprint {https://arxiv.org/abs/1508.02721} {arXiv:1508.02721 [astro-ph.HE]} \BibitemShut {NoStop}%
\bibitem [{\citenamefont {{Mattia}}\ \emph {et~al.}(2023)\citenamefont {{Mattia}}, \citenamefont {{Del Zanna}}, \citenamefont {{Bugli}}, \citenamefont {{Pavan}}, \citenamefont {{Ciolfi}}, \citenamefont {{Bodo}},\ and\ \citenamefont {{Mignone}}}]{2023A&A...679A..49M}%
  \BibitemOpen
  \bibfield  {author} {\bibinfo {author} {\bibfnamefont {G.}~\bibnamefont {{Mattia}}}, \bibinfo {author} {\bibfnamefont {L.}~\bibnamefont {{Del Zanna}}}, \bibinfo {author} {\bibfnamefont {M.}~\bibnamefont {{Bugli}}}, \bibinfo {author} {\bibfnamefont {A.}~\bibnamefont {{Pavan}}}, \bibinfo {author} {\bibfnamefont {R.}~\bibnamefont {{Ciolfi}}}, \bibinfo {author} {\bibfnamefont {G.}~\bibnamefont {{Bodo}}},\ and\ \bibinfo {author} {\bibfnamefont {A.}~\bibnamefont {{Mignone}}},\ }\bibfield  {title} {\bibinfo {title} {{Resistive relativistic MHD simulations of astrophysical jets}},\ }\href {https://doi.org/10.1051/0004-6361/202347126} {\bibfield  {journal} {\bibinfo  {journal} {\aap}\ }\textbf {\bibinfo {volume} {679}},\ \bibinfo {eid} {A49} (\bibinfo {year} {2023})},\ \Eprint {https://arxiv.org/abs/2308.09477} {arXiv:2308.09477 [astro-ph.HE]} \BibitemShut {NoStop}%
\bibitem [{\citenamefont {{Massaglia}}\ \emph {et~al.}(2022)\citenamefont {{Massaglia}}, \citenamefont {{Bodo}}, \citenamefont {{Rossi}}, \citenamefont {{Capetti}},\ and\ \citenamefont {{Mignone}}}]{2022A&A...659A.139M}%
  \BibitemOpen
  \bibfield  {author} {\bibinfo {author} {\bibfnamefont {S.}~\bibnamefont {{Massaglia}}}, \bibinfo {author} {\bibfnamefont {G.}~\bibnamefont {{Bodo}}}, \bibinfo {author} {\bibfnamefont {P.}~\bibnamefont {{Rossi}}}, \bibinfo {author} {\bibfnamefont {A.}~\bibnamefont {{Capetti}}},\ and\ \bibinfo {author} {\bibfnamefont {A.}~\bibnamefont {{Mignone}}},\ }\bibfield  {title} {\bibinfo {title} {{Making Fanaroff-Riley I radio sources. III. The effects of the magnetic field on relativistic jets' propagation and source morphologies}},\ }\href {https://doi.org/10.1051/0004-6361/202038724} {\bibfield  {journal} {\bibinfo  {journal} {\aap}\ }\textbf {\bibinfo {volume} {659}},\ \bibinfo {eid} {A139} (\bibinfo {year} {2022})},\ \Eprint {https://arxiv.org/abs/2112.06827} {arXiv:2112.06827 [astro-ph.HE]} \BibitemShut {NoStop}%
\bibitem [{\citenamefont {{Gottlieb}}\ \emph {et~al.}(2022)\citenamefont {{Gottlieb}}, \citenamefont {{Liska}}, \citenamefont {{Tchekhovskoy}}, \citenamefont {{Bromberg}}, \citenamefont {{Lalakos}}, \citenamefont {{Giannios}},\ and\ \citenamefont {{M{\"o}sta}}}]{2022ApJ...933L...9G}%
  \BibitemOpen
  \bibfield  {author} {\bibinfo {author} {\bibfnamefont {O.}~\bibnamefont {{Gottlieb}}}, \bibinfo {author} {\bibfnamefont {M.}~\bibnamefont {{Liska}}}, \bibinfo {author} {\bibfnamefont {A.}~\bibnamefont {{Tchekhovskoy}}}, \bibinfo {author} {\bibfnamefont {O.}~\bibnamefont {{Bromberg}}}, \bibinfo {author} {\bibfnamefont {A.}~\bibnamefont {{Lalakos}}}, \bibinfo {author} {\bibfnamefont {D.}~\bibnamefont {{Giannios}}},\ and\ \bibinfo {author} {\bibfnamefont {P.}~\bibnamefont {{M{\"o}sta}}},\ }\bibfield  {title} {\bibinfo {title} {{Black Hole to Photosphere: 3D GRMHD Simulations of Collapsars Reveal Wobbling and Hybrid Composition Jets}},\ }\href {https://doi.org/10.3847/2041-8213/ac7530} {\bibfield  {journal} {\bibinfo  {journal} {\apjl}\ }\textbf {\bibinfo {volume} {933}},\ \bibinfo {eid} {L9} (\bibinfo {year} {2022})},\ \Eprint {https://arxiv.org/abs/2204.12501} {arXiv:2204.12501 [astro-ph.HE]} \BibitemShut {NoStop}%
\bibitem [{\citenamefont {{Lalakos}}\ \emph {et~al.}(2024)\citenamefont {{Lalakos}}, \citenamefont {{Tchekhovskoy}}, \citenamefont {{Bromberg}}, \citenamefont {{Gottlieb}}, \citenamefont {{Jacquemin-Ide}}, \citenamefont {{Liska}},\ and\ \citenamefont {{Zhang}}}]{2024ApJ...964...79L}%
  \BibitemOpen
  \bibfield  {author} {\bibinfo {author} {\bibfnamefont {A.}~\bibnamefont {{Lalakos}}}, \bibinfo {author} {\bibfnamefont {A.}~\bibnamefont {{Tchekhovskoy}}}, \bibinfo {author} {\bibfnamefont {O.}~\bibnamefont {{Bromberg}}}, \bibinfo {author} {\bibfnamefont {O.}~\bibnamefont {{Gottlieb}}}, \bibinfo {author} {\bibfnamefont {J.}~\bibnamefont {{Jacquemin-Ide}}}, \bibinfo {author} {\bibfnamefont {M.}~\bibnamefont {{Liska}}},\ and\ \bibinfo {author} {\bibfnamefont {H.}~\bibnamefont {{Zhang}}},\ }\bibfield  {title} {\bibinfo {title} {{Jets with a Twist: The Emergence of FR0 Jets in a 3D GRMHD Simulation of Zero-angular-momentum Black Hole Accretion}},\ }\href {https://doi.org/10.3847/1538-4357/ad0974} {\bibfield  {journal} {\bibinfo  {journal} {\apj}\ }\textbf {\bibinfo {volume} {964}},\ \bibinfo {eid} {79} (\bibinfo {year} {2024})},\ \Eprint {https://arxiv.org/abs/2310.11487} {arXiv:2310.11487 [astro-ph.HE]} \BibitemShut {NoStop}%
\bibitem [{\citenamefont {{Spitkovsky}}(2006)}]{2006ApJ...648L..51S}%
  \BibitemOpen
  \bibfield  {author} {\bibinfo {author} {\bibfnamefont {A.}~\bibnamefont {{Spitkovsky}}},\ }\bibfield  {title} {\bibinfo {title} {{Time-dependent Force-free Pulsar Magnetospheres: Axisymmetric and Oblique Rotators}},\ }\href {https://doi.org/10.1086/507518} {\bibfield  {journal} {\bibinfo  {journal} {\apjl}\ }\textbf {\bibinfo {volume} {648}},\ \bibinfo {pages} {L51} (\bibinfo {year} {2006})},\ \Eprint {https://arxiv.org/abs/astro-ph/0603147} {arXiv:astro-ph/0603147 [astro-ph]} \BibitemShut {NoStop}%
\bibitem [{\citenamefont {Timokhin}(2006)}]{Timokhin_2006}%
  \BibitemOpen
  \bibfield  {author} {\bibinfo {author} {\bibfnamefont {A.~N.}\ \bibnamefont {Timokhin}},\ }\bibfield  {title} {\bibinfo {title} {On the force-free magnetosphere of an aligned rotator},\ }\href {https://doi.org/10.1111/j.1365-2966.2006.10192.x} {\bibfield  {journal} {\bibinfo  {journal} {Monthly Notices of the Royal Astronomical Society}\ }\textbf {\bibinfo {volume} {368}},\ \bibinfo {pages} {1055–1072} (\bibinfo {year} {2006})}\BibitemShut {NoStop}%
\bibitem [{\citenamefont {{Yuan}}\ \emph {et~al.}(2019)\citenamefont {{Yuan}}, \citenamefont {{Spitkovsky}}, \citenamefont {{Blandford}},\ and\ \citenamefont {{Wilkins}}}]{2019MNRAS.487.4114Y}%
  \BibitemOpen
  \bibfield  {author} {\bibinfo {author} {\bibfnamefont {Y.}~\bibnamefont {{Yuan}}}, \bibinfo {author} {\bibfnamefont {A.}~\bibnamefont {{Spitkovsky}}}, \bibinfo {author} {\bibfnamefont {R.~D.}\ \bibnamefont {{Blandford}}},\ and\ \bibinfo {author} {\bibfnamefont {D.~R.}\ \bibnamefont {{Wilkins}}},\ }\bibfield  {title} {\bibinfo {title} {{Black hole magnetosphere with small-scale flux tubes - II. Stability and dynamics}},\ }\href {https://doi.org/10.1093/mnras/stz1599} {\bibfield  {journal} {\bibinfo  {journal} {\mnras}\ }\textbf {\bibinfo {volume} {487}},\ \bibinfo {pages} {4114} (\bibinfo {year} {2019})},\ \Eprint {https://arxiv.org/abs/1901.02834} {arXiv:1901.02834 [astro-ph.HE]} \BibitemShut {NoStop}%
\bibitem [{\citenamefont {{Mahlmann}}\ and\ \citenamefont {{Aloy}}(2022)}]{Mahlmann2022}%
  \BibitemOpen
  \bibfield  {author} {\bibinfo {author} {\bibfnamefont {J.~F.}\ \bibnamefont {{Mahlmann}}}\ and\ \bibinfo {author} {\bibfnamefont {M.~A.}\ \bibnamefont {{Aloy}}},\ }\bibfield  {title} {\bibinfo {title} {{Diffusivity in force-free simulations of global magnetospheres}},\ }\href {https://doi.org/10.1093/mnras/stab2830} {\bibfield  {journal} {\bibinfo  {journal} {\mnras}\ }\textbf {\bibinfo {volume} {509}},\ \bibinfo {pages} {1504} (\bibinfo {year} {2022})},\ \Eprint {https://arxiv.org/abs/2109.13936} {arXiv:2109.13936 [astro-ph.HE]} \BibitemShut {NoStop}%
\bibitem [{\citenamefont {{Sharma}}\ \emph {et~al.}(2023)\citenamefont {{Sharma}}, \citenamefont {{Barkov}},\ and\ \citenamefont {{Lyutikov}}}]{2023MNRAS.524.6024S}%
  \BibitemOpen
  \bibfield  {author} {\bibinfo {author} {\bibfnamefont {P.}~\bibnamefont {{Sharma}}}, \bibinfo {author} {\bibfnamefont {M.~V.}\ \bibnamefont {{Barkov}}},\ and\ \bibinfo {author} {\bibfnamefont {M.}~\bibnamefont {{Lyutikov}}},\ }\bibfield  {title} {\bibinfo {title} {{Relativistic coronal mass ejections from magnetars}},\ }\href {https://doi.org/10.1093/mnras/stad2192} {\bibfield  {journal} {\bibinfo  {journal} {\mnras}\ }\textbf {\bibinfo {volume} {524}},\ \bibinfo {pages} {6024} (\bibinfo {year} {2023})},\ \Eprint {https://arxiv.org/abs/2302.08848} {arXiv:2302.08848 [astro-ph.HE]} \BibitemShut {NoStop}%
\bibitem [{\citenamefont {{Alic}}\ \emph {et~al.}(2012)\citenamefont {{Alic}}, \citenamefont {{Moesta}}, \citenamefont {{Rezzolla}}, \citenamefont {{Zanotti}},\ and\ \citenamefont {{Jaramillo}}}]{Alic2012_FFR}%
  \BibitemOpen
  \bibfield  {author} {\bibinfo {author} {\bibfnamefont {D.}~\bibnamefont {{Alic}}}, \bibinfo {author} {\bibfnamefont {P.}~\bibnamefont {{Moesta}}}, \bibinfo {author} {\bibfnamefont {L.}~\bibnamefont {{Rezzolla}}}, \bibinfo {author} {\bibfnamefont {O.}~\bibnamefont {{Zanotti}}},\ and\ \bibinfo {author} {\bibfnamefont {J.~L.}\ \bibnamefont {{Jaramillo}}},\ }\bibfield  {title} {\bibinfo {title} {{Accurate Simulations of Binary Black Hole Mergers in Force-free Electrodynamics}},\ }\href {https://doi.org/10.1088/0004-637X/754/1/36} {\bibfield  {journal} {\bibinfo  {journal} {\apj}\ }\textbf {\bibinfo {volume} {754}},\ \bibinfo {eid} {36} (\bibinfo {year} {2012})},\ \Eprint {https://arxiv.org/abs/1204.2226} {arXiv:1204.2226 [gr-qc]} \BibitemShut {NoStop}%
\bibitem [{\citenamefont {{Kriel}}\ \emph {et~al.}(2022)\citenamefont {{Kriel}}, \citenamefont {{Beattie}}, \citenamefont {{Seta}},\ and\ \citenamefont {{Federrath}}}]{Kriel2022_turbulence_relation}%
  \BibitemOpen
  \bibfield  {author} {\bibinfo {author} {\bibfnamefont {N.}~\bibnamefont {{Kriel}}}, \bibinfo {author} {\bibfnamefont {J.~R.}\ \bibnamefont {{Beattie}}}, \bibinfo {author} {\bibfnamefont {A.}~\bibnamefont {{Seta}}},\ and\ \bibinfo {author} {\bibfnamefont {C.}~\bibnamefont {{Federrath}}},\ }\bibfield  {title} {\bibinfo {title} {{Fundamental scales in the kinematic phase of the turbulent dynamo}},\ }\href {https://doi.org/10.1093/mnras/stac969} {\bibfield  {journal} {\bibinfo  {journal} {\mnras}\ }\textbf {\bibinfo {volume} {513}},\ \bibinfo {pages} {2457} (\bibinfo {year} {2022})},\ \Eprint {https://arxiv.org/abs/2204.00828} {arXiv:2204.00828 [astro-ph.SR]} \BibitemShut {NoStop}%
\bibitem [{\citenamefont {{Grete}}\ \emph {et~al.}(2023)\citenamefont {{Grete}}, \citenamefont {{O'Shea}},\ and\ \citenamefont {{Beckwith}}}]{Grete2023_transfer_function_dissipation}%
  \BibitemOpen
  \bibfield  {author} {\bibinfo {author} {\bibfnamefont {P.}~\bibnamefont {{Grete}}}, \bibinfo {author} {\bibfnamefont {B.~W.}\ \bibnamefont {{O'Shea}}},\ and\ \bibinfo {author} {\bibfnamefont {K.}~\bibnamefont {{Beckwith}}},\ }\bibfield  {title} {\bibinfo {title} {{As a Matter of Dynamical Range - Scale Dependent Energy Dynamics in MHD Turbulence}},\ }\href {https://doi.org/10.3847/2041-8213/acaea7} {\bibfield  {journal} {\bibinfo  {journal} {\apjl}\ }\textbf {\bibinfo {volume} {942}},\ \bibinfo {eid} {L34} (\bibinfo {year} {2023})},\ \Eprint {https://arxiv.org/abs/2211.09750} {arXiv:2211.09750 [astro-ph.GA]} \BibitemShut {NoStop}%
\bibitem [{\citenamefont {{Shivakumar}}\ and\ \citenamefont {{Federrath}}(2025)}]{Shivakumar2025_numerical_dissipation}%
  \BibitemOpen
  \bibfield  {author} {\bibinfo {author} {\bibfnamefont {L.~M.}\ \bibnamefont {{Shivakumar}}}\ and\ \bibinfo {author} {\bibfnamefont {C.}~\bibnamefont {{Federrath}}},\ }\bibfield  {title} {\bibinfo {title} {{Numerical viscosity and resistivity in MHD turbulence simulations}},\ }\bibfield  {journal} {\bibinfo  {journal} {\mnras}\ }\href {https://doi.org/10.1093/mnras/staf160} {10.1093/mnras/staf160} (\bibinfo {year} {2025}),\ \Eprint {https://arxiv.org/abs/2311.10350} {arXiv:2311.10350 [astro-ph.SR]} \BibitemShut {NoStop}%
\bibitem [{\citenamefont {{Sweet}}(1958)}]{Sweet:1958IAUS....6..123S}%
  \BibitemOpen
  \bibfield  {author} {\bibinfo {author} {\bibfnamefont {P.~A.}\ \bibnamefont {{Sweet}}},\ }\bibfield  {title} {\bibinfo {title} {{The Neutral Point Theory of Solar Flares}},\ }in\ \href@noop {} {\emph {\bibinfo {booktitle} {Electromagnetic Phenomena in Cosmical Physics}}},\ \bibinfo {series} {IAU Symposium}, Vol.~\bibinfo {volume} {6},\ \bibinfo {editor} {edited by\ \bibinfo {editor} {\bibfnamefont {B.}~\bibnamefont {{Lehnert}}}}\ (\bibinfo {year} {1958})\ p.\ \bibinfo {pages} {123}\BibitemShut {NoStop}%
\bibitem [{\citenamefont {{Parker}}(1963)}]{Parker:1963ApJS....8..177P}%
  \BibitemOpen
  \bibfield  {author} {\bibinfo {author} {\bibfnamefont {E.~N.}\ \bibnamefont {{Parker}}},\ }\bibfield  {title} {\bibinfo {title} {{The Solar-Flare Phenomenon and the Theory of Reconnection and Annihiliation of Magnetic Fields.}},\ }\href {https://doi.org/10.1086/190087} {\bibfield  {journal} {\bibinfo  {journal} {\apjs}\ }\textbf {\bibinfo {volume} {8}},\ \bibinfo {pages} {177} (\bibinfo {year} {1963})}\BibitemShut {NoStop}%
\bibitem [{\citenamefont {{Kopp}}\ and\ \citenamefont {{Pneuman}}(1976)}]{1976SoPh...50...85K}%
  \BibitemOpen
  \bibfield  {author} {\bibinfo {author} {\bibfnamefont {R.~A.}\ \bibnamefont {{Kopp}}}\ and\ \bibinfo {author} {\bibfnamefont {G.~W.}\ \bibnamefont {{Pneuman}}},\ }\bibfield  {title} {\bibinfo {title} {{Magnetic reconnection in the corona and the loop prominence phenomenon.}},\ }\href {https://doi.org/10.1007/BF00206193} {\bibfield  {journal} {\bibinfo  {journal} {\solphys}\ }\textbf {\bibinfo {volume} {50}},\ \bibinfo {pages} {85} (\bibinfo {year} {1976})}\BibitemShut {NoStop}%
\bibitem [{\citenamefont {{Liu}}\ \emph {et~al.}(2024)\citenamefont {{Liu}}, \citenamefont {{Angelopoulos}}, \citenamefont {{Nishimura}}, \citenamefont {{Shen}}, \citenamefont {{Shi}},\ and\ \citenamefont {{Hartinger}}}]{2024GeoRL..5112730L}%
  \BibitemOpen
  \bibfield  {author} {\bibinfo {author} {\bibfnamefont {T.~Z.}\ \bibnamefont {{Liu}}}, \bibinfo {author} {\bibfnamefont {V.}~\bibnamefont {{Angelopoulos}}}, \bibinfo {author} {\bibfnamefont {Y.}~\bibnamefont {{Nishimura}}}, \bibinfo {author} {\bibfnamefont {Y.}~\bibnamefont {{Shen}}}, \bibinfo {author} {\bibfnamefont {X.}~\bibnamefont {{Shi}}},\ and\ \bibinfo {author} {\bibfnamefont {M.~D.}\ \bibnamefont {{Hartinger}}},\ }\bibfield  {title} {\bibinfo {title} {{Near-Earth Reconnection Contributing to Recovery Phase of Geomagnetic Storm}},\ }\href {https://doi.org/10.1029/2024GL112730} {\bibfield  {journal} {\bibinfo  {journal} {\grl}\ }\textbf {\bibinfo {volume} {51}},\ \bibinfo {pages} {2024GL112730} (\bibinfo {year} {2024})}\BibitemShut {NoStop}%
\bibitem [{\citenamefont {{Birn}}\ and\ \citenamefont {{Hones}}(1981)}]{1981JGR....86.6802B}%
  \BibitemOpen
  \bibfield  {author} {\bibinfo {author} {\bibfnamefont {J.}~\bibnamefont {{Birn}}}\ and\ \bibinfo {author} {\bibfnamefont {E.~W.}\ \bibnamefont {{Hones}}, \bibfnamefont {Jr.}},\ }\bibfield  {title} {\bibinfo {title} {{Three-dimensional computer modeling of dynamic reconnection in the geomagnetic tail}},\ }\href {https://doi.org/10.1029/JA086iA08p06802} {\bibfield  {journal} {\bibinfo  {journal} {\jgr}\ }\textbf {\bibinfo {volume} {86}},\ \bibinfo {pages} {6802} (\bibinfo {year} {1981})}\BibitemShut {NoStop}%
\bibitem [{\citenamefont {{Lee}}\ and\ \citenamefont {{Fu}}(1985)}]{1985GeoRL..12..105L}%
  \BibitemOpen
  \bibfield  {author} {\bibinfo {author} {\bibfnamefont {L.~C.}\ \bibnamefont {{Lee}}}\ and\ \bibinfo {author} {\bibfnamefont {Z.~F.}\ \bibnamefont {{Fu}}},\ }\bibfield  {title} {\bibinfo {title} {{A theory of magnetic flux transfer at the Earth's magnetopause}},\ }\href {https://doi.org/10.1029/GL012i002p00105} {\bibfield  {journal} {\bibinfo  {journal} {\grl}\ }\textbf {\bibinfo {volume} {12}},\ \bibinfo {pages} {105} (\bibinfo {year} {1985})}\BibitemShut {NoStop}%
\bibitem [{\citenamefont {{Beattie}}\ \emph {et~al.}(2024)\citenamefont {{Beattie}}, \citenamefont {{Federrath}}, \citenamefont {{Klessen}}, \citenamefont {{Cielo}},\ and\ \citenamefont {{Bhattacharjee}}}]{2024arXiv240516626B}%
  \BibitemOpen
  \bibfield  {author} {\bibinfo {author} {\bibfnamefont {J.~R.}\ \bibnamefont {{Beattie}}}, \bibinfo {author} {\bibfnamefont {C.}~\bibnamefont {{Federrath}}}, \bibinfo {author} {\bibfnamefont {R.~S.}\ \bibnamefont {{Klessen}}}, \bibinfo {author} {\bibfnamefont {S.}~\bibnamefont {{Cielo}}},\ and\ \bibinfo {author} {\bibfnamefont {A.}~\bibnamefont {{Bhattacharjee}}},\ }\bibfield  {title} {\bibinfo {title} {{Magnetized compressible turbulence with a fluctuation dynamo and Reynolds numbers over a million}},\ }\href {https://doi.org/10.48550/arXiv.2405.16626} {\bibfield  {journal} {\bibinfo  {journal} {arXiv e-prints}\ ,\ \bibinfo {eid} {arXiv:2405.16626}} (\bibinfo {year} {2024})},\ \Eprint {https://arxiv.org/abs/2405.16626} {arXiv:2405.16626 [astro-ph.GA]} \BibitemShut {NoStop}%
\bibitem [{\citenamefont {{Philippov}}\ and\ \citenamefont {{Kramer}}(2022)}]{2022ARA&A..60..495P}%
  \BibitemOpen
  \bibfield  {author} {\bibinfo {author} {\bibfnamefont {A.}~\bibnamefont {{Philippov}}}\ and\ \bibinfo {author} {\bibfnamefont {M.}~\bibnamefont {{Kramer}}},\ }\bibfield  {title} {\bibinfo {title} {{Pulsar Magnetospheres and Their Radiation}},\ }\href {https://doi.org/10.1146/annurev-astro-052920-112338} {\bibfield  {journal} {\bibinfo  {journal} {\araa}\ }\textbf {\bibinfo {volume} {60}},\ \bibinfo {pages} {495} (\bibinfo {year} {2022})}\BibitemShut {NoStop}%
\bibitem [{\citenamefont {{Tchekhovskoy}}\ \emph {et~al.}(2013)\citenamefont {{Tchekhovskoy}}, \citenamefont {{Spitkovsky}},\ and\ \citenamefont {{Li}}}]{2013MNRAS.435L...1T}%
  \BibitemOpen
  \bibfield  {author} {\bibinfo {author} {\bibfnamefont {A.}~\bibnamefont {{Tchekhovskoy}}}, \bibinfo {author} {\bibfnamefont {A.}~\bibnamefont {{Spitkovsky}}},\ and\ \bibinfo {author} {\bibfnamefont {J.~G.}\ \bibnamefont {{Li}}},\ }\bibfield  {title} {\bibinfo {title} {{Time-dependent 3D magnetohydrodynamic pulsar magnetospheres: oblique rotators.}},\ }\href {https://doi.org/10.1093/mnrasl/slt076} {\bibfield  {journal} {\bibinfo  {journal} {\mnras}\ }\textbf {\bibinfo {volume} {435}},\ \bibinfo {pages} {L1} (\bibinfo {year} {2013})},\ \Eprint {https://arxiv.org/abs/1211.2803} {arXiv:1211.2803 [astro-ph.HE]} \BibitemShut {NoStop}%
\bibitem [{\citenamefont {{Chashkina}}\ \emph {et~al.}(2021)\citenamefont {{Chashkina}}, \citenamefont {{Bromberg}},\ and\ \citenamefont {{Levinson}}}]{2021MNRAS.508.1241C}%
  \BibitemOpen
  \bibfield  {author} {\bibinfo {author} {\bibfnamefont {A.}~\bibnamefont {{Chashkina}}}, \bibinfo {author} {\bibfnamefont {O.}~\bibnamefont {{Bromberg}}},\ and\ \bibinfo {author} {\bibfnamefont {A.}~\bibnamefont {{Levinson}}},\ }\bibfield  {title} {\bibinfo {title} {{GRMHD simulations of BH activation by small scale magnetic loops: formation of striped jets and active coronae}},\ }\href {https://doi.org/10.1093/mnras/stab2513} {\bibfield  {journal} {\bibinfo  {journal} {\mnras}\ }\textbf {\bibinfo {volume} {508}},\ \bibinfo {pages} {1241} (\bibinfo {year} {2021})},\ \Eprint {https://arxiv.org/abs/2106.15738} {arXiv:2106.15738 [astro-ph.HE]} \BibitemShut {NoStop}%
\bibitem [{\citenamefont {{Nathanail}}\ \emph {et~al.}(2022)\citenamefont {{Nathanail}}, \citenamefont {{Mpisketzis}}, \citenamefont {{Porth}}, \citenamefont {{Fromm}},\ and\ \citenamefont {{Rezzolla}}}]{2022MNRAS.513.4267N}%
  \BibitemOpen
  \bibfield  {author} {\bibinfo {author} {\bibfnamefont {A.}~\bibnamefont {{Nathanail}}}, \bibinfo {author} {\bibfnamefont {V.}~\bibnamefont {{Mpisketzis}}}, \bibinfo {author} {\bibfnamefont {O.}~\bibnamefont {{Porth}}}, \bibinfo {author} {\bibfnamefont {C.~M.}\ \bibnamefont {{Fromm}}},\ and\ \bibinfo {author} {\bibfnamefont {L.}~\bibnamefont {{Rezzolla}}},\ }\bibfield  {title} {\bibinfo {title} {{Magnetic reconnection and plasmoid formation in three-dimensional accretion flows around black holes}},\ }\href {https://doi.org/10.1093/mnras/stac1118} {\bibfield  {journal} {\bibinfo  {journal} {\mnras}\ }\textbf {\bibinfo {volume} {513}},\ \bibinfo {pages} {4267} (\bibinfo {year} {2022})},\ \Eprint {https://arxiv.org/abs/2111.03689} {arXiv:2111.03689 [astro-ph.HE]} \BibitemShut {NoStop}%
\bibitem [{\citenamefont {{Sridhar}}\ \emph {et~al.}(2025)\citenamefont {{Sridhar}}, \citenamefont {{Ripperda}}, \citenamefont {{Sironi}}, \citenamefont {{Davelaar}},\ and\ \citenamefont {{Beloborodov}}}]{2025ApJ...979..199S}%
  \BibitemOpen
  \bibfield  {author} {\bibinfo {author} {\bibfnamefont {N.}~\bibnamefont {{Sridhar}}}, \bibinfo {author} {\bibfnamefont {B.}~\bibnamefont {{Ripperda}}}, \bibinfo {author} {\bibfnamefont {L.}~\bibnamefont {{Sironi}}}, \bibinfo {author} {\bibfnamefont {J.}~\bibnamefont {{Davelaar}}},\ and\ \bibinfo {author} {\bibfnamefont {A.~M.}\ \bibnamefont {{Beloborodov}}},\ }\bibfield  {title} {\bibinfo {title} {{Bulk Motions in the Black Hole Jet Sheath as a Candidate for the Comptonizing Corona}},\ }\href {https://doi.org/10.3847/1538-4357/ada385} {\bibfield  {journal} {\bibinfo  {journal} {\apj}\ }\textbf {\bibinfo {volume} {979}},\ \bibinfo {eid} {199} (\bibinfo {year} {2025})},\ \Eprint {https://arxiv.org/abs/2411.10662} {arXiv:2411.10662 [astro-ph.HE]} \BibitemShut {NoStop}%
\bibitem [{\citenamefont {{Lyubarsky}}(2020)}]{2020ApJ...897....1L}%
  \BibitemOpen
  \bibfield  {author} {\bibinfo {author} {\bibfnamefont {Y.}~\bibnamefont {{Lyubarsky}}},\ }\bibfield  {title} {\bibinfo {title} {{Fast Radio Bursts from Reconnection in a Magnetar Magnetosphere}},\ }\href {https://doi.org/10.3847/1538-4357/ab97b5} {\bibfield  {journal} {\bibinfo  {journal} {\apj}\ }\textbf {\bibinfo {volume} {897}},\ \bibinfo {eid} {1} (\bibinfo {year} {2020})},\ \Eprint {https://arxiv.org/abs/2001.02007} {arXiv:2001.02007 [astro-ph.HE]} \BibitemShut {NoStop}%
\bibitem [{\citenamefont {{Mahlmann}}\ \emph {et~al.}(2022)\citenamefont {{Mahlmann}}, \citenamefont {{Philippov}}, \citenamefont {{Levinson}}, \citenamefont {{Spitkovsky}},\ and\ \citenamefont {{Hakobyan}}}]{2022ApJ...932L..20M}%
  \BibitemOpen
  \bibfield  {author} {\bibinfo {author} {\bibfnamefont {J.~F.}\ \bibnamefont {{Mahlmann}}}, \bibinfo {author} {\bibfnamefont {A.~A.}\ \bibnamefont {{Philippov}}}, \bibinfo {author} {\bibfnamefont {A.}~\bibnamefont {{Levinson}}}, \bibinfo {author} {\bibfnamefont {A.}~\bibnamefont {{Spitkovsky}}},\ and\ \bibinfo {author} {\bibfnamefont {H.}~\bibnamefont {{Hakobyan}}},\ }\bibfield  {title} {\bibinfo {title} {{Electromagnetic Fireworks: Fast Radio Bursts from Rapid Reconnection in the Compressed Magnetar Wind}},\ }\href {https://doi.org/10.3847/2041-8213/ac7156} {\bibfield  {journal} {\bibinfo  {journal} {\apjl}\ }\textbf {\bibinfo {volume} {932}},\ \bibinfo {eid} {L20} (\bibinfo {year} {2022})},\ \Eprint {https://arxiv.org/abs/2203.04320} {arXiv:2203.04320 [astro-ph.HE]} \BibitemShut {NoStop}%
\bibitem [{\citenamefont {{Most}}\ and\ \citenamefont {{Philippov}}(2023)}]{2023ApJ...956L..33M}%
  \BibitemOpen
  \bibfield  {author} {\bibinfo {author} {\bibfnamefont {E.~R.}\ \bibnamefont {{Most}}}\ and\ \bibinfo {author} {\bibfnamefont {A.~A.}\ \bibnamefont {{Philippov}}},\ }\bibfield  {title} {\bibinfo {title} {{Electromagnetic Precursors to Black Hole-Neutron Star Gravitational Wave Events: Flares and Reconnection-powered Fast Radio Transients from the Late Inspiral}},\ }\href {https://doi.org/10.3847/2041-8213/acfdae} {\bibfield  {journal} {\bibinfo  {journal} {\apjl}\ }\textbf {\bibinfo {volume} {956}},\ \bibinfo {eid} {L33} (\bibinfo {year} {2023})},\ \Eprint {https://arxiv.org/abs/2309.04271} {arXiv:2309.04271 [astro-ph.HE]} \BibitemShut {NoStop}%
\bibitem [{\citenamefont {{Alfv{\'e}n}}(1942)}]{Alfven:1942Natur.150..405A}%
  \BibitemOpen
  \bibfield  {author} {\bibinfo {author} {\bibfnamefont {H.}~\bibnamefont {{Alfv{\'e}n}}},\ }\bibfield  {title} {\bibinfo {title} {{Existence of Electromagnetic-Hydrodynamic Waves}},\ }\href {https://doi.org/10.1038/150405d0} {\bibfield  {journal} {\bibinfo  {journal} {\nat}\ }\textbf {\bibinfo {volume} {150}},\ \bibinfo {pages} {405} (\bibinfo {year} {1942})}\BibitemShut {NoStop}%
\bibitem [{\citenamefont {{Lyubarsky}}(2005)}]{Lyubarsky_2005MNRAS}%
  \BibitemOpen
  \bibfield  {author} {\bibinfo {author} {\bibfnamefont {Y.~E.}\ \bibnamefont {{Lyubarsky}}},\ }\bibfield  {title} {\bibinfo {title} {{On the relativistic magnetic reconnection}},\ }\href {https://doi.org/10.1111/j.1365-2966.2005.08767.x} {\bibfield  {journal} {\bibinfo  {journal} {\mnras}\ }\textbf {\bibinfo {volume} {358}},\ \bibinfo {pages} {113} (\bibinfo {year} {2005})},\ \Eprint {https://arxiv.org/abs/astro-ph/0501392} {arXiv:astro-ph/0501392 [astro-ph]} \BibitemShut {NoStop}%
\bibitem [{\citenamefont {Biskamp}(1993)}]{Biskamp_1993}%
  \BibitemOpen
  \bibfield  {author} {\bibinfo {author} {\bibfnamefont {D.}~\bibnamefont {Biskamp}},\ }\href@noop {} {\emph {\bibinfo {title} {Nonlinear Magnetohydrodynamics}}},\ Cambridge Monographs on Plasma Physics\ (\bibinfo  {publisher} {Cambridge University Press},\ \bibinfo {year} {1993})\BibitemShut {NoStop}%
\bibitem [{\citenamefont {Lapenta}(2008)}]{Lapenta.PhysRevLett.100.235001}%
  \BibitemOpen
  \bibfield  {author} {\bibinfo {author} {\bibfnamefont {G.}~\bibnamefont {Lapenta}},\ }\bibfield  {title} {\bibinfo {title} {Self-feeding turbulent magnetic reconnection on macroscopic scales},\ }\href {https://doi.org/10.1103/PhysRevLett.100.235001} {\bibfield  {journal} {\bibinfo  {journal} {Phys. Rev. Lett.}\ }\textbf {\bibinfo {volume} {100}},\ \bibinfo {pages} {235001} (\bibinfo {year} {2008})}\BibitemShut {NoStop}%
\bibitem [{\citenamefont {{Loureiro}}\ \emph {et~al.}(2007)\citenamefont {{Loureiro}}, \citenamefont {{Schekochihin}},\ and\ \citenamefont {{Cowley}}}]{2007PhPl...14j0703L}%
  \BibitemOpen
  \bibfield  {author} {\bibinfo {author} {\bibfnamefont {N.~F.}\ \bibnamefont {{Loureiro}}}, \bibinfo {author} {\bibfnamefont {A.~A.}\ \bibnamefont {{Schekochihin}}},\ and\ \bibinfo {author} {\bibfnamefont {S.~C.}\ \bibnamefont {{Cowley}}},\ }\bibfield  {title} {\bibinfo {title} {{Instability of current sheets and formation of plasmoid chains}},\ }\href {https://doi.org/10.1063/1.2783986} {\bibfield  {journal} {\bibinfo  {journal} {Physics of Plasmas}\ }\textbf {\bibinfo {volume} {14}},\ \bibinfo {pages} {100703} (\bibinfo {year} {2007})},\ \Eprint {https://arxiv.org/abs/astro-ph/0703631} {arXiv:astro-ph/0703631 [astro-ph]} \BibitemShut {NoStop}%
\bibitem [{\citenamefont {{Bhattacharjee}}\ \emph {et~al.}(2009)\citenamefont {{Bhattacharjee}}, \citenamefont {{Huang}}, \citenamefont {{Yang}},\ and\ \citenamefont {{Rogers}}}]{Bhattacharjee.2009PhPl...16k2102B}%
  \BibitemOpen
  \bibfield  {author} {\bibinfo {author} {\bibfnamefont {A.}~\bibnamefont {{Bhattacharjee}}}, \bibinfo {author} {\bibfnamefont {Y.-M.}\ \bibnamefont {{Huang}}}, \bibinfo {author} {\bibfnamefont {H.}~\bibnamefont {{Yang}}},\ and\ \bibinfo {author} {\bibfnamefont {B.}~\bibnamefont {{Rogers}}},\ }\bibfield  {title} {\bibinfo {title} {{Fast reconnection in high-Lundquist-number plasmas due to the plasmoid Instability}},\ }\href {https://doi.org/10.1063/1.3264103} {\bibfield  {journal} {\bibinfo  {journal} {Physics of Plasmas}\ }\textbf {\bibinfo {volume} {16}},\ \bibinfo {eid} {112102} (\bibinfo {year} {2009})},\ \Eprint {https://arxiv.org/abs/0906.5599} {arXiv:0906.5599 [physics.plasm-ph]} \BibitemShut {NoStop}%
\bibitem [{\citenamefont {Cassak}\ \emph {et~al.}(2005)\citenamefont {Cassak}, \citenamefont {Shay},\ and\ \citenamefont {Drake}}]{Cassak.PhysRevLett.95.235002}%
  \BibitemOpen
  \bibfield  {author} {\bibinfo {author} {\bibfnamefont {P.~A.}\ \bibnamefont {Cassak}}, \bibinfo {author} {\bibfnamefont {M.~A.}\ \bibnamefont {Shay}},\ and\ \bibinfo {author} {\bibfnamefont {J.~F.}\ \bibnamefont {Drake}},\ }\bibfield  {title} {\bibinfo {title} {Catastrophe model for fast magnetic reconnection onset},\ }\href {https://doi.org/10.1103/PhysRevLett.95.235002} {\bibfield  {journal} {\bibinfo  {journal} {Phys. Rev. Lett.}\ }\textbf {\bibinfo {volume} {95}},\ \bibinfo {pages} {235002} (\bibinfo {year} {2005})}\BibitemShut {NoStop}%
\bibitem [{\citenamefont {Huang}\ and\ \citenamefont {Bhattacharjee}(2010)}]{Huang_2010}%
  \BibitemOpen
  \bibfield  {author} {\bibinfo {author} {\bibfnamefont {Y.-M.}\ \bibnamefont {Huang}}\ and\ \bibinfo {author} {\bibfnamefont {A.}~\bibnamefont {Bhattacharjee}},\ }\bibfield  {title} {\bibinfo {title} {Scaling laws of resistive magnetohydrodynamic reconnection in the high-lundquist-number, plasmoid-unstable regime},\ }\bibfield  {journal} {\bibinfo  {journal} {Physics of Plasmas}\ }\textbf {\bibinfo {volume} {17}},\ \href {https://doi.org/10.1063/1.3420208} {10.1063/1.3420208} (\bibinfo {year} {2010})\BibitemShut {NoStop}%
\bibitem [{\citenamefont {Uzdensky}\ \emph {et~al.}(2010)\citenamefont {Uzdensky}, \citenamefont {Loureiro},\ and\ \citenamefont {Schekochihin}}]{Uzdensky_2010}%
  \BibitemOpen
  \bibfield  {author} {\bibinfo {author} {\bibfnamefont {D.~A.}\ \bibnamefont {Uzdensky}}, \bibinfo {author} {\bibfnamefont {N.~F.}\ \bibnamefont {Loureiro}},\ and\ \bibinfo {author} {\bibfnamefont {A.~A.}\ \bibnamefont {Schekochihin}},\ }\bibfield  {title} {\bibinfo {title} {Fast magnetic reconnection in the plasmoid-dominated regime},\ }\bibfield  {journal} {\bibinfo  {journal} {Physical Review Letters}\ }\textbf {\bibinfo {volume} {105}},\ \href {https://doi.org/10.1103/physrevlett.105.235002} {10.1103/physrevlett.105.235002} (\bibinfo {year} {2010})\BibitemShut {NoStop}%
\bibitem [{\citenamefont {Loureiro}\ \emph {et~al.}(2012)\citenamefont {Loureiro}, \citenamefont {Samtaney}, \citenamefont {Schekochihin},\ and\ \citenamefont {Uzdensky}}]{Loureiro_2012}%
  \BibitemOpen
  \bibfield  {author} {\bibinfo {author} {\bibfnamefont {N.~F.}\ \bibnamefont {Loureiro}}, \bibinfo {author} {\bibfnamefont {R.}~\bibnamefont {Samtaney}}, \bibinfo {author} {\bibfnamefont {A.~A.}\ \bibnamefont {Schekochihin}},\ and\ \bibinfo {author} {\bibfnamefont {D.~A.}\ \bibnamefont {Uzdensky}},\ }\bibfield  {title} {\bibinfo {title} {Magnetic reconnection and stochastic plasmoid chains in high-lundquist-number plasmas},\ }\bibfield  {journal} {\bibinfo  {journal} {Physics of Plasmas}\ }\textbf {\bibinfo {volume} {19}},\ \href {https://doi.org/10.1063/1.3703318} {10.1063/1.3703318} (\bibinfo {year} {2012})\BibitemShut {NoStop}%
\bibitem [{\citenamefont {Ni}\ \emph {et~al.}(2012)\citenamefont {Ni}, \citenamefont {Ziegler}, \citenamefont {Huang}, \citenamefont {Lin},\ and\ \citenamefont {Mei}}]{Ni2012}%
  \BibitemOpen
  \bibfield  {author} {\bibinfo {author} {\bibfnamefont {L.}~\bibnamefont {Ni}}, \bibinfo {author} {\bibfnamefont {U.}~\bibnamefont {Ziegler}}, \bibinfo {author} {\bibfnamefont {Y.-M.}\ \bibnamefont {Huang}}, \bibinfo {author} {\bibfnamefont {J.}~\bibnamefont {Lin}},\ and\ \bibinfo {author} {\bibfnamefont {Z.}~\bibnamefont {Mei}},\ }\bibfield  {title} {\bibinfo {title} {Effects of plasma beta on the plasmoid instability},\ }\href {https://doi.org/10.1063/1.4736993} {\bibfield  {journal} {\bibinfo  {journal} {Physics of Plasmas}\ }\textbf {\bibinfo {volume} {19}},\ \bibinfo {pages} {072902} (\bibinfo {year} {2012})}\BibitemShut {NoStop}%
\bibitem [{\citenamefont {Huang}\ and\ \citenamefont {Bhattacharjee}(2013)}]{Huang.10.1063/1.4802941}%
  \BibitemOpen
  \bibfield  {author} {\bibinfo {author} {\bibfnamefont {Y.-M.}\ \bibnamefont {Huang}}\ and\ \bibinfo {author} {\bibfnamefont {A.}~\bibnamefont {Bhattacharjee}},\ }\bibfield  {title} {\bibinfo {title} {{Plasmoid instability in high-Lundquist-number magnetic reconnection}},\ }\href {https://doi.org/10.1063/1.4802941} {\bibfield  {journal} {\bibinfo  {journal} {Physics of Plasmas}\ }\textbf {\bibinfo {volume} {20}},\ \bibinfo {pages} {055702} (\bibinfo {year} {2013})},\ \Eprint {https://arxiv.org/abs/https://pubs.aip.org/aip/pop/article-pdf/doi/10.1063/1.4802941/13855762/055702\_1\_online.pdf} {https://pubs.aip.org/aip/pop/article-pdf/doi/10.1063/1.4802941/13855762/055702\_1\_online.pdf} \BibitemShut {NoStop}%
\bibitem [{\citenamefont {Takamoto}(2013)}]{Takamoto_2013}%
  \BibitemOpen
  \bibfield  {author} {\bibinfo {author} {\bibfnamefont {M.}~\bibnamefont {Takamoto}},\ }\bibfield  {title} {\bibinfo {title} {Evolution of relativistic plasmoid chains in a poynting-dominated plasma},\ }\href {https://doi.org/10.1088/0004-637X/775/1/50} {\bibfield  {journal} {\bibinfo  {journal} {The Astrophysical Journal}\ }\textbf {\bibinfo {volume} {775}},\ \bibinfo {pages} {50} (\bibinfo {year} {2013})}\BibitemShut {NoStop}%
\bibitem [{\citenamefont {Comisso}\ \emph {et~al.}(2015)\citenamefont {Comisso}, \citenamefont {Grasso},\ and\ \citenamefont {Waelbroeck}}]{Comisso.10.1063/1.4918331}%
  \BibitemOpen
  \bibfield  {author} {\bibinfo {author} {\bibfnamefont {L.}~\bibnamefont {Comisso}}, \bibinfo {author} {\bibfnamefont {D.}~\bibnamefont {Grasso}},\ and\ \bibinfo {author} {\bibfnamefont {F.~L.}\ \bibnamefont {Waelbroeck}},\ }\bibfield  {title} {\bibinfo {title} {{Extended theory of the Taylor problem in the plasmoid-unstable regime}},\ }\href {https://doi.org/10.1063/1.4918331} {\bibfield  {journal} {\bibinfo  {journal} {Physics of Plasmas}\ }\textbf {\bibinfo {volume} {22}},\ \bibinfo {pages} {042109} (\bibinfo {year} {2015})},\ \Eprint {https://arxiv.org/abs/https://pubs.aip.org/aip/pop/article-pdf/doi/10.1063/1.4918331/14015080/042109\_1\_online.pdf} {https://pubs.aip.org/aip/pop/article-pdf/doi/10.1063/1.4918331/14015080/042109\_1\_online.pdf} \BibitemShut {NoStop}%
\bibitem [{\citenamefont {Loureiro}\ and\ \citenamefont {Uzdensky}(2015)}]{Loureiro_2016}%
  \BibitemOpen
  \bibfield  {author} {\bibinfo {author} {\bibfnamefont {N.~F.}\ \bibnamefont {Loureiro}}\ and\ \bibinfo {author} {\bibfnamefont {D.~A.}\ \bibnamefont {Uzdensky}},\ }\bibfield  {title} {\bibinfo {title} {Magnetic reconnection: from the sweet–parker model to stochastic plasmoid chains},\ }\href {https://doi.org/10.1088/0741-3335/58/1/014021} {\bibfield  {journal} {\bibinfo  {journal} {Plasma Physics and Controlled Fusion}\ }\textbf {\bibinfo {volume} {58}},\ \bibinfo {pages} {014021} (\bibinfo {year} {2015})}\BibitemShut {NoStop}%
\bibitem [{\citenamefont {Huang}\ \emph {et~al.}(2017)\citenamefont {Huang}, \citenamefont {Comisso},\ and\ \citenamefont {Bhattacharjee}}]{Huang_2017}%
  \BibitemOpen
  \bibfield  {author} {\bibinfo {author} {\bibfnamefont {Y.-M.}\ \bibnamefont {Huang}}, \bibinfo {author} {\bibfnamefont {L.}~\bibnamefont {Comisso}},\ and\ \bibinfo {author} {\bibfnamefont {A.}~\bibnamefont {Bhattacharjee}},\ }\bibfield  {title} {\bibinfo {title} {Plasmoid instability in evolving current sheets and onset of fast reconnection},\ }\href {https://doi.org/10.3847/1538-4357/aa906d} {\bibfield  {journal} {\bibinfo  {journal} {The Astrophysical Journal}\ }\textbf {\bibinfo {volume} {849}},\ \bibinfo {pages} {75} (\bibinfo {year} {2017})}\BibitemShut {NoStop}%
\bibitem [{\citenamefont {Comisso}\ \emph {et~al.}(2018)\citenamefont {Comisso}, \citenamefont {Huang}, \citenamefont {Lingam}, \citenamefont {Hirvijoki},\ and\ \citenamefont {Bhattacharjee}}]{Comisso_2018}%
  \BibitemOpen
  \bibfield  {author} {\bibinfo {author} {\bibfnamefont {L.}~\bibnamefont {Comisso}}, \bibinfo {author} {\bibfnamefont {Y.-M.}\ \bibnamefont {Huang}}, \bibinfo {author} {\bibfnamefont {M.}~\bibnamefont {Lingam}}, \bibinfo {author} {\bibfnamefont {E.}~\bibnamefont {Hirvijoki}},\ and\ \bibinfo {author} {\bibfnamefont {A.}~\bibnamefont {Bhattacharjee}},\ }\bibfield  {title} {\bibinfo {title} {Magnetohydrodynamic turbulence in the plasmoid-mediated regime},\ }\href {https://doi.org/10.3847/1538-4357/aaac83} {\bibfield  {journal} {\bibinfo  {journal} {The Astrophysical Journal}\ }\textbf {\bibinfo {volume} {854}},\ \bibinfo {pages} {103} (\bibinfo {year} {2018})}\BibitemShut {NoStop}%
\bibitem [{\citenamefont {{Sironi}}\ \emph {et~al.}(2025)\citenamefont {{Sironi}}, \citenamefont {{Uzdensky}},\ and\ \citenamefont {{Giannios}}}]{2025arXiv250602101S}%
  \BibitemOpen
  \bibfield  {author} {\bibinfo {author} {\bibfnamefont {L.}~\bibnamefont {{Sironi}}}, \bibinfo {author} {\bibfnamefont {D.~A.}\ \bibnamefont {{Uzdensky}}},\ and\ \bibinfo {author} {\bibfnamefont {D.}~\bibnamefont {{Giannios}}},\ }\bibfield  {title} {\bibinfo {title} {{Relativistic Magnetic Reconnection in Astrophysical Plasmas: A Powerful Mechanism of Nonthermal Emission}},\ }\href {https://doi.org/10.48550/arXiv.2506.02101} {\bibfield  {journal} {\bibinfo  {journal} {arXiv e-prints}\ ,\ \bibinfo {eid} {arXiv:2506.02101}} (\bibinfo {year} {2025})},\ \Eprint {https://arxiv.org/abs/2506.02101} {arXiv:2506.02101 [astro-ph.HE]} \BibitemShut {NoStop}%
\bibitem [{\citenamefont {{Robbins}}\ and\ \citenamefont {{Spitkovsky}}(2025)}]{2025arXiv250808533R}%
  \BibitemOpen
  \bibfield  {author} {\bibinfo {author} {\bibfnamefont {A.}~\bibnamefont {{Robbins}}}\ and\ \bibinfo {author} {\bibfnamefont {A.}~\bibnamefont {{Spitkovsky}}},\ }\bibfield  {title} {\bibinfo {title} {{Transition to Petschek Reconnection in Subrelativistic Pair Plasmas: Implications for Particle Acceleration}},\ }\href {https://doi.org/10.48550/arXiv.2508.08533} {\bibfield  {journal} {\bibinfo  {journal} {arXiv e-prints}\ ,\ \bibinfo {eid} {arXiv:2508.08533}} (\bibinfo {year} {2025})},\ \Eprint {https://arxiv.org/abs/2508.08533} {arXiv:2508.08533 [physics.plasm-ph]} \BibitemShut {NoStop}%
\bibitem [{\citenamefont {{Zenitani}}\ and\ \citenamefont {{Hoshino}}(2001)}]{2001ApJ...562L..63Z}%
  \BibitemOpen
  \bibfield  {author} {\bibinfo {author} {\bibfnamefont {S.}~\bibnamefont {{Zenitani}}}\ and\ \bibinfo {author} {\bibfnamefont {M.}~\bibnamefont {{Hoshino}}},\ }\bibfield  {title} {\bibinfo {title} {{The Generation of Nonthermal Particles in the Relativistic Magnetic Reconnection of Pair Plasmas}},\ }\href {https://doi.org/10.1086/337972} {\bibfield  {journal} {\bibinfo  {journal} {\apjl}\ }\textbf {\bibinfo {volume} {562}},\ \bibinfo {pages} {L63} (\bibinfo {year} {2001})},\ \Eprint {https://arxiv.org/abs/1402.7139} {arXiv:1402.7139 [astro-ph.HE]} \BibitemShut {NoStop}%
\bibitem [{\citenamefont {{Hoshino}}\ and\ \citenamefont {{Lyubarsky}}(2012)}]{2012SSRv..173..521H}%
  \BibitemOpen
  \bibfield  {author} {\bibinfo {author} {\bibfnamefont {M.}~\bibnamefont {{Hoshino}}}\ and\ \bibinfo {author} {\bibfnamefont {Y.}~\bibnamefont {{Lyubarsky}}},\ }\bibfield  {title} {\bibinfo {title} {{Relativistic Reconnection and Particle Acceleration}},\ }\href {https://doi.org/10.1007/s11214-012-9931-z} {\bibfield  {journal} {\bibinfo  {journal} {\ssr}\ }\textbf {\bibinfo {volume} {173}},\ \bibinfo {pages} {521} (\bibinfo {year} {2012})}\BibitemShut {NoStop}%
\bibitem [{\citenamefont {{Guo}}\ \emph {et~al.}(2014)\citenamefont {{Guo}}, \citenamefont {{Li}}, \citenamefont {{Daughton}},\ and\ \citenamefont {{Liu}}}]{2014PhRvL.113o5005G}%
  \BibitemOpen
  \bibfield  {author} {\bibinfo {author} {\bibfnamefont {F.}~\bibnamefont {{Guo}}}, \bibinfo {author} {\bibfnamefont {H.}~\bibnamefont {{Li}}}, \bibinfo {author} {\bibfnamefont {W.}~\bibnamefont {{Daughton}}},\ and\ \bibinfo {author} {\bibfnamefont {Y.-H.}\ \bibnamefont {{Liu}}},\ }\bibfield  {title} {\bibinfo {title} {{Formation of Hard Power Laws in the Energetic Particle Spectra Resulting from Relativistic Magnetic Reconnection}},\ }\href {https://doi.org/10.1103/PhysRevLett.113.155005} {\bibfield  {journal} {\bibinfo  {journal} {\prl}\ }\textbf {\bibinfo {volume} {113}},\ \bibinfo {eid} {155005} (\bibinfo {year} {2014})},\ \Eprint {https://arxiv.org/abs/1405.4040} {arXiv:1405.4040 [astro-ph.HE]} \BibitemShut {NoStop}%
\bibitem [{\citenamefont {{Sironi}}\ and\ \citenamefont {{Spitkovsky}}(2014)}]{2014ApJ...783L..21S}%
  \BibitemOpen
  \bibfield  {author} {\bibinfo {author} {\bibfnamefont {L.}~\bibnamefont {{Sironi}}}\ and\ \bibinfo {author} {\bibfnamefont {A.}~\bibnamefont {{Spitkovsky}}},\ }\bibfield  {title} {\bibinfo {title} {{Relativistic Reconnection: An Efficient Source of Non-thermal Particles}},\ }\href {https://doi.org/10.1088/2041-8205/783/1/L21} {\bibfield  {journal} {\bibinfo  {journal} {\apjl}\ }\textbf {\bibinfo {volume} {783}},\ \bibinfo {eid} {L21} (\bibinfo {year} {2014})},\ \Eprint {https://arxiv.org/abs/1401.5471} {arXiv:1401.5471 [astro-ph.HE]} \BibitemShut {NoStop}%
\bibitem [{\citenamefont {{Werner}}\ and\ \citenamefont {{Uzdensky}}(2017)}]{Werner2017ApJ}%
  \BibitemOpen
  \bibfield  {author} {\bibinfo {author} {\bibfnamefont {G.~R.}\ \bibnamefont {{Werner}}}\ and\ \bibinfo {author} {\bibfnamefont {D.~A.}\ \bibnamefont {{Uzdensky}}},\ }\bibfield  {title} {\bibinfo {title} {{Nonthermal Particle Acceleration in 3D Relativistic Magnetic Reconnection in Pair Plasma}},\ }\href {https://doi.org/10.3847/2041-8213/aa7892} {\bibfield  {journal} {\bibinfo  {journal} {\apjl}\ }\textbf {\bibinfo {volume} {843}},\ \bibinfo {eid} {L27} (\bibinfo {year} {2017})},\ \Eprint {https://arxiv.org/abs/1705.05507} {arXiv:1705.05507 [astro-ph.HE]} \BibitemShut {NoStop}%
\bibitem [{\citenamefont {{Guo}}\ \emph {et~al.}(2024)\citenamefont {{Guo}}, \citenamefont {{Liu}}, \citenamefont {{Zenitani}},\ and\ \citenamefont {{Hoshino}}}]{2024SSRv..220...43G}%
  \BibitemOpen
  \bibfield  {author} {\bibinfo {author} {\bibfnamefont {F.}~\bibnamefont {{Guo}}}, \bibinfo {author} {\bibfnamefont {Y.-H.}\ \bibnamefont {{Liu}}}, \bibinfo {author} {\bibfnamefont {S.}~\bibnamefont {{Zenitani}}},\ and\ \bibinfo {author} {\bibfnamefont {M.}~\bibnamefont {{Hoshino}}},\ }\bibfield  {title} {\bibinfo {title} {{Magnetic Reconnection and Associated Particle Acceleration in High-Energy Astrophysics}},\ }\href {https://doi.org/10.1007/s11214-024-01073-2} {\bibfield  {journal} {\bibinfo  {journal} {\ssr}\ }\textbf {\bibinfo {volume} {220}},\ \bibinfo {eid} {43} (\bibinfo {year} {2024})},\ \Eprint {https://arxiv.org/abs/2309.13382} {arXiv:2309.13382 [astro-ph.HE]} \BibitemShut {NoStop}%
\bibitem [{\citenamefont {{Lyutikov}}\ and\ \citenamefont {{Blandford}}(2003)}]{2003astro.ph.12347L}%
  \BibitemOpen
  \bibfield  {author} {\bibinfo {author} {\bibfnamefont {M.}~\bibnamefont {{Lyutikov}}}\ and\ \bibinfo {author} {\bibfnamefont {R.}~\bibnamefont {{Blandford}}},\ }\bibfield  {title} {\bibinfo {title} {{Gamma Ray Bursts as Electromagnetic Outflows}},\ }\href {https://doi.org/10.48550/arXiv.astro-ph/0312347} {\bibfield  {journal} {\bibinfo  {journal} {arXiv e-prints}\ ,\ \bibinfo {eid} {astro-ph/0312347}} (\bibinfo {year} {2003})},\ \Eprint {https://arxiv.org/abs/astro-ph/0312347} {arXiv:astro-ph/0312347 [astro-ph]} \BibitemShut {NoStop}%
\bibitem [{\citenamefont {{Sironi}}\ and\ \citenamefont {{Spitkovsky}}(2011)}]{2011ApJ...741...39S}%
  \BibitemOpen
  \bibfield  {author} {\bibinfo {author} {\bibfnamefont {L.}~\bibnamefont {{Sironi}}}\ and\ \bibinfo {author} {\bibfnamefont {A.}~\bibnamefont {{Spitkovsky}}},\ }\bibfield  {title} {\bibinfo {title} {{Acceleration of Particles at the Termination Shock of a Relativistic Striped Wind}},\ }\href {https://doi.org/10.1088/0004-637X/741/1/39} {\bibfield  {journal} {\bibinfo  {journal} {\apj}\ }\textbf {\bibinfo {volume} {741}},\ \bibinfo {eid} {39} (\bibinfo {year} {2011})},\ \Eprint {https://arxiv.org/abs/1107.0977} {arXiv:1107.0977 [astro-ph.HE]} \BibitemShut {NoStop}%
\bibitem [{\citenamefont {{Bucciantini}}\ \emph {et~al.}(2011)\citenamefont {{Bucciantini}}, \citenamefont {{Arons}},\ and\ \citenamefont {{Amato}}}]{2011MNRAS.410..381B}%
  \BibitemOpen
  \bibfield  {author} {\bibinfo {author} {\bibfnamefont {N.}~\bibnamefont {{Bucciantini}}}, \bibinfo {author} {\bibfnamefont {J.}~\bibnamefont {{Arons}}},\ and\ \bibinfo {author} {\bibfnamefont {E.}~\bibnamefont {{Amato}}},\ }\bibfield  {title} {\bibinfo {title} {{Modelling spectral evolution of pulsar wind nebulae inside supernova remnants}},\ }\href {https://doi.org/10.1111/j.1365-2966.2010.17449.x} {\bibfield  {journal} {\bibinfo  {journal} {\mnras}\ }\textbf {\bibinfo {volume} {410}},\ \bibinfo {pages} {381} (\bibinfo {year} {2011})},\ \Eprint {https://arxiv.org/abs/1005.1831} {arXiv:1005.1831 [astro-ph.HE]} \BibitemShut {NoStop}%
\bibitem [{\citenamefont {{Cerutti}}\ \emph {et~al.}(2013)\citenamefont {{Cerutti}}, \citenamefont {{Werner}}, \citenamefont {{Uzdensky}},\ and\ \citenamefont {{Begelman}}}]{2013ApJ...770..147C}%
  \BibitemOpen
  \bibfield  {author} {\bibinfo {author} {\bibfnamefont {B.}~\bibnamefont {{Cerutti}}}, \bibinfo {author} {\bibfnamefont {G.~R.}\ \bibnamefont {{Werner}}}, \bibinfo {author} {\bibfnamefont {D.~A.}\ \bibnamefont {{Uzdensky}}},\ and\ \bibinfo {author} {\bibfnamefont {M.~C.}\ \bibnamefont {{Begelman}}},\ }\bibfield  {title} {\bibinfo {title} {{Simulations of Particle Acceleration beyond the Classical Synchrotron Burnoff Limit in Magnetic Reconnection: An Explanation of the Crab Flares}},\ }\href {https://doi.org/10.1088/0004-637X/770/2/147} {\bibfield  {journal} {\bibinfo  {journal} {\apj}\ }\textbf {\bibinfo {volume} {770}},\ \bibinfo {eid} {147} (\bibinfo {year} {2013})},\ \Eprint {https://arxiv.org/abs/1302.6247} {arXiv:1302.6247 [astro-ph.HE]} \BibitemShut {NoStop}%
\bibitem [{\citenamefont {{Cerutti}}\ \emph {et~al.}(2014)\citenamefont {{Cerutti}}, \citenamefont {{Werner}}, \citenamefont {{Uzdensky}},\ and\ \citenamefont {{Begelman}}}]{2014PhPl...21e6501C}%
  \BibitemOpen
  \bibfield  {author} {\bibinfo {author} {\bibfnamefont {B.}~\bibnamefont {{Cerutti}}}, \bibinfo {author} {\bibfnamefont {G.~R.}\ \bibnamefont {{Werner}}}, \bibinfo {author} {\bibfnamefont {D.~A.}\ \bibnamefont {{Uzdensky}}},\ and\ \bibinfo {author} {\bibfnamefont {M.~C.}\ \bibnamefont {{Begelman}}},\ }\bibfield  {title} {\bibinfo {title} {{Gamma-ray flares in the Crab Nebula: A case of relativistic reconnection?a)}},\ }\href {https://doi.org/10.1063/1.4872024} {\bibfield  {journal} {\bibinfo  {journal} {Physics of Plasmas}\ }\textbf {\bibinfo {volume} {21}},\ \bibinfo {eid} {056501} (\bibinfo {year} {2014})},\ \Eprint {https://arxiv.org/abs/1401.3016} {arXiv:1401.3016 [astro-ph.HE]} \BibitemShut {NoStop}%
\bibitem [{\citenamefont {{Comisso}}\ and\ \citenamefont {{Sironi}}(2019)}]{2019ApJ...886..122C}%
  \BibitemOpen
  \bibfield  {author} {\bibinfo {author} {\bibfnamefont {L.}~\bibnamefont {{Comisso}}}\ and\ \bibinfo {author} {\bibfnamefont {L.}~\bibnamefont {{Sironi}}},\ }\bibfield  {title} {\bibinfo {title} {{The Interplay of Magnetically Dominated Turbulence and Magnetic Reconnection in Producing Nonthermal Particles}},\ }\href {https://doi.org/10.3847/1538-4357/ab4c33} {\bibfield  {journal} {\bibinfo  {journal} {\apj}\ }\textbf {\bibinfo {volume} {886}},\ \bibinfo {eid} {122} (\bibinfo {year} {2019})},\ \Eprint {https://arxiv.org/abs/1909.01420} {arXiv:1909.01420 [astro-ph.HE]} \BibitemShut {NoStop}%
\bibitem [{\citenamefont {{Gupta}}\ \emph {et~al.}(2025)\citenamefont {{Gupta}}, \citenamefont {{Sridhar}},\ and\ \citenamefont {{Sironi}}}]{2025arXiv250100979G}%
  \BibitemOpen
  \bibfield  {author} {\bibinfo {author} {\bibfnamefont {S.}~\bibnamefont {{Gupta}}}, \bibinfo {author} {\bibfnamefont {N.}~\bibnamefont {{Sridhar}}},\ and\ \bibinfo {author} {\bibfnamefont {L.}~\bibnamefont {{Sironi}}},\ }\bibfield  {title} {\bibinfo {title} {{The Role of Electric Dominance for Particle Injection in Relativistic Reconnection}},\ }\href {https://doi.org/10.48550/arXiv.2501.00979} {\bibfield  {journal} {\bibinfo  {journal} {arXiv e-prints}\ ,\ \bibinfo {eid} {arXiv:2501.00979}} (\bibinfo {year} {2025})},\ \Eprint {https://arxiv.org/abs/2501.00979} {arXiv:2501.00979 [astro-ph.HE]} \BibitemShut {NoStop}%
\bibitem [{\citenamefont {{Larrabee}}\ \emph {et~al.}(2003)\citenamefont {{Larrabee}}, \citenamefont {{Lovelace}},\ and\ \citenamefont {{Romanova}}}]{2003ApJ...586...72L}%
  \BibitemOpen
  \bibfield  {author} {\bibinfo {author} {\bibfnamefont {D.~A.}\ \bibnamefont {{Larrabee}}}, \bibinfo {author} {\bibfnamefont {R.~V.~E.}\ \bibnamefont {{Lovelace}}},\ and\ \bibinfo {author} {\bibfnamefont {M.~M.}\ \bibnamefont {{Romanova}}},\ }\bibfield  {title} {\bibinfo {title} {{Lepton Acceleration by Relativistic Collisionless Magnetic Reconnection}},\ }\href {https://doi.org/10.1086/367640} {\bibfield  {journal} {\bibinfo  {journal} {\apj}\ }\textbf {\bibinfo {volume} {586}},\ \bibinfo {pages} {72} (\bibinfo {year} {2003})},\ \Eprint {https://arxiv.org/abs/astro-ph/0210045} {arXiv:astro-ph/0210045 [astro-ph]} \BibitemShut {NoStop}%
\bibitem [{\citenamefont {{Gruzinov}}(1999)}]{1999astro.ph..2288G}%
  \BibitemOpen
  \bibfield  {author} {\bibinfo {author} {\bibfnamefont {A.}~\bibnamefont {{Gruzinov}}},\ }\bibfield  {title} {\bibinfo {title} {{Stability in Force-Free Electrodynamics}},\ }\href {https://doi.org/10.48550/arXiv.astro-ph/9902288} {\bibfield  {journal} {\bibinfo  {journal} {arXiv e-prints}\ ,\ \bibinfo {eid} {astro-ph/9902288}} (\bibinfo {year} {1999})},\ \Eprint {https://arxiv.org/abs/astro-ph/9902288} {arXiv:astro-ph/9902288 [astro-ph]} \BibitemShut {NoStop}%
\bibitem [{\citenamefont {{Gruzinov}}(2006)}]{2006astro.ph..4364G}%
  \BibitemOpen
  \bibfield  {author} {\bibinfo {author} {\bibfnamefont {A.}~\bibnamefont {{Gruzinov}}},\ }\bibfield  {title} {\bibinfo {title} {{Force-Free Electrodynamics of Pulsars}},\ }\href {https://doi.org/10.48550/arXiv.astro-ph/0604364} {\bibfield  {journal} {\bibinfo  {journal} {arXiv e-prints}\ ,\ \bibinfo {eid} {astro-ph/0604364}} (\bibinfo {year} {2006})},\ \Eprint {https://arxiv.org/abs/astro-ph/0604364} {arXiv:astro-ph/0604364 [astro-ph]} \BibitemShut {NoStop}%
\bibitem [{\citenamefont {{McKinney}}(2006)}]{McKinney_2006MNRAS.367.1797M}%
  \BibitemOpen
  \bibfield  {author} {\bibinfo {author} {\bibfnamefont {J.~C.}\ \bibnamefont {{McKinney}}},\ }\bibfield  {title} {\bibinfo {title} {{General relativistic force-free electrodynamics: a new code and applications to black hole magnetospheres}},\ }\href {https://doi.org/10.1111/j.1365-2966.2006.10087.x} {\bibfield  {journal} {\bibinfo  {journal} {\mnras}\ }\textbf {\bibinfo {volume} {367}},\ \bibinfo {pages} {1797} (\bibinfo {year} {2006})},\ \Eprint {https://arxiv.org/abs/astro-ph/0601410} {arXiv:astro-ph/0601410 [astro-ph]} \BibitemShut {NoStop}%
\bibitem [{\citenamefont {{Parfrey}}\ \emph {et~al.}(2012{\natexlab{a}})\citenamefont {{Parfrey}}, \citenamefont {{Beloborodov}},\ and\ \citenamefont {{Hui}}}]{2012MNRAS.423.1416P}%
  \BibitemOpen
  \bibfield  {author} {\bibinfo {author} {\bibfnamefont {K.}~\bibnamefont {{Parfrey}}}, \bibinfo {author} {\bibfnamefont {A.~M.}\ \bibnamefont {{Beloborodov}}},\ and\ \bibinfo {author} {\bibfnamefont {L.}~\bibnamefont {{Hui}}},\ }\bibfield  {title} {\bibinfo {title} {{Introducing PHAEDRA: a new spectral code for simulations of relativistic magnetospheres}},\ }\href {https://doi.org/10.1111/j.1365-2966.2012.20969.x} {\bibfield  {journal} {\bibinfo  {journal} {\mnras}\ }\textbf {\bibinfo {volume} {423}},\ \bibinfo {pages} {1416} (\bibinfo {year} {2012}{\natexlab{a}})},\ \Eprint {https://arxiv.org/abs/1110.6669} {arXiv:1110.6669 [astro-ph.HE]} \BibitemShut {NoStop}%
\bibitem [{\citenamefont {{Paschalidis}}\ and\ \citenamefont {{Shapiro}}(2013)}]{2013PhRvD..88j4031P}%
  \BibitemOpen
  \bibfield  {author} {\bibinfo {author} {\bibfnamefont {V.}~\bibnamefont {{Paschalidis}}}\ and\ \bibinfo {author} {\bibfnamefont {S.~L.}\ \bibnamefont {{Shapiro}}},\ }\bibfield  {title} {\bibinfo {title} {{A new scheme for matching general relativistic ideal magnetohydrodynamics to its force-free limit}},\ }\href {https://doi.org/10.1103/PhysRevD.88.104031} {\bibfield  {journal} {\bibinfo  {journal} {\prd}\ }\textbf {\bibinfo {volume} {88}},\ \bibinfo {eid} {104031} (\bibinfo {year} {2013})},\ \Eprint {https://arxiv.org/abs/1310.3274} {arXiv:1310.3274 [astro-ph.HE]} \BibitemShut {NoStop}%
\bibitem [{\citenamefont {{Etienne}}\ \emph {et~al.}(2017)\citenamefont {{Etienne}}, \citenamefont {{Wan}}, \citenamefont {{Babiuc}}, \citenamefont {{McWilliams}},\ and\ \citenamefont {{Choudhary}}}]{2017CQGra..34u5001E}%
  \BibitemOpen
  \bibfield  {author} {\bibinfo {author} {\bibfnamefont {Z.~B.}\ \bibnamefont {{Etienne}}}, \bibinfo {author} {\bibfnamefont {M.-B.}\ \bibnamefont {{Wan}}}, \bibinfo {author} {\bibfnamefont {M.~C.}\ \bibnamefont {{Babiuc}}}, \bibinfo {author} {\bibfnamefont {S.~T.}\ \bibnamefont {{McWilliams}}},\ and\ \bibinfo {author} {\bibfnamefont {A.}~\bibnamefont {{Choudhary}}},\ }\bibfield  {title} {\bibinfo {title} {{GiRaFFE: an open-source general relativistic force-free electrodynamics code}},\ }\href {https://doi.org/10.1088/1361-6382/aa8ab3} {\bibfield  {journal} {\bibinfo  {journal} {Classical and Quantum Gravity}\ }\textbf {\bibinfo {volume} {34}},\ \bibinfo {eid} {215001} (\bibinfo {year} {2017})},\ \Eprint {https://arxiv.org/abs/1704.00599} {arXiv:1704.00599 [gr-qc]} \BibitemShut {NoStop}%
\bibitem [{\citenamefont {{Mahlmann}}\ \emph {et~al.}(2021)\citenamefont {{Mahlmann}}, \citenamefont {{Aloy}}, \citenamefont {{Mewes}},\ and\ \citenamefont {{Cerd{\'a}-Dur{\'a}n}}}]{Mahlmann2021}%
  \BibitemOpen
  \bibfield  {author} {\bibinfo {author} {\bibfnamefont {J.~F.}\ \bibnamefont {{Mahlmann}}}, \bibinfo {author} {\bibfnamefont {M.~A.}\ \bibnamefont {{Aloy}}}, \bibinfo {author} {\bibfnamefont {V.}~\bibnamefont {{Mewes}}},\ and\ \bibinfo {author} {\bibfnamefont {P.}~\bibnamefont {{Cerd{\'a}-Dur{\'a}n}}},\ }\bibfield  {title} {\bibinfo {title} {{Computational general relativistic force-free electrodynamics. II. Characterization of numerical diffusivity}},\ }\href {https://doi.org/10.1051/0004-6361/202038908} {\bibfield  {journal} {\bibinfo  {journal} {\aap}\ }\textbf {\bibinfo {volume} {647}},\ \bibinfo {eid} {A58} (\bibinfo {year} {2021})},\ \Eprint {https://arxiv.org/abs/2007.06599} {arXiv:2007.06599 [physics.comp-ph]} \BibitemShut {NoStop}%
\bibitem [{\citenamefont {Komissarov}(2002)}]{Komissarov2002}%
  \BibitemOpen
  \bibfield  {author} {\bibinfo {author} {\bibfnamefont {S.~S.}\ \bibnamefont {Komissarov}},\ }\bibfield  {title} {\bibinfo {title} {Time-dependent, force-free, degenerate electrodynamics},\ }\href {https://doi.org/10.1046/j.1365-8711.2002.05313.x} {\bibfield  {journal} {\bibinfo  {journal} {Monthly Notices of the Royal Astronomical Society}\ }\textbf {\bibinfo {volume} {336}},\ \bibinfo {pages} {759} (\bibinfo {year} {2002})},\ \Eprint {https://arxiv.org/abs/astro-ph/0202447} {astro-ph/0202447} \BibitemShut {NoStop}%
\bibitem [{\citenamefont {Komissarov}(2004)}]{Komissarov2004c}%
  \BibitemOpen
  \bibfield  {author} {\bibinfo {author} {\bibfnamefont {S.~S.}\ \bibnamefont {Komissarov}},\ }\bibfield  {title} {\bibinfo {title} {Electrodynamics of black hole magnetospheres},\ }\href {https://doi.org/10.1111/j.1365-2966.2004.07598.x} {\bibfield  {journal} {\bibinfo  {journal} {Monthly Notices of the Royal Astronomical Society}\ }\textbf {\bibinfo {volume} {350}},\ \bibinfo {pages} {427} (\bibinfo {year} {2004})}\BibitemShut {NoStop}%
\bibitem [{\citenamefont {{Del Zanna}}\ \emph {et~al.}(2007)\citenamefont {{Del Zanna}}, \citenamefont {{Zanotti}}, \citenamefont {{Bucciantini}},\ and\ \citenamefont {{Londrillo}}}]{2007A&A...473...11D}%
  \BibitemOpen
  \bibfield  {author} {\bibinfo {author} {\bibfnamefont {L.}~\bibnamefont {{Del Zanna}}}, \bibinfo {author} {\bibfnamefont {O.}~\bibnamefont {{Zanotti}}}, \bibinfo {author} {\bibfnamefont {N.}~\bibnamefont {{Bucciantini}}},\ and\ \bibinfo {author} {\bibfnamefont {P.}~\bibnamefont {{Londrillo}}},\ }\bibfield  {title} {\bibinfo {title} {{ECHO: a Eulerian conservative high-order scheme for general relativistic magnetohydrodynamics and magnetodynamics}},\ }\href {https://doi.org/10.1051/0004-6361:20077093} {\bibfield  {journal} {\bibinfo  {journal} {\aap}\ }\textbf {\bibinfo {volume} {473}},\ \bibinfo {pages} {11} (\bibinfo {year} {2007})},\ \Eprint {https://arxiv.org/abs/0704.3206} {arXiv:0704.3206 [astro-ph]} \BibitemShut {NoStop}%
\bibitem [{\citenamefont {{Noble}}\ \emph {et~al.}(2006)\citenamefont {{Noble}}, \citenamefont {{Gammie}}, \citenamefont {{McKinney}},\ and\ \citenamefont {{Del Zanna}}}]{2006ApJ...641..626N}%
  \BibitemOpen
  \bibfield  {author} {\bibinfo {author} {\bibfnamefont {S.~C.}\ \bibnamefont {{Noble}}}, \bibinfo {author} {\bibfnamefont {C.~F.}\ \bibnamefont {{Gammie}}}, \bibinfo {author} {\bibfnamefont {J.~C.}\ \bibnamefont {{McKinney}}},\ and\ \bibinfo {author} {\bibfnamefont {L.}~\bibnamefont {{Del Zanna}}},\ }\bibfield  {title} {\bibinfo {title} {{Primitive Variable Solvers for Conservative General Relativistic Magnetohydrodynamics}},\ }\href {https://doi.org/10.1086/500349} {\bibfield  {journal} {\bibinfo  {journal} {\apj}\ }\textbf {\bibinfo {volume} {641}},\ \bibinfo {pages} {626} (\bibinfo {year} {2006})},\ \Eprint {https://arxiv.org/abs/astro-ph/0512420} {arXiv:astro-ph/0512420 [astro-ph]} \BibitemShut {NoStop}%
\bibitem [{\citenamefont {{Siegel}}\ \emph {et~al.}(2018)\citenamefont {{Siegel}}, \citenamefont {{M{\"o}sta}}, \citenamefont {{Desai}},\ and\ \citenamefont {{Wu}}}]{2018ApJ...859...71S}%
  \BibitemOpen
  \bibfield  {author} {\bibinfo {author} {\bibfnamefont {D.~M.}\ \bibnamefont {{Siegel}}}, \bibinfo {author} {\bibfnamefont {P.}~\bibnamefont {{M{\"o}sta}}}, \bibinfo {author} {\bibfnamefont {D.}~\bibnamefont {{Desai}}},\ and\ \bibinfo {author} {\bibfnamefont {S.}~\bibnamefont {{Wu}}},\ }\bibfield  {title} {\bibinfo {title} {{Recovery Schemes for Primitive Variables in General-relativistic Magnetohydrodynamics}},\ }\href {https://doi.org/10.3847/1538-4357/aabcc5} {\bibfield  {journal} {\bibinfo  {journal} {\apj}\ }\textbf {\bibinfo {volume} {859}},\ \bibinfo {eid} {71} (\bibinfo {year} {2018})},\ \Eprint {https://arxiv.org/abs/1712.07538} {arXiv:1712.07538 [astro-ph.HE]} \BibitemShut {NoStop}%
\bibitem [{\citenamefont {Ripperda}\ \emph {et~al.}(2019{\natexlab{a}})\citenamefont {Ripperda}, \citenamefont {Bacchini}, \citenamefont {Porth}, \citenamefont {Most}, \citenamefont {Olivares}, \citenamefont {Nathanail}, \citenamefont {Rezzolla}, \citenamefont {Teunissen},\ and\ \citenamefont {Keppens}}]{Ripperda:2019lsi}%
  \BibitemOpen
  \bibfield  {author} {\bibinfo {author} {\bibfnamefont {B.}~\bibnamefont {Ripperda}}, \bibinfo {author} {\bibfnamefont {F.}~\bibnamefont {Bacchini}}, \bibinfo {author} {\bibfnamefont {O.}~\bibnamefont {Porth}}, \bibinfo {author} {\bibfnamefont {E.~R.}\ \bibnamefont {Most}}, \bibinfo {author} {\bibfnamefont {H.}~\bibnamefont {Olivares}}, \bibinfo {author} {\bibfnamefont {A.}~\bibnamefont {Nathanail}}, \bibinfo {author} {\bibfnamefont {L.}~\bibnamefont {Rezzolla}}, \bibinfo {author} {\bibfnamefont {J.}~\bibnamefont {Teunissen}},\ and\ \bibinfo {author} {\bibfnamefont {R.}~\bibnamefont {Keppens}},\ }\bibfield  {title} {\bibinfo {title} {{General relativistic resistive magnetohydrodynamics with robust primitive variable recovery for accretion disk simulations}},\ }\href {https://doi.org/10.3847/1538-4365/ab3922} {\bibfield  {journal} {\bibinfo  {journal} {Astrophys. J. Suppl.}\ }\textbf {\bibinfo {volume} {244}},\ \bibinfo {pages} {10} (\bibinfo {year} {2019}{\natexlab{a}})},\ \Eprint
  {https://arxiv.org/abs/1907.07197} {arXiv:1907.07197 [physics.comp-ph]} \BibitemShut {NoStop}%
\bibitem [{\citenamefont {{Kastaun}}\ \emph {et~al.}(2021)\citenamefont {{Kastaun}}, \citenamefont {{Kalinani}},\ and\ \citenamefont {{Ciolfi}}}]{2021PhRvD.103b3018K}%
  \BibitemOpen
  \bibfield  {author} {\bibinfo {author} {\bibfnamefont {W.}~\bibnamefont {{Kastaun}}}, \bibinfo {author} {\bibfnamefont {J.~V.}\ \bibnamefont {{Kalinani}}},\ and\ \bibinfo {author} {\bibfnamefont {R.}~\bibnamefont {{Ciolfi}}},\ }\bibfield  {title} {\bibinfo {title} {{Robust recovery of primitive variables in relativistic ideal magnetohydrodynamics}},\ }\href {https://doi.org/10.1103/PhysRevD.103.023018} {\bibfield  {journal} {\bibinfo  {journal} {\prd}\ }\textbf {\bibinfo {volume} {103}},\ \bibinfo {eid} {023018} (\bibinfo {year} {2021})},\ \Eprint {https://arxiv.org/abs/2005.01821} {arXiv:2005.01821 [gr-qc]} \BibitemShut {NoStop}%
\bibitem [{\citenamefont {{Ripperda}}\ \emph {et~al.}(2021)\citenamefont {{Ripperda}}, \citenamefont {{Mahlmann}}, \citenamefont {{Chernoglazov}}, \citenamefont {{TenBarge}}, \citenamefont {{Most}}, \citenamefont {{Juno}}, \citenamefont {{Yuan}}, \citenamefont {{Philippov}},\ and\ \citenamefont {{Bhattacharjee}}}]{Ripperda2021_FFR}%
  \BibitemOpen
  \bibfield  {author} {\bibinfo {author} {\bibfnamefont {B.}~\bibnamefont {{Ripperda}}}, \bibinfo {author} {\bibfnamefont {J.~F.}\ \bibnamefont {{Mahlmann}}}, \bibinfo {author} {\bibfnamefont {A.}~\bibnamefont {{Chernoglazov}}}, \bibinfo {author} {\bibfnamefont {J.~M.}\ \bibnamefont {{TenBarge}}}, \bibinfo {author} {\bibfnamefont {E.~R.}\ \bibnamefont {{Most}}}, \bibinfo {author} {\bibfnamefont {J.}~\bibnamefont {{Juno}}}, \bibinfo {author} {\bibfnamefont {Y.}~\bibnamefont {{Yuan}}}, \bibinfo {author} {\bibfnamefont {A.~A.}\ \bibnamefont {{Philippov}}},\ and\ \bibinfo {author} {\bibfnamefont {A.}~\bibnamefont {{Bhattacharjee}}},\ }\bibfield  {title} {\bibinfo {title} {{Weak Alfv{\'e}nic turbulence in relativistic plasmas. Part 2. current sheets and dissipation}},\ }\href {https://doi.org/10.1017/S0022377821000957} {\bibfield  {journal} {\bibinfo  {journal} {Journal of Plasma Physics}\ }\textbf {\bibinfo {volume} {87}},\ \bibinfo {eid} {905870512} (\bibinfo {year} {2021})},\ \Eprint
  {https://arxiv.org/abs/2105.01145} {arXiv:2105.01145 [astro-ph.HE]} \BibitemShut {NoStop}%
\bibitem [{\citenamefont {{Komissarov}}\ and\ \citenamefont {{Phillips}}(2025)}]{2025MNRAS.536.1268K}%
  \BibitemOpen
  \bibfield  {author} {\bibinfo {author} {\bibfnamefont {S.~S.}\ \bibnamefont {{Komissarov}}}\ and\ \bibinfo {author} {\bibfnamefont {D.}~\bibnamefont {{Phillips}}},\ }\bibfield  {title} {\bibinfo {title} {{A splitting method for numerical relativistic magnetohydrodynamics}},\ }\href {https://doi.org/10.1093/mnras/stae2620} {\bibfield  {journal} {\bibinfo  {journal} {\mnras}\ }\textbf {\bibinfo {volume} {536}},\ \bibinfo {pages} {1268} (\bibinfo {year} {2025})},\ \Eprint {https://arxiv.org/abs/2409.03637} {arXiv:2409.03637 [astro-ph.HE]} \BibitemShut {NoStop}%
\bibitem [{\citenamefont {Nathanail}\ \emph {et~al.}(2020)\citenamefont {Nathanail}, \citenamefont {Fromm}, \citenamefont {Porth}, \citenamefont {Olivares}, \citenamefont {Younsi}, \citenamefont {Mizuno},\ and\ \citenamefont {Rezzolla}}]{Nathanail_2020}%
  \BibitemOpen
  \bibfield  {author} {\bibinfo {author} {\bibfnamefont {A.}~\bibnamefont {Nathanail}}, \bibinfo {author} {\bibfnamefont {C.~M.}\ \bibnamefont {Fromm}}, \bibinfo {author} {\bibfnamefont {O.}~\bibnamefont {Porth}}, \bibinfo {author} {\bibfnamefont {H.}~\bibnamefont {Olivares}}, \bibinfo {author} {\bibfnamefont {Z.}~\bibnamefont {Younsi}}, \bibinfo {author} {\bibfnamefont {Y.}~\bibnamefont {Mizuno}},\ and\ \bibinfo {author} {\bibfnamefont {L.}~\bibnamefont {Rezzolla}},\ }\bibfield  {title} {\bibinfo {title} {Plasmoid formation in global grmhd simulations and agn flares},\ }\href {https://doi.org/10.1093/mnras/staa1165} {\bibfield  {journal} {\bibinfo  {journal} {Monthly Notices of the Royal Astronomical Society}\ }\textbf {\bibinfo {volume} {495}},\ \bibinfo {pages} {1549–1565} (\bibinfo {year} {2020})}\BibitemShut {NoStop}%
\bibitem [{\citenamefont {Chashkina}\ \emph {et~al.}(2021)\citenamefont {Chashkina}, \citenamefont {Bromberg},\ and\ \citenamefont {Levinson}}]{Chashkina_2021}%
  \BibitemOpen
  \bibfield  {author} {\bibinfo {author} {\bibfnamefont {A.}~\bibnamefont {Chashkina}}, \bibinfo {author} {\bibfnamefont {O.}~\bibnamefont {Bromberg}},\ and\ \bibinfo {author} {\bibfnamefont {A.}~\bibnamefont {Levinson}},\ }\bibfield  {title} {\bibinfo {title} {Grmhd simulations of bh activation by small scale magnetic loops: formation of striped jets and active coronae},\ }\href {https://doi.org/10.1093/mnras/stab2513} {\bibfield  {journal} {\bibinfo  {journal} {Monthly Notices of the Royal Astronomical Society}\ }\textbf {\bibinfo {volume} {508}},\ \bibinfo {pages} {1241–1252} (\bibinfo {year} {2021})}\BibitemShut {NoStop}%
\bibitem [{\citenamefont {Nathanail}\ \emph {et~al.}(2022)\citenamefont {Nathanail}, \citenamefont {Mpisketzis}, \citenamefont {Porth}, \citenamefont {Fromm},\ and\ \citenamefont {Rezzolla}}]{Nathanail_2022}%
  \BibitemOpen
  \bibfield  {author} {\bibinfo {author} {\bibfnamefont {A.}~\bibnamefont {Nathanail}}, \bibinfo {author} {\bibfnamefont {V.}~\bibnamefont {Mpisketzis}}, \bibinfo {author} {\bibfnamefont {O.}~\bibnamefont {Porth}}, \bibinfo {author} {\bibfnamefont {C.~M.}\ \bibnamefont {Fromm}},\ and\ \bibinfo {author} {\bibfnamefont {L.}~\bibnamefont {Rezzolla}},\ }\bibfield  {title} {\bibinfo {title} {Magnetic reconnection and plasmoid formation in three-dimensional accretion flows around black holes},\ }\href {https://doi.org/10.1093/mnras/stac1118} {\bibfield  {journal} {\bibinfo  {journal} {Monthly Notices of the Royal Astronomical Society}\ }\textbf {\bibinfo {volume} {513}},\ \bibinfo {pages} {4267–4277} (\bibinfo {year} {2022})}\BibitemShut {NoStop}%
\bibitem [{\citenamefont {Salas}\ \emph {et~al.}(2024)\citenamefont {Salas}, \citenamefont {Musoke}, \citenamefont {Chatterjee}, \citenamefont {Markoff}, \citenamefont {Porth}, \citenamefont {Liska},\ and\ \citenamefont {Ripperda}}]{salas2024resolutionanalysismagneticallyarrested}%
  \BibitemOpen
  \bibfield  {author} {\bibinfo {author} {\bibfnamefont {L.}~\bibnamefont {Salas}}, \bibinfo {author} {\bibfnamefont {G.}~\bibnamefont {Musoke}}, \bibinfo {author} {\bibfnamefont {K.}~\bibnamefont {Chatterjee}}, \bibinfo {author} {\bibfnamefont {S.}~\bibnamefont {Markoff}}, \bibinfo {author} {\bibfnamefont {O.}~\bibnamefont {Porth}}, \bibinfo {author} {\bibfnamefont {M.}~\bibnamefont {Liska}},\ and\ \bibinfo {author} {\bibfnamefont {B.}~\bibnamefont {Ripperda}},\ }\href {https://arxiv.org/abs/2405.00564} {\bibinfo {title} {Resolution analysis of magnetically arrested disk simulations}} (\bibinfo {year} {2024}),\ \Eprint {https://arxiv.org/abs/2405.00564} {arXiv:2405.00564 [astro-ph.HE]} \BibitemShut {NoStop}%
\bibitem [{\citenamefont {{Parfrey}}\ \emph {et~al.}(2012{\natexlab{b}})\citenamefont {{Parfrey}}, \citenamefont {{Beloborodov}},\ and\ \citenamefont {{Hui}}}]{2012ApJ...754L..12P}%
  \BibitemOpen
  \bibfield  {author} {\bibinfo {author} {\bibfnamefont {K.}~\bibnamefont {{Parfrey}}}, \bibinfo {author} {\bibfnamefont {A.~M.}\ \bibnamefont {{Beloborodov}}},\ and\ \bibinfo {author} {\bibfnamefont {L.}~\bibnamefont {{Hui}}},\ }\bibfield  {title} {\bibinfo {title} {{Twisting, Reconnecting Magnetospheres and Magnetar Spindown}},\ }\href {https://doi.org/10.1088/2041-8205/754/1/L12} {\bibfield  {journal} {\bibinfo  {journal} {\apjl}\ }\textbf {\bibinfo {volume} {754}},\ \bibinfo {eid} {L12} (\bibinfo {year} {2012}{\natexlab{b}})},\ \Eprint {https://arxiv.org/abs/1201.3635} {arXiv:1201.3635 [astro-ph.HE]} \BibitemShut {NoStop}%
\bibitem [{\citenamefont {{Tchekhovskoy}}\ \emph {et~al.}(2016)\citenamefont {{Tchekhovskoy}}, \citenamefont {{Philippov}},\ and\ \citenamefont {{Spitkovsky}}}]{2016MNRAS.457.3384T}%
  \BibitemOpen
  \bibfield  {author} {\bibinfo {author} {\bibfnamefont {A.}~\bibnamefont {{Tchekhovskoy}}}, \bibinfo {author} {\bibfnamefont {A.}~\bibnamefont {{Philippov}}},\ and\ \bibinfo {author} {\bibfnamefont {A.}~\bibnamefont {{Spitkovsky}}},\ }\bibfield  {title} {\bibinfo {title} {{Three-dimensional analytical description of magnetized winds from oblique pulsars}},\ }\href {https://doi.org/10.1093/mnras/stv2869} {\bibfield  {journal} {\bibinfo  {journal} {\mnras}\ }\textbf {\bibinfo {volume} {457}},\ \bibinfo {pages} {3384} (\bibinfo {year} {2016})},\ \Eprint {https://arxiv.org/abs/1503.01467} {arXiv:1503.01467 [astro-ph.HE]} \BibitemShut {NoStop}%
\bibitem [{\citenamefont {{Mahlmann}}\ \emph {et~al.}(2023)\citenamefont {{Mahlmann}}, \citenamefont {{Philippov}}, \citenamefont {{Mewes}}, \citenamefont {{Ripperda}}, \citenamefont {{Most}},\ and\ \citenamefont {{Sironi}}}]{2023ApJ...947L..34M}%
  \BibitemOpen
  \bibfield  {author} {\bibinfo {author} {\bibfnamefont {J.~F.}\ \bibnamefont {{Mahlmann}}}, \bibinfo {author} {\bibfnamefont {A.~A.}\ \bibnamefont {{Philippov}}}, \bibinfo {author} {\bibfnamefont {V.}~\bibnamefont {{Mewes}}}, \bibinfo {author} {\bibfnamefont {B.}~\bibnamefont {{Ripperda}}}, \bibinfo {author} {\bibfnamefont {E.~R.}\ \bibnamefont {{Most}}},\ and\ \bibinfo {author} {\bibfnamefont {L.}~\bibnamefont {{Sironi}}},\ }\bibfield  {title} {\bibinfo {title} {{Three-dimensional Dynamics of Strongly Twisted Magnetar Magnetospheres: Kinking Flux Tubes and Global Eruptions}},\ }\href {https://doi.org/10.3847/2041-8213/accada} {\bibfield  {journal} {\bibinfo  {journal} {\apjl}\ }\textbf {\bibinfo {volume} {947}},\ \bibinfo {eid} {L34} (\bibinfo {year} {2023})},\ \Eprint {https://arxiv.org/abs/2302.07273} {arXiv:2302.07273 [astro-ph.HE]} \BibitemShut {NoStop}%
\bibitem [{\citenamefont {Porth}\ \emph {et~al.}(2017)\citenamefont {Porth}, \citenamefont {Olivares}, \citenamefont {Mizuno}, \citenamefont {Younsi}, \citenamefont {Rezzolla}, \citenamefont {Moscibrodzka}, \citenamefont {Falcke},\ and\ \citenamefont {Kramer}}]{Porth:2016rfi}%
  \BibitemOpen
  \bibfield  {author} {\bibinfo {author} {\bibfnamefont {O.}~\bibnamefont {Porth}}, \bibinfo {author} {\bibfnamefont {H.}~\bibnamefont {Olivares}}, \bibinfo {author} {\bibfnamefont {Y.}~\bibnamefont {Mizuno}}, \bibinfo {author} {\bibfnamefont {Z.}~\bibnamefont {Younsi}}, \bibinfo {author} {\bibfnamefont {L.}~\bibnamefont {Rezzolla}}, \bibinfo {author} {\bibfnamefont {M.}~\bibnamefont {Moscibrodzka}}, \bibinfo {author} {\bibfnamefont {H.}~\bibnamefont {Falcke}},\ and\ \bibinfo {author} {\bibfnamefont {M.}~\bibnamefont {Kramer}},\ }\bibfield  {title} {\bibinfo {title} {The black hole accretion code},\ }\href {https://doi.org/10.1186/s40668-017-0020-2} {\bibfield  {journal} {\bibinfo  {journal} {Computational Astrophysics and Cosmology}\ }\textbf {\bibinfo {volume} {4}},\ \bibinfo {eid} {1} (\bibinfo {year} {2017})},\ \Eprint {https://arxiv.org/abs/1611.09720} {arXiv:1611.09720 [gr-qc]} \BibitemShut {NoStop}%
\bibitem [{\citenamefont {{Olivares, Hector}}\ \emph {et~al.}(2019)\citenamefont {{Olivares, Hector}}, \citenamefont {{Porth, Oliver}}, \citenamefont {{Davelaar, Jordy}}, \citenamefont {{Most, Elias R.}}, \citenamefont {{Fromm, Christian M.}}, \citenamefont {{Mizuno, Yosuke}}, \citenamefont {{Younsi, Ziri}},\ and\ \citenamefont {{Rezzolla, Luciano}}}]{Olivares2019}%
  \BibitemOpen
  \bibfield  {author} {\bibinfo {author} {\bibnamefont {{Olivares, Hector}}}, \bibinfo {author} {\bibnamefont {{Porth, Oliver}}}, \bibinfo {author} {\bibnamefont {{Davelaar, Jordy}}}, \bibinfo {author} {\bibnamefont {{Most, Elias R.}}}, \bibinfo {author} {\bibnamefont {{Fromm, Christian M.}}}, \bibinfo {author} {\bibnamefont {{Mizuno, Yosuke}}}, \bibinfo {author} {\bibnamefont {{Younsi, Ziri}}},\ and\ \bibinfo {author} {\bibnamefont {{Rezzolla, Luciano}}},\ }\bibfield  {title} {\bibinfo {title} {Constrained transport and adaptive mesh refinement in the black hole accretion code},\ }\href {https://doi.org/10.1051/0004-6361/201935559} {\bibfield  {journal} {\bibinfo  {journal} {Astronomy and Astrophysics}\ }\textbf {\bibinfo {volume} {629}},\ \bibinfo {pages} {A61} (\bibinfo {year} {2019})}\BibitemShut {NoStop}%
\bibitem [{\citenamefont {{Arnowitt}}\ \emph {et~al.}(1962)\citenamefont {{Arnowitt}}, \citenamefont {{Deser}},\ and\ \citenamefont {{Misner}}}]{1962gicr.book..227A}%
  \BibitemOpen
  \bibfield  {author} {\bibinfo {author} {\bibfnamefont {R.}~\bibnamefont {{Arnowitt}}}, \bibinfo {author} {\bibfnamefont {S.}~\bibnamefont {{Deser}}},\ and\ \bibinfo {author} {\bibfnamefont {C.~W.}\ \bibnamefont {{Misner}}},\ }\bibfield  {title} {\bibinfo {title} {{The Dynamics of General Relativity}},\ }in\ \href@noop {} {\emph {\bibinfo {booktitle} {in Gravitation: An Introduction to Current Research (Chap. 7). Edited by Louis Witten. John Wiley \& Sons Inc}}}\ (\bibinfo {year} {1962})\ p.\ \bibinfo {pages} {227}\BibitemShut {NoStop}%
\bibitem [{\citenamefont {{Blackman}}\ and\ \citenamefont {{Field}}(1993)}]{1993PhRvL..71.3481B}%
  \BibitemOpen
  \bibfield  {author} {\bibinfo {author} {\bibfnamefont {E.~G.}\ \bibnamefont {{Blackman}}}\ and\ \bibinfo {author} {\bibfnamefont {G.~B.}\ \bibnamefont {{Field}}},\ }\bibfield  {title} {\bibinfo {title} {{Ohm's law for a relativistic pair plasma}},\ }\href {https://doi.org/10.1103/PhysRevLett.71.3481} {\bibfield  {journal} {\bibinfo  {journal} {\prl}\ }\textbf {\bibinfo {volume} {71}},\ \bibinfo {pages} {3481} (\bibinfo {year} {1993})},\ \Eprint {https://arxiv.org/abs/astro-ph/9402068} {arXiv:astro-ph/9402068 [astro-ph]} \BibitemShut {NoStop}%
\bibitem [{\citenamefont {{Lyutikov}}\ and\ \citenamefont {{Uzdensky}}(2003)}]{2003ApJ...589..893L}%
  \BibitemOpen
  \bibfield  {author} {\bibinfo {author} {\bibfnamefont {M.}~\bibnamefont {{Lyutikov}}}\ and\ \bibinfo {author} {\bibfnamefont {D.}~\bibnamefont {{Uzdensky}}},\ }\bibfield  {title} {\bibinfo {title} {{Dynamics of Relativistic Reconnection}},\ }\href {https://doi.org/10.1086/374808} {\bibfield  {journal} {\bibinfo  {journal} {\apj}\ }\textbf {\bibinfo {volume} {589}},\ \bibinfo {pages} {893} (\bibinfo {year} {2003})},\ \Eprint {https://arxiv.org/abs/astro-ph/0210206} {arXiv:astro-ph/0210206 [astro-ph]} \BibitemShut {NoStop}%
\bibitem [{\citenamefont {{Komissarov}}(2007)}]{2007MNRAS.382..995K}%
  \BibitemOpen
  \bibfield  {author} {\bibinfo {author} {\bibfnamefont {S.~S.}\ \bibnamefont {{Komissarov}}},\ }\bibfield  {title} {\bibinfo {title} {{Multidimensional numerical scheme for resistive relativistic magnetohydrodynamics}},\ }\href {https://doi.org/10.1111/j.1365-2966.2007.12448.x} {\bibfield  {journal} {\bibinfo  {journal} {\mnras}\ }\textbf {\bibinfo {volume} {382}},\ \bibinfo {pages} {995} (\bibinfo {year} {2007})},\ \Eprint {https://arxiv.org/abs/0708.0323} {arXiv:0708.0323 [astro-ph]} \BibitemShut {NoStop}%
\bibitem [{\citenamefont {{Mignone}}\ \emph {et~al.}(2019)\citenamefont {{Mignone}}, \citenamefont {{Mattia}}, \citenamefont {{Bodo}},\ and\ \citenamefont {{Del Zanna}}}]{2019MNRAS.486.4252M}%
  \BibitemOpen
  \bibfield  {author} {\bibinfo {author} {\bibfnamefont {A.}~\bibnamefont {{Mignone}}}, \bibinfo {author} {\bibfnamefont {G.}~\bibnamefont {{Mattia}}}, \bibinfo {author} {\bibfnamefont {G.}~\bibnamefont {{Bodo}}},\ and\ \bibinfo {author} {\bibfnamefont {L.}~\bibnamefont {{Del Zanna}}},\ }\bibfield  {title} {\bibinfo {title} {{A constrained transport method for the solution of the resistive relativistic MHD equations}},\ }\href {https://doi.org/10.1093/mnras/stz1015} {\bibfield  {journal} {\bibinfo  {journal} {\mnras}\ }\textbf {\bibinfo {volume} {486}},\ \bibinfo {pages} {4252} (\bibinfo {year} {2019})},\ \Eprint {https://arxiv.org/abs/1904.01530} {arXiv:1904.01530 [physics.comp-ph]} \BibitemShut {NoStop}%
\bibitem [{\citenamefont {{Lyutikov}}(2003)}]{Lyutikov2003}%
  \BibitemOpen
  \bibfield  {author} {\bibinfo {author} {\bibfnamefont {M.}~\bibnamefont {{Lyutikov}}},\ }\bibfield  {title} {\bibinfo {title} {{Explosive reconnection in magnetars}},\ }\href {https://doi.org/10.1046/j.1365-2966.2003.07110.x} {\bibfield  {journal} {\bibinfo  {journal} {\mnras}\ }\textbf {\bibinfo {volume} {346}},\ \bibinfo {pages} {540} (\bibinfo {year} {2003})},\ \Eprint {https://arxiv.org/abs/astro-ph/0303384} {arXiv:astro-ph/0303384 [astro-ph]} \BibitemShut {NoStop}%
\bibitem [{\citenamefont {Gruzinov}(2007)}]{gruzinov2007dissipativestrongfieldelectrodynamics}%
  \BibitemOpen
  \bibfield  {author} {\bibinfo {author} {\bibfnamefont {A.}~\bibnamefont {Gruzinov}},\ }\href {https://arxiv.org/abs/0710.1875} {\bibinfo {title} {Dissipative strong-field electrodynamics}} (\bibinfo {year} {2007}),\ \Eprint {https://arxiv.org/abs/0710.1875} {arXiv:0710.1875 [astro-ph]} \BibitemShut {NoStop}%
\bibitem [{\citenamefont {{Li}}\ \emph {et~al.}(2012)\citenamefont {{Li}}, \citenamefont {{Spitkovsky}},\ and\ \citenamefont {{Tchekhovskoy}}}]{Li2012}%
  \BibitemOpen
  \bibfield  {author} {\bibinfo {author} {\bibfnamefont {J.}~\bibnamefont {{Li}}}, \bibinfo {author} {\bibfnamefont {A.}~\bibnamefont {{Spitkovsky}}},\ and\ \bibinfo {author} {\bibfnamefont {A.}~\bibnamefont {{Tchekhovskoy}}},\ }\bibfield  {title} {\bibinfo {title} {{Resistive Solutions for Pulsar Magnetospheres}},\ }\href {https://doi.org/10.1088/0004-637X/746/1/60} {\bibfield  {journal} {\bibinfo  {journal} {\apj}\ }\textbf {\bibinfo {volume} {746}},\ \bibinfo {eid} {60} (\bibinfo {year} {2012})},\ \Eprint {https://arxiv.org/abs/1107.0979} {arXiv:1107.0979 [astro-ph.HE]} \BibitemShut {NoStop}%
\bibitem [{\citenamefont {{Parfrey}}\ \emph {et~al.}(2017)\citenamefont {{Parfrey}}, \citenamefont {{Spitkovsky}},\ and\ \citenamefont {{Beloborodov}}}]{Parfrey2017}%
  \BibitemOpen
  \bibfield  {author} {\bibinfo {author} {\bibfnamefont {K.}~\bibnamefont {{Parfrey}}}, \bibinfo {author} {\bibfnamefont {A.}~\bibnamefont {{Spitkovsky}}},\ and\ \bibinfo {author} {\bibfnamefont {A.~M.}\ \bibnamefont {{Beloborodov}}},\ }\bibfield  {title} {\bibinfo {title} {{Simulations of the magnetospheres of accreting millisecond pulsars}},\ }\href {https://doi.org/10.1093/mnras/stx950} {\bibfield  {journal} {\bibinfo  {journal} {\mnras}\ }\textbf {\bibinfo {volume} {469}},\ \bibinfo {pages} {3656} (\bibinfo {year} {2017})},\ \Eprint {https://arxiv.org/abs/1608.04159} {arXiv:1608.04159 [astro-ph.HE]} \BibitemShut {NoStop}%
\bibitem [{\citenamefont {{TenBarge}}\ \emph {et~al.}(2021)\citenamefont {{TenBarge}}, \citenamefont {{Ripperda}}, \citenamefont {{Chernoglazov}}, \citenamefont {{Bhattacharjee}}, \citenamefont {{Mahlmann}}, \citenamefont {{Most}}, \citenamefont {{Juno}}, \citenamefont {{Yuan}},\ and\ \citenamefont {{Philippov}}}]{2021JPlPh..87f9014T}%
  \BibitemOpen
  \bibfield  {author} {\bibinfo {author} {\bibfnamefont {J.~M.}\ \bibnamefont {{TenBarge}}}, \bibinfo {author} {\bibfnamefont {B.}~\bibnamefont {{Ripperda}}}, \bibinfo {author} {\bibfnamefont {A.}~\bibnamefont {{Chernoglazov}}}, \bibinfo {author} {\bibfnamefont {A.}~\bibnamefont {{Bhattacharjee}}}, \bibinfo {author} {\bibfnamefont {J.~F.}\ \bibnamefont {{Mahlmann}}}, \bibinfo {author} {\bibfnamefont {E.~R.}\ \bibnamefont {{Most}}}, \bibinfo {author} {\bibfnamefont {J.}~\bibnamefont {{Juno}}}, \bibinfo {author} {\bibfnamefont {Y.}~\bibnamefont {{Yuan}}},\ and\ \bibinfo {author} {\bibfnamefont {A.~A.}\ \bibnamefont {{Philippov}}},\ }\bibfield  {title} {\bibinfo {title} {{Weak Alfv{\'e}nic turbulence in relativistic plasmas. Part 1. Dynamical equations and basic dynamics of interacting resonant triads}},\ }\href {https://doi.org/10.1017/S002237782100115X} {\bibfield  {journal} {\bibinfo  {journal} {Journal of Plasma Physics}\ }\textbf {\bibinfo {volume} {87}},\ \bibinfo {eid} {905870614} (\bibinfo {year}
  {2021})},\ \Eprint {https://arxiv.org/abs/2105.01146} {arXiv:2105.01146 [astro-ph.HE]} \BibitemShut {NoStop}%
\bibitem [{\citenamefont {{Harris}}(1962)}]{1962NCim...23..115H}%
  \BibitemOpen
  \bibfield  {author} {\bibinfo {author} {\bibfnamefont {E.~G.}\ \bibnamefont {{Harris}}},\ }\bibfield  {title} {\bibinfo {title} {{On a plasma sheath separating regions of oppositely directed magnetic field}},\ }\href {https://doi.org/10.1007/BF02733547} {\bibfield  {journal} {\bibinfo  {journal} {Il Nuovo Cimento}\ }\textbf {\bibinfo {volume} {23}},\ \bibinfo {pages} {115} (\bibinfo {year} {1962})}\BibitemShut {NoStop}%
\bibitem [{\citenamefont {{Angstmann}}\ \emph {et~al.}(2017)\citenamefont {{Angstmann}}, \citenamefont {{Henry}},\ and\ \citenamefont {{McGann}}}]{2017PhRvE..96d2153A}%
  \BibitemOpen
  \bibfield  {author} {\bibinfo {author} {\bibfnamefont {C.~N.}\ \bibnamefont {{Angstmann}}}, \bibinfo {author} {\bibfnamefont {B.~I.}\ \bibnamefont {{Henry}}},\ and\ \bibinfo {author} {\bibfnamefont {A.~V.}\ \bibnamefont {{McGann}}},\ }\bibfield  {title} {\bibinfo {title} {{Generalized fractional diffusion equations for subdiffusion in arbitrarily growing domains}},\ }\href {https://doi.org/10.1103/PhysRevE.96.042153} {\bibfield  {journal} {\bibinfo  {journal} {\pre}\ }\textbf {\bibinfo {volume} {96}},\ \bibinfo {eid} {042153} (\bibinfo {year} {2017})}\BibitemShut {NoStop}%
\bibitem [{\citenamefont {{Werner}}\ and\ \citenamefont {{Uzdensky}}(2021)}]{Werner2021JPlPh}%
  \BibitemOpen
  \bibfield  {author} {\bibinfo {author} {\bibfnamefont {G.~R.}\ \bibnamefont {{Werner}}}\ and\ \bibinfo {author} {\bibfnamefont {D.~A.}\ \bibnamefont {{Uzdensky}}},\ }\bibfield  {title} {\bibinfo {title} {{Reconnection and particle acceleration in three-dimensional current sheet evolution in moderately magnetized astrophysical pair plasma}},\ }\href {https://doi.org/10.1017/S0022377821001185} {\bibfield  {journal} {\bibinfo  {journal} {Journal of Plasma Physics}\ }\textbf {\bibinfo {volume} {87}},\ \bibinfo {eid} {905870613} (\bibinfo {year} {2021})},\ \Eprint {https://arxiv.org/abs/2106.02790} {arXiv:2106.02790 [astro-ph.HE]} \BibitemShut {NoStop}%
\bibitem [{\citenamefont {{Salas}}\ \emph {et~al.}(2024)\citenamefont {{Salas}}, \citenamefont {{Musoke}}, \citenamefont {{Chatterjee}}, \citenamefont {{Markoff}}, \citenamefont {{Porth}}, \citenamefont {{Liska}},\ and\ \citenamefont {{Ripperda}}}]{2024MNRAS.533..254S}%
  \BibitemOpen
  \bibfield  {author} {\bibinfo {author} {\bibfnamefont {L.~D.~S.}\ \bibnamefont {{Salas}}}, \bibinfo {author} {\bibfnamefont {G.}~\bibnamefont {{Musoke}}}, \bibinfo {author} {\bibfnamefont {K.}~\bibnamefont {{Chatterjee}}}, \bibinfo {author} {\bibfnamefont {S.~B.}\ \bibnamefont {{Markoff}}}, \bibinfo {author} {\bibfnamefont {O.}~\bibnamefont {{Porth}}}, \bibinfo {author} {\bibfnamefont {M.~T.~P.}\ \bibnamefont {{Liska}}},\ and\ \bibinfo {author} {\bibfnamefont {B.}~\bibnamefont {{Ripperda}}},\ }\bibfield  {title} {\bibinfo {title} {{Resolution analysis of magnetically arrested disc simulations}},\ }\href {https://doi.org/10.1093/mnras/stae1834} {\bibfield  {journal} {\bibinfo  {journal} {\mnras}\ }\textbf {\bibinfo {volume} {533}},\ \bibinfo {pages} {254} (\bibinfo {year} {2024})},\ \Eprint {https://arxiv.org/abs/2405.00564} {arXiv:2405.00564 [astro-ph.HE]} \BibitemShut {NoStop}%
\bibitem [{\citenamefont {{McKinney}}\ and\ \citenamefont {{Gammie}}(2004)}]{2004ApJ...611..977M}%
  \BibitemOpen
  \bibfield  {author} {\bibinfo {author} {\bibfnamefont {J.~C.}\ \bibnamefont {{McKinney}}}\ and\ \bibinfo {author} {\bibfnamefont {C.~F.}\ \bibnamefont {{Gammie}}},\ }\bibfield  {title} {\bibinfo {title} {{A Measurement of the Electromagnetic Luminosity of a Kerr Black Hole}},\ }\href {https://doi.org/10.1086/422244} {\bibfield  {journal} {\bibinfo  {journal} {\apj}\ }\textbf {\bibinfo {volume} {611}},\ \bibinfo {pages} {977} (\bibinfo {year} {2004})},\ \Eprint {https://arxiv.org/abs/astro-ph/0404512} {arXiv:astro-ph/0404512 [astro-ph]} \BibitemShut {NoStop}%
\bibitem [{\citenamefont {{Sironi}}\ \emph {et~al.}(2021)\citenamefont {{Sironi}}, \citenamefont {{Rowan}},\ and\ \citenamefont {{Narayan}}}]{2021ApJ...907L..44S}%
  \BibitemOpen
  \bibfield  {author} {\bibinfo {author} {\bibfnamefont {L.}~\bibnamefont {{Sironi}}}, \bibinfo {author} {\bibfnamefont {M.~E.}\ \bibnamefont {{Rowan}}},\ and\ \bibinfo {author} {\bibfnamefont {R.}~\bibnamefont {{Narayan}}},\ }\bibfield  {title} {\bibinfo {title} {{Reconnection-driven Particle Acceleration in Relativistic Shear Flows}},\ }\href {https://doi.org/10.3847/2041-8213/abd9bc} {\bibfield  {journal} {\bibinfo  {journal} {\apjl}\ }\textbf {\bibinfo {volume} {907}},\ \bibinfo {eid} {L44} (\bibinfo {year} {2021})},\ \Eprint {https://arxiv.org/abs/2009.11877} {arXiv:2009.11877 [astro-ph.HE]} \BibitemShut {NoStop}%
\bibitem [{\citenamefont {{Mbarek}}\ \emph {et~al.}(2022)\citenamefont {{Mbarek}}, \citenamefont {{Haggerty}}, \citenamefont {{Sironi}}, \citenamefont {{Shay}},\ and\ \citenamefont {{Caprioli}}}]{2022PhRvL.128n5101M}%
  \BibitemOpen
  \bibfield  {author} {\bibinfo {author} {\bibfnamefont {R.}~\bibnamefont {{Mbarek}}}, \bibinfo {author} {\bibfnamefont {C.}~\bibnamefont {{Haggerty}}}, \bibinfo {author} {\bibfnamefont {L.}~\bibnamefont {{Sironi}}}, \bibinfo {author} {\bibfnamefont {M.}~\bibnamefont {{Shay}}},\ and\ \bibinfo {author} {\bibfnamefont {D.}~\bibnamefont {{Caprioli}}},\ }\bibfield  {title} {\bibinfo {title} {{Relativistic Asymmetric Magnetic Reconnection}},\ }\href {https://doi.org/10.1103/PhysRevLett.128.145101} {\bibfield  {journal} {\bibinfo  {journal} {\prl}\ }\textbf {\bibinfo {volume} {128}},\ \bibinfo {eid} {145101} (\bibinfo {year} {2022})},\ \Eprint {https://arxiv.org/abs/2109.12125} {arXiv:2109.12125 [physics.plasm-ph]} \BibitemShut {NoStop}%
\bibitem [{\citenamefont {{Bransgrove}}\ \emph {et~al.}(2021)\citenamefont {{Bransgrove}}, \citenamefont {{Ripperda}},\ and\ \citenamefont {{Philippov}}}]{2021PhRvL.127e5101B}%
  \BibitemOpen
  \bibfield  {author} {\bibinfo {author} {\bibfnamefont {A.}~\bibnamefont {{Bransgrove}}}, \bibinfo {author} {\bibfnamefont {B.}~\bibnamefont {{Ripperda}}},\ and\ \bibinfo {author} {\bibfnamefont {A.}~\bibnamefont {{Philippov}}},\ }\bibfield  {title} {\bibinfo {title} {{Magnetic Hair and Reconnection in Black Hole Magnetospheres}},\ }\href {https://doi.org/10.1103/PhysRevLett.127.055101} {\bibfield  {journal} {\bibinfo  {journal} {\prl}\ }\textbf {\bibinfo {volume} {127}},\ \bibinfo {eid} {055101} (\bibinfo {year} {2021})},\ \Eprint {https://arxiv.org/abs/2109.14620} {arXiv:2109.14620 [astro-ph.HE]} \BibitemShut {NoStop}%
\bibitem [{\citenamefont {{Chernoglazov}}\ \emph {et~al.}(2021)\citenamefont {{Chernoglazov}}, \citenamefont {{Ripperda}},\ and\ \citenamefont {{Philippov}}}]{2021ApJ...923L..13C}%
  \BibitemOpen
  \bibfield  {author} {\bibinfo {author} {\bibfnamefont {A.}~\bibnamefont {{Chernoglazov}}}, \bibinfo {author} {\bibfnamefont {B.}~\bibnamefont {{Ripperda}}},\ and\ \bibinfo {author} {\bibfnamefont {A.}~\bibnamefont {{Philippov}}},\ }\bibfield  {title} {\bibinfo {title} {{Dynamic Alignment and Plasmoid Formation in Relativistic Magnetohydrodynamic Turbulence}},\ }\href {https://doi.org/10.3847/2041-8213/ac3afa} {\bibfield  {journal} {\bibinfo  {journal} {\apjl}\ }\textbf {\bibinfo {volume} {923}},\ \bibinfo {eid} {L13} (\bibinfo {year} {2021})},\ \Eprint {https://arxiv.org/abs/2111.08188} {arXiv:2111.08188 [astro-ph.HE]} \BibitemShut {NoStop}%
\bibitem [{\citenamefont {{Dong}}\ \emph {et~al.}(2022)\citenamefont {{Dong}}, \citenamefont {{Wang}}, \citenamefont {{Huang}}, \citenamefont {{Comisso}}, \citenamefont {{Sandstrom}},\ and\ \citenamefont {{Bhattacharjee}}}]{2022SciA....8N7627D}%
  \BibitemOpen
  \bibfield  {author} {\bibinfo {author} {\bibfnamefont {C.}~\bibnamefont {{Dong}}}, \bibinfo {author} {\bibfnamefont {L.}~\bibnamefont {{Wang}}}, \bibinfo {author} {\bibfnamefont {Y.-M.}\ \bibnamefont {{Huang}}}, \bibinfo {author} {\bibfnamefont {L.}~\bibnamefont {{Comisso}}}, \bibinfo {author} {\bibfnamefont {T.~A.}\ \bibnamefont {{Sandstrom}}},\ and\ \bibinfo {author} {\bibfnamefont {A.}~\bibnamefont {{Bhattacharjee}}},\ }\bibfield  {title} {\bibinfo {title} {{Reconnection-driven energy cascade in magnetohydrodynamic turbulence}},\ }\href {https://doi.org/10.1126/sciadv.abn7627} {\bibfield  {journal} {\bibinfo  {journal} {Science Advances}\ }\textbf {\bibinfo {volume} {8}},\ \bibinfo {eid} {eabn7627} (\bibinfo {year} {2022})},\ \Eprint {https://arxiv.org/abs/2210.10736} {arXiv:2210.10736 [astro-ph.SR]} \BibitemShut {NoStop}%
\bibitem [{\citenamefont {{Galishnikova}}\ \emph {et~al.}(2023)\citenamefont {{Galishnikova}}, \citenamefont {{Philippov}}, \citenamefont {{Quataert}}, \citenamefont {{Bacchini}}, \citenamefont {{Parfrey}},\ and\ \citenamefont {{Ripperda}}}]{2023PhRvL.130k5201G}%
  \BibitemOpen
  \bibfield  {author} {\bibinfo {author} {\bibfnamefont {A.}~\bibnamefont {{Galishnikova}}}, \bibinfo {author} {\bibfnamefont {A.}~\bibnamefont {{Philippov}}}, \bibinfo {author} {\bibfnamefont {E.}~\bibnamefont {{Quataert}}}, \bibinfo {author} {\bibfnamefont {F.}~\bibnamefont {{Bacchini}}}, \bibinfo {author} {\bibfnamefont {K.}~\bibnamefont {{Parfrey}}},\ and\ \bibinfo {author} {\bibfnamefont {B.}~\bibnamefont {{Ripperda}}},\ }\bibfield  {title} {\bibinfo {title} {{Collisionless Accretion onto Black Holes: Dynamics and Flares}},\ }\href {https://doi.org/10.1103/PhysRevLett.130.115201} {\bibfield  {journal} {\bibinfo  {journal} {\prl}\ }\textbf {\bibinfo {volume} {130}},\ \bibinfo {eid} {115201} (\bibinfo {year} {2023})},\ \Eprint {https://arxiv.org/abs/2212.02583} {arXiv:2212.02583 [astro-ph.HE]} \BibitemShut {NoStop}%
\bibitem [{\citenamefont {{Chandra}}\ \emph {et~al.}(2015)\citenamefont {{Chandra}}, \citenamefont {{Gammie}}, \citenamefont {{Foucart}},\ and\ \citenamefont {{Quataert}}}]{2015ApJ...810..162C}%
  \BibitemOpen
  \bibfield  {author} {\bibinfo {author} {\bibfnamefont {M.}~\bibnamefont {{Chandra}}}, \bibinfo {author} {\bibfnamefont {C.~F.}\ \bibnamefont {{Gammie}}}, \bibinfo {author} {\bibfnamefont {F.}~\bibnamefont {{Foucart}}},\ and\ \bibinfo {author} {\bibfnamefont {E.}~\bibnamefont {{Quataert}}},\ }\bibfield  {title} {\bibinfo {title} {{An Extended Magnetohydrodynamics Model for Relativistic Weakly Collisional Plasmas}},\ }\href {https://doi.org/10.1088/0004-637X/810/2/162} {\bibfield  {journal} {\bibinfo  {journal} {\apj}\ }\textbf {\bibinfo {volume} {810}},\ \bibinfo {eid} {162} (\bibinfo {year} {2015})},\ \Eprint {https://arxiv.org/abs/1508.00878} {arXiv:1508.00878 [astro-ph.HE]} \BibitemShut {NoStop}%
\bibitem [{\citenamefont {{Comisso}}\ and\ \citenamefont {{Grasso}}(2016)}]{2016PhPl...23c2111C}%
  \BibitemOpen
  \bibfield  {author} {\bibinfo {author} {\bibfnamefont {L.}~\bibnamefont {{Comisso}}}\ and\ \bibinfo {author} {\bibfnamefont {D.}~\bibnamefont {{Grasso}}},\ }\bibfield  {title} {\bibinfo {title} {{Visco-resistive plasmoid instability}},\ }\href {https://doi.org/10.1063/1.4942940} {\bibfield  {journal} {\bibinfo  {journal} {Physics of Plasmas}\ }\textbf {\bibinfo {volume} {23}},\ \bibinfo {eid} {032111} (\bibinfo {year} {2016})},\ \Eprint {https://arxiv.org/abs/1603.00090} {arXiv:1603.00090 [physics.plasm-ph]} \BibitemShut {NoStop}%
\bibitem [{\citenamefont {{Komissarov}}(2006)}]{Komissarov_2006MNRAS.367...19K}%
  \BibitemOpen
  \bibfield  {author} {\bibinfo {author} {\bibfnamefont {S.~S.}\ \bibnamefont {{Komissarov}}},\ }\bibfield  {title} {\bibinfo {title} {{Simulations of the axisymmetric magnetospheres of neutron stars}},\ }\href {https://doi.org/10.1111/j.1365-2966.2005.09932.x} {\bibfield  {journal} {\bibinfo  {journal} {\mnras}\ }\textbf {\bibinfo {volume} {367}},\ \bibinfo {pages} {19} (\bibinfo {year} {2006})},\ \Eprint {https://arxiv.org/abs/astro-ph/0510310} {arXiv:astro-ph/0510310 [astro-ph]} \BibitemShut {NoStop}%
\bibitem [{\citenamefont {Ripperda}\ \emph {et~al.}(2019{\natexlab{b}})\citenamefont {Ripperda}, \citenamefont {Porth}, \citenamefont {Sironi},\ and\ \citenamefont {Keppens}}]{Ripperda_2019}%
  \BibitemOpen
  \bibfield  {author} {\bibinfo {author} {\bibfnamefont {B.}~\bibnamefont {Ripperda}}, \bibinfo {author} {\bibfnamefont {O.}~\bibnamefont {Porth}}, \bibinfo {author} {\bibfnamefont {L.}~\bibnamefont {Sironi}},\ and\ \bibinfo {author} {\bibfnamefont {R.}~\bibnamefont {Keppens}},\ }\bibfield  {title} {\bibinfo {title} {Relativistic resistive magnetohydrodynamic reconnection and plasmoid formation in merging flux tubes},\ }\href {https://doi.org/10.1093/mnras/stz387} {\bibfield  {journal} {\bibinfo  {journal} {Monthly Notices of the Royal Astronomical Society}\ }\textbf {\bibinfo {volume} {485}},\ \bibinfo {pages} {299–314} (\bibinfo {year} {2019}{\natexlab{b}})}\BibitemShut {NoStop}%
\bibitem [{\citenamefont {{Pareschi}}\ and\ \citenamefont {{Russo}}(2010)}]{2010arXiv1009.2757P}%
  \BibitemOpen
  \bibfield  {author} {\bibinfo {author} {\bibfnamefont {L.}~\bibnamefont {{Pareschi}}}\ and\ \bibinfo {author} {\bibfnamefont {G.}~\bibnamefont {{Russo}}},\ }\bibfield  {title} {\bibinfo {title} {{Implicit-explicit Runge-Kutta schemes and applications to hyperbolic systems with relaxation}},\ }\href {https://doi.org/10.48550/arXiv.1009.2757} {\bibfield  {journal} {\bibinfo  {journal} {arXiv e-prints}\ ,\ \bibinfo {eid} {arXiv:1009.2757}} (\bibinfo {year} {2010})},\ \Eprint {https://arxiv.org/abs/1009.2757} {arXiv:1009.2757 [math.NA]} \BibitemShut {NoStop}%
\bibitem [{\citenamefont {{Palenzuela}}\ \emph {et~al.}(2009)\citenamefont {{Palenzuela}}, \citenamefont {{Lehner}}, \citenamefont {{Reula}},\ and\ \citenamefont {{Rezzolla}}}]{2009MNRAS.394.1727P}%
  \BibitemOpen
  \bibfield  {author} {\bibinfo {author} {\bibfnamefont {C.}~\bibnamefont {{Palenzuela}}}, \bibinfo {author} {\bibfnamefont {L.}~\bibnamefont {{Lehner}}}, \bibinfo {author} {\bibfnamefont {O.}~\bibnamefont {{Reula}}},\ and\ \bibinfo {author} {\bibfnamefont {L.}~\bibnamefont {{Rezzolla}}},\ }\bibfield  {title} {\bibinfo {title} {{Beyond ideal MHD: towards a more realistic modelling of relativistic astrophysical plasmas}},\ }\href {https://doi.org/10.1111/j.1365-2966.2009.14454.x} {\bibfield  {journal} {\bibinfo  {journal} {\mnras}\ }\textbf {\bibinfo {volume} {394}},\ \bibinfo {pages} {1727} (\bibinfo {year} {2009})},\ \Eprint {https://arxiv.org/abs/0810.1838} {arXiv:0810.1838 [astro-ph]} \BibitemShut {NoStop}%
\bibitem [{\citenamefont {Kelley}(1995)}]{Kelley1995}%
  \BibitemOpen
  \bibfield  {author} {\bibinfo {author} {\bibfnamefont {C.~T.}\ \bibnamefont {Kelley}},\ }\href {https://doi.org/10.1137/1.9781611970944} {\emph {\bibinfo {title} {Iterative Methods for Linear and Nonlinear Equations}}},\ \bibinfo {series} {Frontiers in Applied Mathematics}, Vol.~\bibinfo {volume} {16}\ (\bibinfo  {publisher} {Society for Industrial and Applied Mathematics},\ \bibinfo {address} {Philadelphia},\ \bibinfo {year} {1995})\BibitemShut {NoStop}%
\end{thebibliography}%

\end{document}